\documentclass[prl,amsmath,amssymb,twocolumn,superscriptaddress,10pt]{revtex4-2}

\usepackage{amsmath,amssymb,amsfonts,amsthm}
\usepackage{graphicx}
\usepackage{bm}
\usepackage{color}
\usepackage{hyperref}
\usepackage[normalem]{ulem}
\usepackage{booktabs}
\usepackage{array}
\usepackage{tikz}
\usetikzlibrary{positioning}
\usepackage{mathrsfs}

\definecolor{Blue}{rgb}{0.00,0.00,1.00}
\definecolor{Red}{rgb}{1.00,0.00,0.00}
\definecolor{Darkgreen}{rgb}{0.00,0.45,0.20}

\definecolor{Lilac}{rgb}{0.58,0.29,0.75}

\definecolor{Teal}{rgb}{0.00,0.26,0.32}

\hypersetup{colorlinks=true,linkcolor=red,citecolor=blue,urlcolor=cyan}

\newcommand{\aver}[1]{\langle{#1}\rangle}
\newcommand{\be}{\begin{equation}}
\newcommand{\ee}{\end{equation}}
\newcommand{\bea}{\begin{eqnarray}}
\newcommand{\eea}{\end{eqnarray}}
\newcommand{\Tr}{\operatorname{Tr}}

\newcommand{\FP}{\operatorname{FP}}
\newcommand{\PV}{\operatorname{PV}}
\newcommand{\Ai}{\operatorname{Ai}}
\newcommand{\Bi}{\operatorname{Bi}}
\newcommand{\Imn}{\operatorname{Im}}

\newcommand{\cA}{\mathcal A}

\newcommand{\cH}{\mathcal H}
\newcommand{\cL}{\mathcal L}

\newcommand{\cZ}{\mathcal Z}
\newcommand{\dd}{\mathrm d}

\newcommand{\ii}{\mathrm i}
\newcommand{\eps}{\varepsilon}
\newcommand{\R}{\mathbb R}

\newcommand{\bml}{\bm\lambda}

\begin{document}

\title{
Exact spectral integrability of a finite-channel random Schr\"odinger operator:\\ from fermionization of the Dyson--Riccati gas to Kontsevich integral}

\author{Yan V. Fyodorov}
\affiliation{Department of Mathematics, King's College London, Strand, London WC2R 2LS, United Kingdom}

\author{Pierre Le Doussal}
\affiliation{Laboratoire de Physique de l'\'Ecole Normale Sup\'erieure, CNRS, ENS and Universit\'e PSL, Sorbonne Universit\'e, Universit\'e Paris Cit\'e, 24 rue Lhomond, 75005 Paris, France}

\date{\today}

\begin{abstract}
 We solve exactly, for arbitrary finite channel number $N$, the Riccati dynamics of a quasi-one-dimensional random matrix Schr\"odinger operator with Hermitian white-noise potential in the $\beta=2$ unitary symmetry class.
The associated Dyson--Riccati eigenvalue gas admits an exact free-fermion representation at any sample length $L$, which in the limit $L\to \infty$ expresses the density of states and the total Lyapunov growth rate exactly through a single Airy Hankel $\tau$-function  providing multichannel generalization of $N=1$ Halperin--Thouless formulas. The $\tau-$function further admits an outgoing-contour Kontsevich matrix Airy integral representation. Its saddle-point and double-scaling analyses give, in a unified way, the finite-channel density Lifshitz tail, and the large-$N$ spectral edge. In particular, for $N\gg 1$ the sharp $N=\infty$ edge is softened  within an $N^{-4/5}$ spectral window, with its profile governed by the tritronqu\'ee solution of Painlev\'e~I. We conjecture universality of this quasi-one-dimensional scaling edge
 profile for a broad class of $\beta=2$ random band matrices.
\end{abstract}

\maketitle

The one-dimensional (1D) Schr\"odinger operator with Gaussian
white noise is one of the few disordered spectral problems
that can be treated nonperturbatively. The exact integrated density of states (IDOS), expressible in
terms of Airy functions, follows from the pioneering work
of Frisch and Lloyd and Halperin
\cite{FrischLLoyd1960,Halperin1965}. The IDOS and the Lyapunov exponent are encoded in the
imaginary and real parts, respectively, of a complex
characteristic exponent, related through the
Herbert--Jones--Thouless relation
\cite{HerbertJones1971,Thouless1972}.
We refer to this quantity as the
\emph{Halperin--Thouless exponent}.
Its real part determines the inverse localization
length~\cite{ComtetTexierTourigny2013,ComtetTourigny2017},
a fundamental scale governing transport and eigenfunction
statistics in 1D
\cite{Berezinskii1974,AbrikosovRyzhkin1978,Melnikov1980,
Kolokolov1993,LifshitsGredeskulPastur}.

Rigorous work has further developed a detailed probabilistic picture of localized eigenfunctions and spectral fluctuations~\cite{GoldsheidMolchanovPastur1977,DumazLabbe2020Localization,DumazLabbe2023Delocalized,DumazLabbe2024WhiteNoise,DumazLabbe2024Crossover}.  
Since strictly 1D systems are highly idealized, a natural question is how much of this solvability survives for an $N$-channel, matrix-valued random 
Schr\"odinger operator interpolating between the scalar problem at $N=1$ and quasi-one-dimensional multichannel localization as $N$ grows.

Multichannel localization is usually approached through the DMPK transport equation ~\cite{Dorokhov1982,MelloPereyraKumar1988}. In the unitary class the DMPK equation is known to admit a mapping to  free fermions~\cite{BeenakkerRejaei1994,Caselle1995} and is therefore exactly solvable for any length $L$ and number of channels $N$. This framework is, however, designed to address scattering and transport characteristics, such as transmission and reflection, rather than spectral properties. Recently DMPK exact solvability in the unitary class has been further extended to reflection in the presence of absorption, the density of resonance poles and the statistics of time delays \cite{FyoRes}. 
At the same time, the microscopic spectral problem for the multichannel disordered Hamiltonian $\cH_L$ of the closed system detached from the leads remains seemingly less studied \cite{footnoteB},
apart from its infinite-channel limit equivalent to the supersymmetric nonlinear sigma model~\cite{EfetovLarkin1983,Efetov1983}.

In the last decade, the study of random matrix Schr\"odinger operators has also been motivated by their relation to Hessians at equilibria of directed elastic manifolds in random potentials and in Kac--Rice calculations of their complexity~\cite{FLRT2018,FyodorovLeDoussal2020,BenArousBourgadeMcKenna2024}. 
In this context, 
Fyodorov and Le Doussal obtained the shape of the spectral density in the limit $N\to \infty$~\cite{FyodorovLeDoussal2020}.  More recently Fyodorov, Le Doussal, and Ossipov~\cite{FyodorovLeDoussalOssipov2025} studied the statistics of the spectral determinant $\det(\cH_L-E)$ for $N\gg 1$,
 using the matrix Gel'fand--Yaglom relation
\cite{GelfandYaglom1960,Forman1987,Ossipov2018}
and a matrix Riccati formulation.
This approach revealed a striking connection between the matrix Schr\"odinger problem and the problem of interacting particles in nonconfining potentials studied earlier by Allez and Dumaz~\cite{AllezDumaz2015}, who showed that a large-$N$ Dyson gas in a cubic potential undergoes a sharp transition from a confined/pinned regime to a stationary flowing regime.
 Its stationary winding current maps to the integrated density of states (IDOS), and vanishes
below the infinite-$N$ edge $E^*$. Ref.  \cite{FyodorovLeDoussalOssipov2025} further estimated the finite-$N$ leakage below $E^*$ by an Arrhenius-type single-eigenvalue activation argument, with the remaining eigenvalues treated in the $N=\infty$ mean field. The resulting barrier interpolates between $\propto N(-E)^{3/2}$ deep in the tail and $\propto N(E^*-E)^{5/4}$ near the edge, the latter implying a rounding window $|E-E^*|\propto N^{-4/5}$. This places the problem next to recent macroscopic-fluctuation approaches
to activation and currents of interacting diffusive
systems~\cite{KumarPalShpielbergPRE,KumarPalShpielbergJCP,
DandekarKrapivskyMallick,GrabschVenturelliBenichou}, and to mesoscopic
SNS junctions, where sigma-model instantons 
yield the same $5/4$ subgap leakage
exponent~\cite{OstrovskySkvortsovFeigelman2001}.

Here we solve the stationary finite-$N$ Dyson--Riccati gas exactly at $\beta=2$ by mapping it to free fermions and determine its full large-$N$ critical scaling. The closest precedents, the chiral multichannel Dirac model~\cite{GrabschTexier2016} and matrix Kesten/Riccati processes~\cite{GautieBouchaudLeDoussal2021}, relax to an equilibrium (Gibbs) state, in the former case at imaginary energy. In our case no such equilibrium exists (the gas carries a current), 
yet a Hankel $\tau$-function emerges: the solution is encoded by an Airy Hankel determinant, 
equivalently by a cubic matrix integral on a complex contour. 

Consider the spectral problem for the following $N\times N$ matrix-valued Schr\"odinger operator 
$\cH_{L}$ acting 
on $N$-component complex functions  $\psi(x), \, x\in [0,L]$ via
\be
\cH_{L}=-\partial_x^2 I_N+W(x),
\qquad \cH_{L} \psi=E\psi
\label{eq:Hmain}
\ee
 where $\psi$  satisfies the Dirichlet boundary conditions $\psi(0)=\psi(L)=0$.
 Here $I_N$ is the $N\times N$ identity matrix and we choose the $N\times N$ matrix potential $W(x)=W^\dagger(x)$ to be the standard complex Hermitian white noise:  
$\aver{W_{ij}(x)}=0,\,\aver{W_{ij}(x)W_{kl}(x')}
=2T\,\delta_{il}\delta_{jk}\delta(x-x')$ and  $T=J^2/(2N)$. Let $Y$ be the fundamental solution of \eqref{eq:Hmain}: $-Y''+WY=EY$, $Y(0)=0$, $Y'(0)=I_N$. 
The Hermitian Riccati matrix $Z:=Y'Y^{-1}$, whose real
eigenvalue windings encode the spectrum of $\cH_L$ (see below),
then evolves as
\be
\begin{aligned}
 Z'=-EI_N-Z^2+W.
\end{aligned}
\label{eq:GYRiccati}
\ee
The eigenvalues $\lambda_i$ of $Z$ obey
\be
\dd\lambda_i=
\left[
 -(E+\lambda_i^2)
 +2T\sum_{j\ne i}\frac{1}{\lambda_i-\lambda_j}
\right]\dd x
+\sqrt{2T}\,\dd B_i ,
\label{eq:DBMmain}
\ee
where the $B_i$ are independent standard Brownian motions.
This is a Dyson diffusion~\cite{Dyson1962} driven by the nonconfining cubic
potential $U_E(\lambda)=E\lambda+\frac{\lambda^3}{3}$.
The forward Fokker--Planck equation for the joint density
$P_x(\bm\lambda)$ on the ordered chamber $\mathcal C_N=\{\lambda_1<\lambda_2<\cdots<\lambda_N\}$
is most conveniently written in continuity form: $\partial_xP_x(\bm\lambda)=\cL_N^{\rm FP}P_x(\bm\lambda)=
\sum_{i=1}^N \partial_i\mathcal J_i[P_x](\bm\lambda)$, 
where \cite{footnote1}
\be
\mathcal J_i[P](\bm\lambda)
:=
T\partial_iP(\bm\lambda)
+
\left[
E+\lambda_i^2
-2T\sum_{j\ne i}\frac{1}{\lambda_i-\lambda_j}
\right]P(\bm\lambda),
\label{eq:FPcurrent}
\ee
with $\partial_i:=\frac{\partial}{\partial\lambda_i}$.
 The apparent explosions of the Riccati eigenvalues are ordinary
projective windings.  Indeed, writing 
$\lambda_i=\kappa\cot\theta_i,\, \, \theta_i\equiv\theta_i+\pi$, identifies $\lambda_i=-\infty$ and $\lambda_i=+\infty$ as the same point
of the phase circle.  The Jacobian $\left|\frac{\dd\theta_i}{\dd\lambda_i}\right|=\frac{\kappa}{\lambda_i^2+\kappa^2}
\sim\frac{\kappa}{\lambda_i^2}$~\cite{footnoteAsympt}
shows that a density smooth in $\theta_i$ induces the universal
$\lambda_i^{-2}$ tail in the corresponding one-point density. For $N>1$, when the smallest ordered eigenvalue reaches $-\infty$, it is
reinserted at $+\infty$ and becomes the largest; equivalently, the
ordered labels undergo a cyclic permutation.  This cyclic projective
gluing is the global boundary condition obeyed by the determinantal
solution constructed below, see \cite{SM}. 

Matrix oscillation theory identifies the total number of Riccati windings
with the number $\mathcal N_L(E):=\#\{\mathcal E_\alpha<E\}$
of Schr\"odinger energy levels below $E$~\cite{BaurKratz1989,Kratz1995,RiderValko2024}. Each winding contributes one level; hence the
stationary probability flux through the cyclic projective boundary is the
IDOS per unit length:
\be
\nu_N(E)
:=
\lim_{L\to\infty}\frac{1}{L}\,
\aver{\mathcal N_L(E)}
=
\mathcal J_{\mathrm{st}}^{\mathrm{wind}},
\,\,
\rho_N(E)=\partial_E\nu_N(E).
\label{eq:currentIDOS}
\ee
Here $\mathcal J_{i,\mathrm{st}}:=\mathcal J_i[P_{\mathrm{st}}]$ and \cite{footnoteCurrent}
\be 
\mathcal J_{\mathrm{st}}^{\mathrm{wind}}
:=
\int_{\lambda_2<\cdots<\lambda_N}
\mathcal J_{1,\mathrm{st}}
(-\infty,\lambda_2,\ldots,\lambda_N)
\prod_{j=2}^N\mathrm d\lambda_j .
\ee

We now exploit a special property of~\eqref{eq:FPcurrent}. Writing, in $\mathcal C_N$, $P_x(\bm \lambda)= V_N(\bm \lambda)\Psi_x(\bm \lambda)$, with $V_N(\bm\lambda)=\prod_{i<j}(\lambda_j-\lambda_i)$ 
yields the usual $\beta=2$ cancellation of the Calogero interaction. This
cancellation alone is not sufficient for a generic nonlinear drift: a divided-difference many-body term remains. Here the drift is quadratic, and
\be
\sum_{i<j}\frac{-(E+\lambda_i^2)+(E+\lambda_j^2)}{\lambda_i-\lambda_j}
=-(N-1)\sum_i\lambda_i,
\label{eq:secondcancellation}
\ee
so the residual term collapses exactly to a one-body contribution. One finds
\be
\cL_N^{\rm FP}\left[V_N(\bm \lambda)\Psi_x(\bm \lambda)\right]
=V_N(\bm \lambda)\sum_{i=1}^N\ell_N^{(i)}\Psi_x(\bm \lambda),
\label{eq:fermionizationmain}
\ee
Upon continuation between ordering chambers, $\Psi_x=P_x/V_N$ is
antisymmetric, so the resulting sum of one-body operators describes a genuinely free-fermion
dynamics for any $N$.
Here $\ell_N^{(i)}$ denotes $\ell_N$ acting on $\lambda_i$, with the adjoint pair
\be
\begin{aligned}
\ell_N&=T\partial_\lambda^2+(E+\lambda^2)\partial_\lambda+(N+1)\lambda,\\
\ell_N^\dagger&=T\partial_\lambda^2-(E+\lambda^2)\partial_\lambda+(N-1)\lambda.
\end{aligned}
\label{eq:ellpair}
\ee
This second cancellation, rather than $\beta=2$ alone, is the special structural
feature of the cubic-potential Riccati problem. The resulting fermionization is
also compatible with the global projective boundary condition: when one ordered
eigenvalue winds from $-\infty$ to $+\infty$, cyclic relabelling moves it from
first to last and produces a factor $(-1)^{N-1}$ in the Slater determinant.
The corresponding one-body orbital acquires the same factor upon continuation
through the projective point, so the two signs cancel and the physical
many-body solution is single-valued across reinjection, see \cite{SM}.

The key finite-$N$ simplification is that the adjoint one-particle
operator $\ell_N^\dagger$ preserves the $N$-dimensional polynomial sector
$\mathcal P_{N-1}=\mathrm{span}\{1,\lambda,\ldots,\lambda^{N-1}\}$.
Indeed,
\be 
\ell_N^\dagger\lambda^n
=
Tn(n-1)\lambda^{n-2}
-En\lambda^{n-1}
+(N-1-n)\lambda^{n+1}, \label{eq:monomialAction}
\ee 
so the raising term vanishes at $n=N-1$. Moreover, this action contains
no term proportional to $\lambda^n$; hence, in this monomial basis, the
finite matrix representing $\ell_N^\dagger|_{\mathcal P_{N-1}}$ has zero
diagonal and therefore
$\operatorname{Tr}_{\mathcal P_{N-1}}\ell_N^\dagger=0$.
For generic parameters, diagonalizing this finite-dimensional sector gives
$N$ polynomial modes
\be
\ell_N^\dagger p_a=\eps_a p_a,\qquad
p_a\in\mathcal P_{N-1},\qquad a=1,\ldots,N,
\label{eq:QESmain}
\ee
whose eigenvalues satisfy $\sum_{a=1}^N\eps_a=0$.
Let $v_a$ denote the corresponding physical right modes,
\be \label{eigenva} 
\ell_Nv_a=\eps_av_a .
\ee 
Their Slater determinant then obeys
\be
\left(\sum_{i=1}^N\ell_N^{(i)}\right)
\det_{a,i}[v_a(\lambda_i)]
=
\left(\sum_{a=1}^N\eps_a\right)
\det_{a,i}[v_a(\lambda_i)]
=0.
\label{eq:manybodyzeromode}
\ee
Since the $p_a$ span $\mathcal P_{N-1}$,
$\det[p_a(\lambda_i)]\propto V_N(\bm\lambda)$, so by
Eqs.~\eqref{eq:fermionizationmain} and~\eqref{eq:manybodyzeromode}
the stationary density takes the biorthogonal form 
\cite{footnoteZ} 
\be
 P_{\rm st}(\bml;E)
 =\frac{1}{\cZ_N}
 \det_{a,i}[p_a(\lambda_i)]
 \det_{a,i}[v_a(\lambda_i)].
 \label{eq:stationaryJPD}
\ee

 For each polynomial left mode $p_a$, projective continuity selects a
physical right solution $v_a$ of~\eqref{eigenva} up to normalization.
We fix the normalization by setting the Lagrange
concomitant 
\be
\mathcal B[p_a,v_a]
\equiv
T\bigl(p_av_a'-p_a'v_a\bigr)
 +(E+\lambda^2)p_av_a=1 .
\label{eq:concomitantmain}
\ee
The Lagrange identity
$p\,\ell_Nv-v\,\ell_N^\dagger p=\partial_\lambda\mathcal B[p,v]$
shows that $\mathcal B[p_a,v_a]$ is independent of $\lambda$.  With this
normalization,
$p_a(\lambda)v_a(\lambda)\sim(E+\lambda^2)^{-1}\sim\lambda^{-2}$,
and explicitly~\cite{footnote2} 
\be
v_a(\lambda)=\frac1T e^{-U_E(\lambda)/T}p_a(\lambda)\,
\FP\!\int_{-\infty}^{\lambda}
\frac{e^{U_E(\lambda')/T}}{p_a(\lambda')^2}\,\dd \lambda' .
\label{eq:vamain}
\ee
For $N=1$, $p_1=1$ and Eq.~\eqref{eq:vamain} is Halperin's
constant-current solution normalized to unit current.

Both the IDOS and the total volume growth
$\Gamma_N(E):=\lim_{L\to\infty}L^{-1}\log|\det Y(L)|=\sum_{j}\gamma_j(E)$,
i.e., the sum of the Lyapunov exponents, are encoded in the one-point Riccati density
$n(\lambda):=\aver{\sum_{j}\delta(\lambda-\lambda_j)}_{\rm st}$, normalized to $N$. 
As $\lambda_1\to-\infty$ the drift dominates the current~\eqref{eq:FPcurrent},
$\mathcal J_{1,\rm st}\sim\lambda_1^2P_{\rm st}$. In the ordered chamber
only $\lambda_1$ contributes to $n(\lambda)$ in this limit; integrating
over the remaining $N-1$ eigenvalues therefore gives
$\mathcal J_{\rm st}^{\rm wind}\sim\lambda^2n(\lambda)$. Hence,
by Eq.~\eqref{eq:currentIDOS},
$n(\lambda)\sim\nu_N(E)/\lambda^2$; by
continuity through the projective point the same tail holds as
$\lambda\to+\infty$. The IDOS is thus the coefficient of this universal tail. 
Since $\partial_x\log\det Y=\Tr Z$ between explosions, the total volume growth is the
first moment $\Gamma_N(E)=\PV\!\int_{\R}\lambda\,n(\lambda)\,\dd\lambda$, understood as a principal-value.
It is then convenient to introduce the one-sided characteristic function, with $s=-E/T^{2/3}$ 
and $q \geq 0$,
\be
\mathcal F_N(q;s):=
\int_{\mathbb R^N}
P_{\mathrm{st}}(\bm\lambda;E)\,
\exp\!\left[
-\frac{\mathrm i q}{T^{1/3}}
\sum_{j=1}^N\lambda_j
\right]\dd^N\lambda,
\label{eq:genfun}
\ee
normalized by $\mathcal F_N(0;s)=1$. Here $P_{\mathrm{st}}$ denotes the
normalized symmetric extension of the stationary density from the ordered
chamber to $\mathbb R^N$. One can show that our two
physical observables can be read off from 
its small-$q$ behavior, as 
\be
\mathcal F_N(q;s)
=
1-\frac{q}{T^{1/3}}
\left[\pi\nu_N(E)+\mathrm i\Gamma_N(E)\right]
+o(q),
\qquad q\to0_+ .
\label{eq:genfunsmallq}
\ee
Indeed, the leading small-$q$ behavior is fixed by the one-point
Riccati density,
\be
1-\mathcal F_N(q;s)
\simeq
\int_{\mathbb R}n(\lambda)
\left(
1-e^{-\mathrm i q\lambda/T^{1/3}}
\right)\dd\lambda .
\label{eq:genfuntail}
\ee
Using $n(\lambda)\sim\nu_N(E)/\lambda^2$, the even part gives
$\pi\nu_N(E)q/T^{1/3}$, since
$\int_{\mathbb R}(1-\cos u)u^{-2}\dd u=\pi$.
The equality of the projective tails at $+\infty$ and $-\infty$ makes the
odd part well defined in the principal-value sense and yields
$\mathrm i q\,\Gamma_N(E)/T^{1/3}$.

\begin{figure}[t]
\centering
\includegraphics[width=\linewidth]{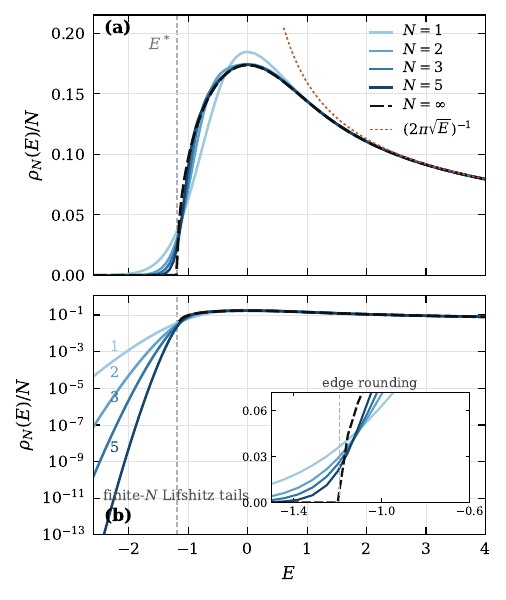}
\caption{{Finite-channel density of states and the rounding of the collective
edge, at fixed disorder $J=1$ (so $T=J^2/2N$), evaluated from the exact expressions.
 (a) $\rho_N(E)/N$ for $N=1,2,3,5$ together with
the $N=\infty$ mean-field density, obtained in closed form in
Ref.~\cite{FyodorovLeDoussal2020}; the vanishing of the associated cubic
discriminant reproduces $E^*=-\tfrac34 s_J^2$ (dashed
vertical line).
At high energy all curves approach $(2\pi\sqrt E)^{-1}$ with the correction of Eq.~\eqref{eq:highE}.
(b) Logarithmic representation: at every finite $N$ the density retains a
nonzero Lifshitz tail below the limiting edge, in agreement with
Eq.~\eqref{eq:Lifmain}, including its full $N$-dependent prefactor and first
correction, whereas the $N=\infty$ density vanishes identically for $E<E^*$.
Inset: the edge region, where increasing $N$ sharpens the rounded
finite-channel density onto the square-root edge of the 
large-$N$ solution.}}
\label{fig:DOS}
\end{figure}

Inserting the biorthogonal form~\eqref{eq:stationaryJPD} of
$P_{\mathrm{st}}$ into \eqref{eq:genfun} 
Andr\'eief's identity reduces the $N$-body integral to 
\be
\mathcal F_N(q;s)=\frac{\det I(q)}{\det I(0)},\qquad
I_{ab}(q)=\int_{\R}p_a(x)v_b(x)e^{-\ii qx}\,\dd x ,
\label{eq:mainoverlap}
\ee
where $x=\lambda/T^{1/3}$. The determinant $\det I(q)$ can be evaluated
exactly, see End Matter and \cite{SM}, giving
\be
\mathcal F_N(q;s)
=
\frac{\tau_N(s-\ii q)}{\tau_N(s)},
\qquad q>0,
\label{eq:characteristicTauRatio}
\ee
in terms of the Hankel determinant
\be
\tau_N(z)
=
\det_{0\leq j,k\leq N-1}
\left[\mathcal A_-^{(j+k)}(z)\right].
\label{eq:AiryTauDefinition}
\ee
where $\mathcal A_-(z):=\Ai(z)-\ii\Bi(z)$ and
$\mathcal A_-^{(m)}(z):=\partial_z^m\mathcal A_-(z)$. On the other hand, Eq.~\eqref{eq:characteristicTauRatio} gives
$\mathcal F_N(q;s)
=
1-\mathrm i q\,\partial_s\log\tau_N(s)+o(q)$. From \eqref{eq:genfunsmallq}
we thus obtain the exact multichannel analogue of Halperin -- Thouless complex characteristic exponent for any $N$
\be
\Xi_N(E):=\Gamma_N(E)-\mathrm i\pi\nu_N(E)
=
T^{1/3}
\left.\partial_s\log\tau_N(s)\right|_{s=-E/T^{2/3}}, 
\label{eq:KontsevichHalperinMain}
\ee
From it one extracts the eigenvalue density $\rho_N(E)=\frac{1}{\pi T^{1/3}}\left.
\Imn\,\partial_s^2\log\tau_N(s)\right|_{s=-E/T^{2/3}}$, which is plotted in Fig.~\ref{fig:DOS}). For $N=1$,
$\tau_1=\cA_-$ recovers Halperin's density~\cite{Halperin1965} and the
associated Lyapunov exponent~\cite{ComtetTourigny2017}.

Using $\mathcal A_-(z)=2e^{-\ii\pi/3}\Ai(ze^{2\pi\ii/3})$ together
with the standard contour representation of $\Ai$~\cite{DLMF};
see Eqs.~(9.2.11) and (9.5.4), one obtains
\be
\mathcal A_-(z)
=
\frac{\ii}{\pi}\int_{\mathcal C_-}
\exp\!\left(\frac{w^3}{3}-zw\right)\dd w,
\label{eq:outgoingAiryContourRepresentation}
\ee
where $\mathcal C_-:\,e^{\ii\pi/3}\infty\longrightarrow
e^{\ii\pi}\infty$ runs between two decay sectors of the cubic
exponential. Under $w=-\lambda/T^{1/3}$, this contour is the image of
$\Gamma_-:\,e^{-2\pi\ii/3}\infty\longrightarrow+\infty$.

Applying Andr\'eief's identity in reverse to the Hankel
determinant~\eqref{eq:AiryTauDefinition} and using
\eqref{eq:outgoingAiryContourRepresentation} gives a scalar-source
version of the Kontsevich matrix Airy integral~\cite{Kontsevich1992}
on the contour $\Gamma_-$~\cite{footnote3},
\be
\begin{aligned}
\mathcal Z_N^{(-)}(E)
&:=\frac{1}{N!}\int_{\Gamma_-^N}
V_N(\bm\lambda)^2
\prod_{j=1}^N e^{-NV_E(\lambda_j)}\,\dd^N\lambda\\
&=C_N T^{N^2/3}\,
\tau_N\big(-E/T^{2/3}\big),
\end{aligned}
\label{eq:overviewZ}
\ee
with $V_E(\lambda)=\frac{2}{J^2}U_E(\lambda)$ and $C_N$ independent of
$E$. 
Such cubic one-matrix integrals have a long history in the
description of random triangulated surfaces and 2D quantum
gravity~\cite{DiFrancescoGinspargZinnJustin1995}; in particular, David
emphasized the role of complex integration contours in defining the unstable
cubic model nonperturbatively~\cite{David1991}. Here 
however, the contour is selected by the projective boundary condition 
of the Dyson--Riccati dynamics. We thus obtain a distinct microscopic physical
realization of the cubic matrix integral, for which
$\Xi_N(E)=-T\partial_E\log\mathcal Z_N^{(-)}(E)$ encodes both the IDOS
$\nu_N(E)$ and the total Lyapunov growth $\Gamma_N(E)$.

Let us now obtain the large-$|E|$ asymptotics at fixed $N$. From the
contour representation,
$\tau_N\propto\int_{\mathcal C_-^N}V_N^2(\bm w)
\prod_j e^{\varphi(w_j)}\dd w_j$, with
$\varphi(w)=w^3/3-sw$ and saddles $w_\pm=\pm\sqrt{s}$.
The contour $\mathcal C_-$ selects $w_0=-\sqrt{s}$ for $E<0$
($s>0$), while for $E>0$ ($s<0$) its analytic continuation selects
$w_0=\ii\sqrt{|s|}$. Expanding about the selected saddle gives
(see End Matter)
\be
\log\tau_N\simeq N\varphi(w_0)-\tfrac{N^2}{2}\log w_0
-\tfrac{N(4N^2+1)}{48\,w_0^3}+\cdots,
\label{eq:genus}
\ee
up to $s$-independent constants.

For $E\gg T^{2/3}$, Eq.~\eqref{eq:genus} yields
\be
\frac{\rho_N(E)}N=\frac1{2\pi\sqrt E}
\Big[1-\frac{5(4N^2+1)}{32}\frac{T^2}{E^3}+\cdots\Big],
\label{eq:highE}
\ee
and $\Gamma_N=N^2T/4E+\dots$. For $-E\gg T^{2/3}$, the dominant
saddle $w_-$ gives only $\Gamma_N\simeq N\sqrt{-E}$, while the density
is generated by one-eigenvalue tunnelling to $w_+$. The barrier, cross
Vandermonde, and Gaussian fluctuations of the remaining $N-1$
eigenvalues give the finite-channel Lifshitz tail (End Matter)
\be
\begin{aligned}
\nu_N(E)\simeq{}&
\frac{2^{3N-3}}{\pi(N-1)!}
\frac{(-E)^{(3N-2)/2}}{T^{N-1}}
e^{-\frac{4(-E)^{3/2}}{3T}}\\
&\times\Big[1-\frac{21N^2-33N+17}{24}
\frac{T}{(-E)^{3/2}}+\cdots\Big].
\end{aligned}
\label{eq:Lifmain}
\ee

At fixed $J$ the exponent $8N(-E)^{3/2}/3J^2$ of the Lifshitz
factor is $O(N)$. The prefactor and its correction encode the
multichannel information, and for $N=1$ both reduce to the Airy
expansions. As $N$
grows at fixed $J$, this tail gives way to a sharp edge, whose location $E^*$
follows from the planar limit below.

Near the edge we use the monic polynomials $\pi_n$ orthogonal on $\Gamma_-$
with the complex weight $e^{-NV_E}$~\cite{DiFrancescoGinspargZinnJustin1995}.
Their three-term recurrence defines the coefficients $\beta_n,r_n$, $
\lambda\pi_n=\pi_{n+1}+\beta_n\pi_n+r_n\pi_{n-1}$, 
where their $N$-dependence is suppressed for now. Integration by parts
along $\Gamma_-$ (End Matter) gives the two exact ``string equations''
\be
\beta_n^2+r_n+r_{n+1}+E=0,\qquad
r_n(\beta_n+\beta_{n-1})=\frac{nJ^2}{2N}.
\label{eq:string}
\ee

In the planar $N\to\infty$ limit, with $t=n/N$ fixed, the recurrence
coefficients become smooth functions of $t$:
$\beta_{n,N}\to d(t)$ and $r_{n,N}\to r(t)$, so neighbouring
coefficients have the same leading limit. At the endpoint $n=N$, i.e.
$t=1$, Eq.~\eqref{eq:string} then reduces to $d^2+2r+E=0$ and
$2dr=J^2/2$, hence
$d^3+Ed+J^2/2=0$. The edge occurs when this cubic acquires a double root,
so that $3d_*^2+E^*=0$. Hence
$d_*=s_J/2$ and
$E^*=-3s_J^2/4$, with $s_J=(2J^2)^{1/3}$, in agreement with
Refs.~\cite{FyodorovLeDoussal2020,FyodorovLeDoussalOssipov2025}
and with the confined-to-flowing transition point of~\cite{AllezDumaz2015}.
Near this point $E-E^*\simeq-3(d-d_*)^2$ is quadratic rather than linear
in $d-d_*$, and the regular large-$N$ expansion about the planar solution
breaks down.

Near the double root, let $\ell$ neighbouring polynomial degrees
participate in the critical region, $n-N=O(\ell)$, and let $A$ measure
the dimensionless deviations
$\beta_{n,N}-d_*=O(s_JA)$ and
$r_{n,N}-r_*=O(s_J^2A)$.
After the leading planar terms cancel, the string equations contain three
competing effects: the quadratic critical nonlinearity, of order $A^2$;
the discreteness of $n$, through second differences, of order
$A/\ell^2$; and the degree shift $(n-N)/N$, of order $\ell/N$.
Balancing
$A^2\sim A/\ell^2\sim\ell/N$ gives
$\ell\sim N^{1/5}$ and $A\sim N^{-2/5}$. Since near the double root
$E-E^*=O(s_J^2A^2)$, the energy window is
$\delta=E-E^*\sim N^{-4/5}$.  A complementary Thouless argument gives the
same scale from $L_\delta N\delta^{3/2}\sim1$, where
$L_\delta\sim\delta^{-1/4}$ becomes of order the localization length
at the edge, hence $\delta\sim N^{-4/5}$ ~\cite{SM}.  In this double-scaling window
Eq.~\eqref{eq:string} reduces to Painlev\'e~I. As shown in End Matter, the endpoint recurrence coefficient is related exactly to
the spectral observables by
$\partial_E\Xi_N=-r_{N,N}/T$, hence
$\rho_N=(\pi T)^{-1}\Imn r_{N,N}$.
Since
$r_{N,N}-r_*=O(s_J^2N^{-2/5})$ while
$T=s_J^3/(4N)$, the critical density is of order
$N^{3/5}/s_J$. For
$E=E^*+s_J^2N^{-4/5}X$ one obtains
\be
\rho_N(E)\sim \frac{N^{3/5}}{s_J}\Phi'(X),\qquad
\Phi'(X)=-\frac{2^{4/5}}{\pi}
\Imn y_1(2^{7/5}X).
\label{eq:criticalDensity}
\ee
where $y_1(\mathfrak t)$ is the so-called {\it tritronqu\'ee} solution \cite{BleherDeano2016} of  Painlev\'e I equation $
y''(\mathfrak t)=6y(\mathfrak t)^2+\mathfrak t$. The local equation alone does not fix the solution: tronqu\'ee solutions differ
by exponentially small terms, and these carry the whole density on the
forbidden side. It is the outgoing contour that selects $y_1$: it maps onto
the contour of the critical cubic model of
Refs.~\cite{BleherDeano2013,BleherDeano2016} (End Matter). This solution
connects the density smoothly through the threshold $E=E^*$.
Namely, for  $X\to+\infty$, the scaling function returns to the sharp
large-$N$ edge:
\be \label{eq:PhiPlusAsym}
\Phi'(X)\sim\frac{2}{\pi\sqrt3}\,X^{1/2},
\ee
while for $X\to-\infty$, every algebraic term in $y_1$ is real, and the density
comes from an exponentially small imaginary part, invisible in the power
series~\cite{Kapaev2004}:
\be
\log\Phi'(X)\sim
-\frac{16}{5}\sqrt2\,3^{1/4}|X|^{5/4},
\label {eq:Phiasym}
\ee
This exponent is the action of small fluctuations around the real
algebraic branch of Painlev\'e~I.
In physical units it equals $\mathcal B/T$ with $\mathcal B=\frac45\sqrt2\,3^{1/4}s_J^{1/2}(E^*-E)^{5/4}$, exactly the single-eigenvalue activation barrier of Ref.~\cite{FyodorovLeDoussalOssipov2025} (at $\beta=2$), 
see also \cite{OstrovskySkvortsovFeigelman2001},
whose deep-tail limit $\mathcal B\to\frac43(-E)^{3/2}$ reproduces the exponent of Eq.~\eqref{eq:Lifmain}.

The critical-window scale $E-E^*\sim N^{-4/5}$ and the forbidden-side exponent $5/4$ in~\eqref{eq:Phiasym} arose earlier in studies of other classes of quasi-one-dimensional disordered models, such as random banded matrices (RBM), see \cite{CasatiMolinariIzrailev1990,FyodorovMirlin1991,FyodorovMirlin1994} for early physics insights into their spectral bulk properties and \cite{ErdosRiabov2025,Yau2026} for recent mathematical progress.
For an $M\times M$ complex-Hermitian RBM in the unitary class
($\beta=2$), of bandwidth $W$ and normalized to have spectral edge
$2$, let $\sigma_2(\xi)$ denote the localized-edge cumulative intensity,
$\sigma_2(\xi)=\lim W^{6/5}M^{-1}
\aver{\#\{\alpha:(2-\lambda_\alpha)W^{4/5}<\xi\}}$.
Silvestrov~\cite{Silvestrov1997} identified the corresponding edge
crossover diagrammatically, and Sodin~\cite{Sodin2010} subsequently
proved the $W^{-4/5}$ edge scale, the allowed-side law
$\sigma_2(\xi)\sim2\xi^{3/2}/(3\pi)$, and a forbidden-side
stretched-exponential bound on the $|\xi|^{5/4}$ scale; see also
Refs.~\cite{LiuZou2024,LiuZou2026} for the modern $\phi^3$
diagrammatic description of the edge regimes.

We conjecture the one-point correspondence
\be
\Phi(c\xi)=\mathcal A\,\sigma_2(\xi),
\qquad
c=\left(\frac{3}{64}\right)^{1/5},
\qquad
\mathcal A=48^{-1/5},
\label{eq:RBMconjectureMain}
\ee
where $\Phi(X):=\int_{-\infty}^{X}\Phi'(Y)\,\dd Y$.
The two metric factors are fixed (without fitting) by the leading
allowed-side normalization and a regularized first moment; with these
already fixed, the first nontrivial $\xi^{-7/2}$ coefficient obtained
from the unitary RBM ribbon diagrams agrees exactly with the
Painlev\'e-I expansion (Supplemental Material).  Conditional on
Eq.~\eqref{eq:RBMconjectureMain}, the forbidden tail is therefore
predicted sharply as
$-\log\sigma_2(\xi)\sim(8\sqrt3/5)|\xi|^{5/4}
=2.771281\ldots|\xi|^{5/4}$ for $\xi\to-\infty$.

Fermionization also applies to the finite-length evolution.  Assign a
factor $z$ to each projective passage of a Riccati eigenvalue, and let
$G_x^{[z]}(\lambda,\mu)$ be the corresponding forward one-body
propagator of $\ell_N$.  The counting field changes only the projective
boundary matching, not the bulk operator; see
Supplemental Material~\cite{SM}.  Matrix oscillation then gives the
finite-length Dirichlet level-counting generating function via an ordered chamber integral:
\begin{equation}
 \begin{split}
 \mathcal F_N(z;E,L)
 :=\left\langle z^{\mathcal N_L(E)}\right\rangle
 &=\lim_{\bm\mu\to+\infty}^{\rm D}
 \int_{\mathcal C_N}
 \frac{V_N(\bm\lambda)}{V_N(\bm\mu)}\\
 \quad\times
 &\underline{}\det_{i,j}\!\left[
 G_L^{[z]}(\lambda_i,\mu_j)
 \right]\,\dd^N\lambda .
 \end{split}
 \label{eq:FCS}
\end{equation}
Here $\lim^{\rm D}_{\bm\mu\to+\infty}$ is the confluent entrance
corresponding to the Dirichlet behaviour $Z(x)\simeq I_N/x$.
For $z=e^{\ii\chi}$, Eq.~\eqref{eq:FCS} is the characteristic
function of the level count.  At $z=0$ it selects trajectories with
 no current, hence
$\mathcal F_N(0;E,L)=\Pr\{\mathcal E_0(L)>E\}$.
In this zero-count sector reinjection is suppressed and the gauge
transform of $\ell_N$ reduces the problem to $N$ free fermions in the
ordinary square-integrable, self-adjoint realization of the quartic
Hamiltonian
$H_N=-T\partial_\lambda^2+(E+\lambda^2)^2/(4T)-N\lambda$.
Their filled ground-state energy controls the large-$L$ decay of this
survival probability. The same fermionic framework extends the one-channel formulation of
generalized Lyapunov exponents~\cite{FLRT2018} to the multichannel
problem; see ~\cite{SM}.

In summary, the unitary matrix white-noise Schr\"odinger operator remains
fully integrable for any length $L$ and any number of channels $N$.
For $L\to\infty$ the solution provides an exact microscopic interpolation
from Halperin--Thouless scalar theory to the collective Ricatti cubic edge.  
The departure from the standard random-matrix Airy/Tracy--Widom
soft-edge scaling~\cite{Forrester2010} reflects the fact that near the
edge spectral counting is controlled by a localization volume rather
than by the full system; see~\cite{SM}.
The same solution is simultaneously an exact finite-$N$
stationary-transport solution for logarithmically interacting particles
in a nonconfining cubic potential, i.e., for the Dyson gas of
Refs.~\cite{AllezDumaz2015,FyodorovLeDoussalOssipov2025} at $\beta=2$,
thereby connecting the spectral and depinning/activation viewpoints at
the microscopic level.
The finite-length lowest-eigenvalue distribution and its collective edge scaling are further studied in the companion paper Ref.~\cite{FLDGap}, while $L\to \infty$ analysis is extended to $\beta=1,4$ in \cite{Riccati_allbeta}.

\begin{acknowledgments}
Both authors acknowledge support from CNRS IRP- Hidings. 
YVF has been supported by grant UKRI1015, ``Non-Hermitian random
matrices: theory and applications.''  and 
PLD by ANR Grant No. ANR-23-CE30-0020-01 EDIPS.
The authors acknowledge the use of AI-based large language model assistants for help with asymptotic analysis,  numerical computations and 
 understanding the content of \cite{LiuZou2024}. All scientific arguments,
derivations, numerical results, and conclusions are the responsibility of the
authors.
\end{acknowledgments}

\section*{End Matter}

We explain the main steps connecting the stationary Riccati
solution to the Airy determinant and its asymptotic regimes.
Detailed derivations are given in the Supplemental
Material~\cite{SM}.

\noindent\textit{A. From the stationary modes to the Airy
determinant.}---
In the dimensionless variables $x=\lambda/T^{1/3}$ and
$s=-E/T^{2/3}$, the characteristic function is
$\mathcal F_N(q;s)=\det I(q)/\det I(0)$, where
$I_{ab}(q)=\int_{\mathbb R}p_a(x)v_b(x)e^{-\ii qx}\dd x$.
The left modes $p_a$ span the polynomials of degree at most
$N-1$. The key observation is that the Fourier transform
of each physical right mode is an Airy transform of its
polynomial left partner.

Indeed, writing $e_b=\eps_b/T^{1/3}$ and $z=s-\ii q$,
Fourier transformation of the right-mode equation gives
\[
\big[(z-s)(\partial_z^2-z)-(N-1)\partial_z+e_b\big]
\widetilde v_b=0, \quad q>0.
\]
The contour transform
$\mathcal T_-p=\int_{\Gamma_-}p(x)e^{-x^3/3+zx}\dd x$
maps the left polynomial equation to this equation:
integration by parts establishes the correspondence,
while multiplication by $x$ becomes $\partial_z$.
Since $\mathcal T_-1$ is proportional to
$\mathcal A_-=\Ai-\ii\Bi$, writing
$p_b(x)=\sum_{n<N}c_{bn}x^n$ gives
\[
\widetilde v_b(q)=
d_b\sum_{n<N}c_{bn}\mathcal A_-^{(n)}(s-\ii q).
\]
Here the outgoing Airy combination is fixed by the physical
boundary condition. The tails $v_b=O(x^{-N-1})$ imply
integrability, hence $\widetilde v_b(q)\to0$ as
$q\to+\infty$. Along $z=s-\ii q$, the transform built
from $\mathcal A_-$ decays, whereas the independent
one built from $\Ai+\ii\Bi$ grows.

Since multiplication by $x$ under the Fourier integral
corresponds to $\ii\partial_q=\partial_z$, the overlap is
$I_{ab}(q)=p_a(\partial_z)\widetilde v_b(q)$.
Using the preceding expression for $\widetilde v_b$ gives
\[
I_{ab}(q)=d_b\sum_{m,n<N}
c_{am}c_{bn}\mathcal A_-^{(m+n)}(s-\ii q).
\]
Thus $I(q)=C H(s-\ii q)C^{\mathsf T}D$, with
$C=(c_{an})$, $D=\operatorname{diag}(d_b)$, and
$H_{mn}(z)=\mathcal A_-^{(m+n)}(z)$.
The $q$-independent factors cancel in the determinant
ratio, yielding Eq.~\eqref{eq:characteristicTauRatio}.
Continuity supplies $q=0$ and extends the result through
degeneracies of the polynomial eigenvalues.

\medskip
\noindent\textit{B. Fixed-$N$ fluctuations and the Lifshitz
tail.}---
The cubic integral has phase
$\varphi(w)=w^3/3-sw$ and saddles $w_\pm=\pm\sqrt{s}$.
Around the selected saddle $w_0$, put
$w_j=w_0+(-2w_0)^{-1/2}h_j$.
The quadratic fluctuations form a standard GUE matrix $H$,
with weight $e^{-\Tr H^2/2}$, while the remaining cubic
term is treated perturbatively. Its first cumulant
vanishes; the second is determined by
$\langle(\Tr H^3)^2\rangle_{\rm GUE}=3N(4N^2+1)$
and gives Eq.~\eqref{eq:genus}.
For $E>0$, the outgoing contour selects
$w_0=\ii\sqrt{|s|}$, giving Eq.~\eqref{eq:highE}.

For $E<0$, all variables near $w_-=-\sqrt{s}$ give,
up to an $s$-independent phase, a real expansion.
Its logarithmic derivative contributes to $\Gamma_N$
but produces no IDOS at any algebraic order.
The density instead comes from a configuration with one
integration variable near $w_+=+\sqrt{s}$ and the
remaining $N-1$ near $w_-$.
Its exponential cost is $e^{-4s^{3/2}/3}$.
The outgoing scalar Airy asymptotics fix its relative
phase and weight, including the factor $\ii/2$ on the
positive real $s$ axis. Including the $N$ choices of
the displaced variable, its Vandermonde interaction
with the others, and the Gaussian normalizations gives
\begin{equation}
\frac{\tau_N^{(1)}}{\tau_N^{(0)}}\simeq
\frac{\ii\,2^{3N-4}}{(N-1)!}
s^{3(N-1)/2}e^{-4s^{3/2}/3}.
\label{eq:EMinstanton}
\end{equation}
Here the superscripts denote the one-variable and
dominant saddle contributions.
Using
$\nu_N=-(T^{1/3}/\pi)\Imn\partial_s\log\tau_N$
gives the leading Lifshitz tail.
The next fluctuation and Vandermonde corrections give
the bracket in Eq.~\eqref{eq:Lifmain}; their evaluation
and an independent norm-based derivation are given
in~\cite{SM}. This is a fixed-$N$ expansion, distinct
from the collective scaling below.

\medskip
\noindent\textit{C. Recurrence coefficients and the collective
edge.}---
The monic polynomials $\pi_n$ on $\Gamma_-$ are orthogonal
with respect to the bilinear product: $\langle\pi_n,\pi_m\rangle_-=h_n\delta_{nm}$, with
\[
\langle f,g\rangle_-:=
\int_{\Gamma_-}f(\lambda)g(\lambda)
e^{-NV_E(\lambda)}\,\dd\lambda
\]
without complex conjugation. Then $r_n=h_n/h_{n-1}$ and
$\mathcal Z_N^{(-)}=\prod_{n<N}h_n$.
Integrating derivatives of
$\pi_n^2e^{-NV_E}$ and
$\pi_n\pi_{n-1}e^{-NV_E}$ gives the two string
equations~\eqref{eq:string}; boundary terms vanish
in the cubic decay sectors.

To connect the recurrence problem to the physical density,
set $\theta=-E/T$. Differentiating orthogonality gives
$\partial_\theta\log h_n=\beta_n$ and
$\partial_\theta\beta_n=r_{n+1}-r_n$.
Summing telescopes, yielding
$\partial_E^2\log\mathcal Z_N^{(-)}=r_{N,N}/T^2$.
Consequently,
\[
\partial_E\Xi_N=-\frac{r_{N,N}}{T},
\qquad
\rho_N(E)=\frac{\Imn r_{N,N}}{\pi T}.
\]
Throughout, the physical size $N$, hence the weight,
is fixed while the polynomial degree $n$ varies.

Near the planar double root, use
$E=E^*+s_J^2N^{-4/5}X$ and
$\mathfrak t=2^{7/5}X+4^{1/5}(n-N)N^{-1/5}$.
The leading recurrence coefficients take the form
\begin{equation}
\begin{aligned}
\beta_{n,N}&\simeq \frac{s_J}{2}
 +\frac{s_J\,y(\mathfrak t)}{2^{1/5}N^{2/5}},\quad 
r_{n,N}&\simeq \frac{s_J^2}{4}
 -\frac{s_J^2\,y(\mathfrak t)}{2^{6/5}N^{2/5}}.
\end{aligned}
\label{eq:EMansatz}
\end{equation}
A unit shift of $n$ becomes a small shift of
$\mathfrak t$. Expanding the string equations and
eliminating the subleading corrections leaves the
second-derivative term, the quadratic nonlinearity,
and the critical displacement:
$y''=6y^2+\mathfrak t$.
At $n=N$, the exact density identity above immediately
gives Eq.~\eqref{eq:criticalDensity}.

The equation alone does not fix the exponentially small
imaginary part responsible for the forbidden-side density;
this is determined by the contour $\Gamma_-$.
The change of variables
$\lambda=\sqrt{-E}-(J/2)(-E)^{-1/4}z$
maps the integral, up to an overall prefactor, to the cubic
model with potential $z^2/2-\mathfrak u z^3$,
where $\mathfrak u=J/[12(-E)^{3/4}]$.
At $E=E^*$, the coupling reaches the critical value
of Refs.~\cite{BleherDeano2013,BleherDeano2016},
and the mapped outgoing contour is their $\alpha=1$
contour, selecting the tritronqu\'ee solution $y_1$.
Its allowed-side behavior
$y_1\sim-\ii\sqrt{\mathfrak t/6}$ gives the
square-root density. On the forbidden side, linearization
about $y_0=\sqrt{-\mathfrak t/6}$ gives
$\delta y''=12y_0\delta y$ and the decaying action
$S=(4/5)24^{1/4}(-\mathfrak t)^{5/4}$.
This yields Eq.~\eqref{eq:Phiasym}; the contour fixes
the amplitude and sign. The full double-scaling
expansion and contour identification are given
in~\cite{SM}.

\clearpage
\onecolumngrid
\setcounter{section}{0}
\setcounter{subsection}{0}
\setcounter{secnumdepth}{3}
\renewcommand{\thesection}{\Roman{section}}
\setcounter{equation}{0}
\renewcommand{\theequation}{S\arabic{equation}}
\setcounter{figure}{0}
\renewcommand{\thefigure}{S\arabic{figure}}
\setcounter{table}{0}
\renewcommand{\thetable}{S\arabic{table}}

\begin{center}
{\large\bfseries Supplemental Material for\\[0.5ex]
``Exact spectral integrability of a finite-channel random Schr\"odinger operator:\\
from fermionization of the Dyson--Riccati gas to Kontsevich integral''}
\end{center}

\makeatletter
\let\supplementsection\section
\newcommand{\supplementalcontents}{
  \supplementsection*{Contents}
  \@starttoc{stoc}
}
\renewcommand{\section}[1]{
  \supplementsection{#1}
  \addcontentsline{stoc}{section}{\protect\numberline{\thesection}#1}
}
\makeatother

\supplementalcontents
\clearpage

\section{Model, conventions, Gelfand--Yaglom formula, and matrix oscillation}
\label{sec:SMmodel}

\subsection{White-noise normalization and the matrix Riccati equation}

We first make explicit the white-noise convention used in
Eq.~\eqref{eq:Hmain} of the Main Text.  We write
\be
\cH=-\partial_x^2I_N+W(x),
\qquad
W(x)\dd x=\frac{J}{\sqrt N}\,\dd B(x),
\label{eq:SMH}
\ee
where $B$ is standard Hermitian Brownian motion normalized by
\be
\mathbb E[\dd B_{ij}(x)\dd B_{kl}(x)]
=
\delta_{il}\delta_{jk}\,\dd x .
\label{eq:SMBrownianCov}
\ee
Equivalently, the Hermitian Brownian increment may be parametrized as
\be
\dd B_{ii}=\dd X_{ii},
\qquad
\dd B_{ij}
=
\frac{\dd X_{ij}+\ii\,\dd Y_{ij}}{\sqrt{2}},
\qquad
\dd B_{ji}
=
\dd B_{ij}^{*}
=
\frac{\dd X_{ij}-\ii\,\dd Y_{ij}}{\sqrt{2}},
\quad i<j,
\label{eq:HermitianBrownian}
\ee
where all $\dd X_{ii}$, $\dd X_{ij}$, and $\dd Y_{ij}$ are independent
real Brownian increments with variance $\dd x$.  This makes explicit
that $B$ is Hermitian Brownian motion in the unitary ($\beta=2$)
symmetry class.  Since $W\dd x=(J/\sqrt N)\dd B$, the covariance is
\[
\mathbb E[W_{ij}(x)W_{kl}(x')]
=
\frac{J^2}{N}\,
\delta_{il}\delta_{jk}\delta(x-x')
=
2T\,\delta_{il}\delta_{jk}\delta(x-x'),
\]
and hence
\be
T=\frac{J^2}{2N},
\label{eq:SMTconvention}
\ee
as used throughout the Main Text. Let $Y$ be the fundamental matrix solving the Schr\"odinger equation
with
\[
Y(0)=0,
\qquad
Y'(0)=I_N .
\]
The corresponding first-order Hamiltonian system is formally
\be
\partial_x
\binom{Y}{Y'}
=
\begin{pmatrix}
0&I_N\\
W-EI_N&0
\end{pmatrix}
\binom{Y}{Y'} .
\label{eq:SMfirstorder}
\ee
For white noise, $W$ is the distributional derivative of the Brownian
primitive, and Eq.~\eqref{eq:SMfirstorder} is understood through its
integrated stochastic form.  Whenever $Y$ is invertible, the Riccati
matrix
\be
Z:=Y'Y^{-1}
\label{eq:SMZdef}
\ee
satisfies
\be
\dd Z
=
-(EI_N+Z^2)\dd x
+
\frac{J}{\sqrt N}\,\dd B .
\label{eq:SMmatrixRiccati}
\ee
This is the stochastic form of Eq.~\eqref{eq:GYRiccati} of the Main
Text.  Unitary invariance then gives the interacting eigenvalue
diffusion Eq.~\eqref{eq:DBMmain}, i.e. the Dyson Brownian-motion
reduction~\cite{Dyson1962}.  With the convention above, the scalar
$N=1$ noise covariance is $2T\delta(x-x')$.

\subsection{Gelfand--Yaglom and determinant growth}

For finite $L$ with Dirichlet boundary conditions, the
Gel'fand--Yaglom formula and its matrix generalizations
\cite{GelfandYaglom1960,Forman1987,Ossipov2018} give
\be
\frac{\det(\cH_L-E)}
{\det(-\partial_x^2I_N-E)}
=
\frac{\det Y(L;E)}
{\det Y_0(L;E)},
\label{eq:SMGYratio}
\ee
where $Y_0$ is the fundamental matrix of the free reference problem,
with $Y_0(0)=0$ and $Y_0'(0)=I_N$, and the functional determinants on
the two sides are understood with the same regularization.  In
particular, for the Dirichlet problem,
\be
\det Y(L;E)=0
\quad\Longleftrightarrow\quad
E\in\operatorname{Spec}(\cH_L).
\label{eq:SMGYzeros}
\ee
Thus singularities of $Z=Y'Y^{-1}$ occur when $Y$ becomes singular,
namely at conjugate points associated with the Dirichlet spectral
problem. Between such singularities,
\be
\partial_x\log\det Y(x)
=
\Tr Z(x)
=
\sum_{i=1}^N\lambda_i(x).
\label{eq:SMlogdet}
\ee
Hence the same Riccati process controls both spectral counting and
longitudinal volume growth.  In particular,
Eq.~\eqref{eq:SMlogdet} is the identity used in the Main Text to express
the total growth rate
$\Gamma_N(E)=\lim_{L\to\infty}L^{-1}\log|\det Y(L)|$,
equivalently the sum of the Lyapunov exponents, in terms of the
stationary one-point density.

\subsection{Matrix oscillation and the IDOS}

For a smooth Hermitian matrix potential, the matrix Sturm oscillation
theorem states that the number of Dirichlet eigenvalues of $\cH_L$
below $E$ is equal, up to the convention for endpoint events, to the
number of conjugate points encountered by the fundamental solution
$Y(x;E)$ on $0<x\leq L$, counted with multiplicity
\cite{BaurKratz1989,Kratz1995}.  A conjugate point is a point where
$\det Y(x;E)=0$; equivalently, the Riccati matrix
$Z=Y'Y^{-1}$ becomes singular and one or more of its eigenvalues pass
through infinity.  In the projective angle representation introduced below
in Sec.~\ref{sec:SMprojective}, these singular passages become ordinary
phase windings. The white-noise problem is obtained by
smooth approximation of the Brownian primitive, with the limiting
first-order dynamics described by Eq.~\eqref{eq:SMfirstorder}.

Let $K_L(E)$ denote the total projective winding count of the Riccati
eigenvalues on $[0,L]$.  Up to a convention-dependent endpoint term,
matrix oscillation gives
\be
\mathcal N_L(E)
=
K_L(E)+O(1),
\label{eq:SMfiniteOscillation}
\ee
where $\mathcal N_L(E)$ is the Dirichlet level-counting function defined
in the Main Text.  The $O(1)$ contribution depends only on the endpoint
convention and therefore disappears after division by $L$.  With the
Dirichlet entrance convention used explicitly in the finite-length
construction below, the convention can moreover be fixed so that
$K_L(E)=\mathcal N_L(E)$ exactly. Consequently,
\be
\nu_N(E)
=
\lim_{L\to\infty}
\frac{\mathbb E\,\mathcal N_L(E)}{L}
=
\lim_{L\to\infty}
\frac{\mathbb E\,K_L(E)}{L}.
\label{eq:SMoscillation}
\ee
This establishes the spectral-counting--winding part of
Eq.~\eqref{eq:currentIDOS} of the Main Text.  In
Sec.~\ref{sec:SMprojective} below we identify the limiting winding rate
with the stationary flux through the projective boundary, thereby
completing the current interpretation of Eq.~\eqref{eq:currentIDOS}.

For $N>1$, individual Riccati labels are cyclically relabelled when a
particle winds through the projective point and are therefore not
globally distinguished.  The total winding count and all spectral
observables are symmetric under this relabelling.

\section{Projective compactification and cyclic boundary condition}
\label{sec:SMprojective}

We spell out the global boundary condition underlying the Riccati
dynamics and, in particular, its stationary determinantal solution.
To make passages of the Riccati eigenvalues through infinity into
ordinary phase windings, we introduce the Pr\"ufer angle $\theta$ by
setting the inessential scale $\kappa=1$,
\be
\lambda=\cot\theta,\qquad 0<\theta<\pi,\qquad
\theta\equiv\theta+\pi .
\label{eq:SMprufer}
\ee
Thus $\lambda=+\infty$ and $\lambda=-\infty$ correspond respectively to
$\theta=0$ and $\theta=\pi$.  Since $\theta$ is defined modulo $\pi$,
these two endpoints are identified: the Riccati line is thereby
compactified to the real projective line $\mathbb{RP}^1$, which is
topologically a circle. Moreover
\be
\left|\frac{\dd\theta}{\dd\lambda}\right|
=\sin^2\theta=\frac{1}{1+\lambda^2},
\label{eq:SMpruferJacobian}
\ee
so a density regular in $\theta$ has the universal $\lambda^{-2}$ tail
on the Riccati line. It is useful to see directly why the coefficient of this tail is the
winding current.  Applying It\^o's formula to $\lambda_i=\cot\theta_i$ gives
\be
\begin{aligned}
\dd\theta_i
&=
A_i(\bm\theta)\,\dd x
-\sqrt{2T}\sin^2\theta_i\,\dd B_i,\\
A_i(\bm\theta)
&=
\cos^2\theta_i+E\sin^2\theta_i
+2T\sin^3\theta_i
\left(
\cos\theta_i+
\sum_{j\ne i}
\frac{\sin\theta_j}{\sin(\theta_i-\theta_j)}
\right),
\qquad
A_i\longrightarrow1
\quad(\theta_i\to0,\pi).
\end{aligned}
\label{eq:SMthetaSDE}
\ee
If $\widehat P_{\rm st}(\bm\theta)$ denotes the normalized stationary
density with respect to $\dd^N\theta$ for
$(\theta_1,\ldots,\theta_N)$, extended symmetrically from the ordered
chamber to all permutations of the particle labels, the corresponding
probability current in the $\theta_i$ direction is
\be
\widehat{\mathcal J}_i
=
A_i\widehat P_{\rm st}
-
T\partial_{\theta_i}
\left[\sin^4\theta_i\,\widehat P_{\rm st}\right].
\label{eq:SMthetaCurrent}
\ee
Define the one-point phase density and the associated one-point current by
\be
\widehat n(\theta)
:=
\left\langle\sum_{i=1}^N\delta(\theta-\theta_i)\right\rangle_{\rm st},
\qquad
j(\theta)
:=
\sum_{i=1}^N
\int \widehat{\mathcal J}_i(\bm\theta)\,
\delta(\theta-\theta_i)\,
\prod_{k\ne i}\dd\theta_k .
\label{eq:SMphaseDensityCurrent}
\ee
At the projective point $\theta=0\equiv\pi$, the diffusion coefficient
$\sin^4\theta$ vanishes while $A_i\to1$.  Hence the boundary current is
equal to the boundary one-point phase density.  Stationarity together
with reinjection through the identified endpoints gives
\be
j(0)=\widehat n(0)
=
j(\pi)=\widehat n(\pi)
=
\mathcal J_{\rm st}^{\rm wind},
\label{eq:SMboundaryCurrent}
\ee
where $\mathcal J_{\rm st}^{\rm wind}$ is the projective winding current
defined in the Main Text below Eq.~\eqref{eq:currentIDOS}.

The line and phase one-point densities are related by
$n(\lambda)|\dd\lambda|=\widehat n(\theta)\dd\theta$, or
$n(\lambda)=\widehat n(\theta)|\dd\theta/\dd\lambda|$.
Using Eq.~\eqref{eq:SMpruferJacobian} and
$\mathcal J_{\rm st}^{\rm wind}=\nu_N(E)$ from
Eq.~\eqref{eq:currentIDOS}, we therefore obtain
\be
n(\lambda)\sim
\frac{\mathcal J_{\rm st}^{\rm wind}}{\lambda^2}
=
\frac{\nu_N(E)}{\lambda^2},
\qquad |\lambda|\to\infty .
\label{eq:SMprojectiveTail}
\ee
Thus the universal $\lambda^{-2}$ tail is simply the line-coordinate
image of a regular current-carrying density at the identified projective
endpoint.

With the ordering convention of the main text,
$\lambda_1<\cdots<\lambda_N$, the corresponding phase chamber is
\be
0<\theta_N<\cdots<\theta_1<\pi .
\label{eq:SMphaseChamber}
\ee
When the smallest eigenvalue $\lambda_1$ reaches $-\infty$, $\theta_1$ reaches
$\pi$ and is continued through the projective point to $\theta=0$,
corresponding to reinjection at $+\infty$. Restoring the ordered
eigenvalue labels gives the cyclic permutation
\be
(\lambda_1,\lambda_2,\ldots,\lambda_N)
\longrightarrow
(\lambda_2,\ldots,\lambda_N,+\infty).
\label{eq:SMcyclicReinjection}
\ee
This cyclic gluing, rather than separate boundary conditions at
$\lambda=\pm\infty$, is part of the global definition of the Riccati
dynamics.

The same geometry is visible in the Vandermonde.  With
$V_N(\bm\lambda)=\prod_{i<j}(\lambda_j-\lambda_i)$,
\be
V_N(\bm\lambda)
=
D_N(\bm\theta)
\prod_{i=1}^N\sin^{-(N-1)}\theta_i,
\qquad
D_N(\bm\theta)
:=
\prod_{i<j}\sin(\theta_i-\theta_j).
\label{eq:SMcircularVandermonde}
\ee

At the projective boundary,
\be
D_N(\pi,\theta_2,\ldots,\theta_N)
=
D_N(\theta_2,\ldots,\theta_N,0),
\label{eq:SMcircularGluing}
\ee
so the circular Vandermonde is continuous under reinjection followed by
the cyclic relabelling~\eqref{eq:SMcyclicReinjection}.  The remaining
factor in Eq.~\eqref{eq:SMcircularVandermonde} matches as well: just
before and after reinjection the escaping angle is respectively
$\pi-\varepsilon$ and $\varepsilon$, and
$\sin(\pi-\varepsilon)=\sin\varepsilon$, while the product over all
angles is invariant under cyclic relabelling.  Thus the full
Vandermonde representation has the same boundary asymptotics on the two
sides of the projective identification.

We now show that the biorthogonal stationary solution
Eq.~\eqref{eq:stationaryJPD} obeys this gluing.  Every polynomial
eigenmode $p_a$ has degree exactly $N-1$: otherwise its highest-degree
term would be raised by $\ell_N^\dagger$, according to
Eq.~\eqref{eq:monomialAction}.  The unit-concomitant condition
Eq.~\eqref{eq:concomitantmain} then implies
$v_a(\lambda)=O(\lambda^{-N-1})$ as $\lambda\to\pm\infty$.  Define the
compactified left and right orbitals
\be
\hat p_a(\theta)
=
\sin^{N-1}\!\theta\,p_a(\cot\theta),
\qquad
\hat v_a(\theta)
=
\sin^{-N-1}\!\theta\,v_a(\cot\theta).
\label{eq:SMcompactOrbitals}
\ee
Both are regular at the projective point.  Since
$\cot(\theta+\pi)=\cot\theta$ and
$\sin(\theta+\pi)=-\sin\theta$, their continuation once around the
projective circle obeys
\be
\hat p_a(\theta+\pi)=(-1)^{N-1}\hat p_a(\theta),
\qquad
\hat v_a(\theta+\pi)=(-1)^{N-1}\hat v_a(\theta).
\label{eq:SMfermionicTwist}
\ee

The compactification is also natural directly at the level of the
stationary measure.  Using
$|\dd\lambda|=\dd\theta/\sin^2\theta$ in
Eq.~\eqref{eq:stationaryJPD}, all powers of $\sin\theta_i$ cancel and
one obtains
\be
P_{\rm st}(\bm\lambda)\,\dd^N\lambda
\propto
\det_{a,i}[\hat p_a(\theta_i)]
\det_{a,i}[\hat v_a(\theta_i)]\,\dd^N\theta .
\label{eq:SMcompactMeasure}
\ee
Under one reinjection, restoring the ordered labels cyclically permutes
the columns of either Slater determinant,
$(1,2,\ldots,N)\mapsto(2,\ldots,N,1)$, and therefore produces the
factor $(-1)^{N-1}$.  At the same time the orbital carried through the
projective point acquires the factor $(-1)^{N-1}$ from
Eq.~\eqref{eq:SMfermionicTwist}.  The two factors cancel, so each
Slater determinant, and hence the stationary measure and current, is
single-valued under cyclic reinjection.

This also identifies the global boundary condition selecting the physical
right mode $v_a$.  Of the two local solutions of
$\ell_Nv_a=\eps_av_a$, the physical one is that for which
$\hat v_a$ is regular at the projective point and obeys the continuation
law~\eqref{eq:SMfermionicTwist}.  Expressed in the $\lambda$ variable,
this boundary condition is implemented by the current-carrying solution
Eq.~\eqref{eq:vamain}.  Consequently, the free-fermion representation
derived in Eq.~\eqref{eq:fermionizationmain} is not only valid locally
in the bulk: its Slater determinants also satisfy the cyclic
reinjection condition at infinity.

For comparison with the matrix Pr\"ufer formulation, restore an arbitrary
scale $\kappa>0$ and write $\lambda_i=\kappa\cot\theta_i$.  The Cayley
transform
\be
U_{\rm P}(x)
=
[Z(x)+\ii\kappa I_N][Z(x)-\ii\kappa I_N]^{-1}
\label{eq:SMCayley}
\ee
is unitary because $Z$ is Hermitian.  If $\lambda_i$ is an eigenvalue
of $Z$, the corresponding eigenvalue of $U_{\rm P}$ is
$e^{2\ii\theta_i}$.  Thus the Pr\"ufer angles introduced above are,
up to the factor of two, the phases of the eigenvalues of the unitary
matrix $U_{\rm P}$.  When a Riccati eigenvalue passes through infinity,
$e^{2\ii\theta_i}$ makes one complete turn around the unit circle.

The same Cayley transform has a standard scattering interpretation
\cite{Beenakker1997,FyodorovSavinSommers2005}.  To see it, consider as
an auxiliary construction the disordered interval $[0,L]$, with the
Dirichlet condition at $x=0$, attached at $x=L$ to an ideal $N$-channel
lead with wave number $\kappa$.  If $\bm a$ and $\bm b$ denote,
respectively, the incoming and reflected wave amplitudes in the lead,
then $\bm b=S_L\bm a$ defines the reflection matrix $S_L$.  Writing the
lead wave as
$\psi(x)=e^{-\ii\kappa(x-L)}\bm a+
e^{\ii\kappa(x-L)}\bm b$
and matching its logarithmic derivative at $x=L$ to
$Z_L:=Z(L;E)$ gives
\be
S_L
=
-[Z_L-\ii\kappa I_N]^{-1}[Z_L+\ii\kappa I_N]
=
-U_{\rm P}(L).
\label{eq:SMreflectionCayley}
\ee
Thus a Riccati winding can equivalently be viewed as a winding of a
reflection eigenvalue around the unit circle.  This scattering
interpretation is auxiliary; the projective compactification itself
does not require attaching leads.

The same projective boundary condition also provides a natural way to
count Riccati windings: assigning a factor $z$ to each cyclic passage
changes only the matching across the identified endpoints, while leaving
the bulk differential operator unchanged.  The corresponding forward
and backward matching conditions, and their use for finite-length
spectral counting, are given in Sec.~\ref{sec:SMsurvival}; the associated
long-length counting statistics are discussed in
Sec.~\ref{sec:SMFCS}.

\section{Exact $\beta=2$ fermionization}
\label{sec:SMfermionization}

We give the details of the cancellation leading to
Eq.~\eqref{eq:fermionizationmain}.  In the ordered chamber
$\mathcal C_N$ the forward generator associated with
Eq.~\eqref{eq:FPcurrent} is
\be
\cL_N^{\rm FP}P
=
\sum_{i=1}^N
\left\{
T\partial_i^2P
+\partial_i\!\left[(E+\lambda_i^2)P\right]
-2T\partial_i\!\left[
P\sum_{j\ne i}\frac{1}{\lambda_i-\lambda_j}
\right]
\right\}.
\label{eq:SMFPoperator}
\ee
Set $P=V_N\Psi$, with
$V_N(\bm\lambda)=\prod_{i<j}(\lambda_j-\lambda_i)$.
With our ordering convention
$\lambda_1<\cdots<\lambda_N$, one has $V_N>0$ in
$\mathcal C_N$, while the probability density $P$ is non-negative.
Thus no sign is introduced by this factorization inside the ordered
chamber.  Upon extending $P$ symmetrically to the other ordering
chambers, $P$ is unchanged by permutations whereas $V_N$ changes by
their parity; consequently $\Psi=P/V_N$ is antisymmetric under exchange of two eigenvalues.
The elementary
identities
\be
\frac{\partial_iV_N}{V_N}
=
\sum_{j\ne i}\frac{1}{\lambda_i-\lambda_j},
\qquad
\sum_{i=1}^N\partial_i^2V_N=0,
\label{eq:SMVandermondeIdentities}
\ee
produce the usual $\beta=2$ cancellation of the singular derivative
interaction.  For a generic nonlinear drift this does not yet leave a
sum of one-body operators: a divided-difference term survives.  In the
present problem $U_E'(\lambda)=E+\lambda^2$, and
\be
\sum_{i=1}^N
U_E'(\lambda_i)
\sum_{j\ne i}\frac{1}{\lambda_i-\lambda_j}
=
\sum_{i<j}
\frac{U_E'(\lambda_i)-U_E'(\lambda_j)}
{\lambda_i-\lambda_j}
=
(N-1)\sum_{i=1}^N\lambda_i.
\label{eq:SMsecondCancellation}
\ee
Together with $U_E''(\lambda)=2\lambda$, this converts the whole
remainder into the one-body term $(N+1)\lambda$.  Hence
\be
\cL_N^{\rm FP}[V_N\Psi]
=
V_N\sum_{i=1}^N\ell_N^{(i)}\Psi,
\qquad
\ell_N
=
T\partial_\lambda^2
+(E+\lambda^2)\partial_\lambda
+(N+1)\lambda,
\label{eq:SMfermionization}
\ee
with adjoint
\be
\ell_N^\dagger
=
T\partial_\lambda^2
-(E+\lambda^2)\partial_\lambda
+(N-1)\lambda.
\label{eq:SMellAdjoint}
\ee
Thus the transformed bulk generator is exactly a sum of one-body
operators at every finite $N$.  The first cancellation in
Eq.~\eqref{eq:SMVandermondeIdentities} is the familiar unitary-class
fermionization, whereas Eq.~\eqref{eq:SMsecondCancellation}, which uses
the quadratic Riccati drift, is the additional special property of the
present model.  The compatibility of this bulk reduction with the cyclic
projective boundary condition, including the compensating fermionic
sign under reinjection, was established in the preceding section.

\section{Polynomial sector and biorthogonal stationary process}
\label{sec:SMpolynomial}

\subsection{Finite-dimensional polynomial sector and electrostatic roots}
\label{subsec:SMfinitepoly}

The adjoint operator~\eqref{eq:SMellAdjoint} preserves
$\mathcal P_{N-1}=\mathrm{span}\{1,\lambda,\ldots,\lambda^{N-1}\}$.
Indeed,
\be
\ell_N^\dagger\lambda^n
=
Tn(n-1)\lambda^{n-2}
-En\lambda^{n-1}
+(N-1-n)\lambda^{n+1},
\qquad 0\leq n\leq N-1.
\label{eq:SMpolyAction}
\ee
In the monomial basis $(1,\lambda,\ldots,\lambda^{N-1})$, let $M_N$
denote the matrix of $\ell_N^\dagger|_{\mathcal P_{N-1}}$.  With row and
column indices $m,n=0,\ldots,N-1$,
\be
(M_N)_{mn}
=
Tn(n-1)\delta_{m,n-2}
-En\delta_{m,n-1}
+(N-1-n)\delta_{m,n+1}.
\label{eq:SMfiniteMatrix}
\ee
Thus the finite-$N$ algebraic spectral problem is the ordinary matrix
eigenvalue problem for $M_N$.  Defining
\be
Q_N(\eps):=\det(\eps I_N-M_N),
\label{eq:SMcharacteristicQ}
\ee
the first cases are
\be
\begin{aligned}
Q_1(\eps)&=\eps,\\
Q_2(\eps)&=\eps^2+E,\\
Q_3(\eps)&=\eps^3+4E\eps-4T,\\
Q_4(\eps)&=\eps^4+10E\eps^2-24T\eps+9E^2.
\end{aligned}
\label{eq:SMfirstQ}
\ee
A useful feature is visible directly in Eq.~\eqref{eq:SMpolyAction}:
there is no term proportional to $\lambda^n$.  Consequently $M_N$ has
zero diagonal and
\be
\sum_{a=1}^N\eps_a=\Tr M_N=0,
\label{eq:SMsumEpsZero}
\ee
where the eigenvalues are counted with algebraic multiplicity.  For
generic parameters $M_N$ is diagonalizable and its $N$ eigenvectors give
the polynomial modes $p_a$ of Eq.~\eqref{eq:QESmain}.  Equation
\eqref{eq:SMsumEpsZero} is precisely what makes the filled $N$-fermion
Slater determinant stationary: its many-body eigenvalue is the sum of
all one-body eigenvalues and therefore vanishes, as in
Eq.~\eqref{eq:manybodyzeromode}.

Using the monic normalization of the polynomial modes, with the
corresponding physical right partners subsequently normalized by
$\mathcal B[p_a,v_a]=1$, the first cases are
\be
\begin{aligned}
N=1:\qquad&
\eps_1=0,
& p_1(\lambda)&=1,
\\[1mm]
N=2:\qquad&
\eps_a^2+E=0,
& p_a(\lambda)&=\lambda+\eps_a,
\qquad a=1,2,
\\[1mm]
N=3:\qquad&
\eps_a^3+4E\eps_a-4T=0,
& p_a(\lambda)&=
\lambda^2+\eps_a\lambda+\frac{\eps_a^2+2E}{2},
\qquad a=1,2,3.
\end{aligned}
\label{eq:SMlowNpModes}
\ee
Thus, at fixed $N$, the $N$ functions $p_a$ are not a degree-graded
family: generically they are $N$ linearly independent eigenmodes, all
of degree $N-1$, and together they span $\mathcal P_{N-1}$.

Every nonzero polynomial eigenmode has degree exactly $N-1$.  Indeed, if
$p(\lambda)=c_m\lambda^m+\cdots$ with $m<N-1$, the last term in
Eq.~\eqref{eq:SMpolyAction} produces a nonzero contribution of degree
$m+1$, incompatible with $\ell_N^\dagger p=\eps p$.  We may therefore
write a monic mode as
\be
p_a(\lambda)
=
\prod_{j=1}^{N-1}
\bigl(\lambda-r_j^{(a)}\bigr).
\label{eq:SMrootFactorization}
\ee
At a simple zero $r_j^{(a)}$, the eigenvalue equation gives
$Tp_a''-(E+\lambda^2)p_a'=0$.  Using
$p_a''(r_j)/p_a'(r_j)=2\sum_{k\ne j}(r_j-r_k)^{-1}$ yields the exact
Stieltjes equations
\be
E+\bigl(r_j^{(a)}\bigr)^2
=
2T\sum_{k\ne j}
\frac{1}{r_j^{(a)}-r_k^{(a)}},
\qquad j=1,\ldots,N-1.
\label{eq:SMStieltjesRoots}
\ee
The roots $r_j^{(a)}$ need not be real.  For generic parameters the
$N$ polynomial eigenmodes $p_a$ give $N$ distinct unordered sets
$\{r_1^{(a)},\ldots,r_{N-1}^{(a)}\}$ solving
Eq.~\eqref{eq:SMStieltjesRoots}; the label $a$ distinguishes these
different solutions.  Already for $N=2$ there is a single root satisfying
$E+r^2=0$: for $E<0$ the two modes have
$r=\pm\sqrt{-E}$, whereas for $E>0$ they form the complex pair
$r=\pm\ii\sqrt E$.
Comparison of the coefficient of $\lambda^{N-1}$ in
$\ell_N^\dagger p_a=\eps_ap_a$ gives in addition
\be
\eps_a
=-\sum_{j=1}^{N-1}r_j^{(a)}
=-(N-1)\,\bar r_a.
\label{eq:SMEpsRoots}
\ee

Equations~\eqref{eq:SMStieltjesRoots} are the stationary-point equations
of the finite-dimensional logarithmic energy
\be
\mathcal W_E(\bm r)
=
\sum_{j=1}^{N-1}U_E(r_j)
-2T\sum_{j<k}\log(r_j-r_k),
\qquad
U_E(r)=Er+\frac{r^3}{3}.
\label{eq:SMrootElectrostaticEnergy}
\ee
For real $r_i$'s  this has the usual Coulomb-gas interpretation, with
$\log(r_j-r_k)$ replaced by $\log|r_j-r_k|$.  In general, however, the
roots $r_j^{(a)}$ and the corresponding stationary configurations need
not be real, and Eq.~\eqref{eq:SMrootElectrostaticEnergy} is then
understood by analytic continuation.  There is consequently no ordering
of such complex configurations by a real ``electrostatic energy.''
Moreover, even on the real axis the cubic external potential is
nonconfining, so the solutions are stationary configurations rather than
global minima.

For each exact polynomial mode $p_a$, the unordered root set
$\{r_1^{(a)},\ldots,r_{N-1}^{(a)}\}$ defines one such finite-$N$
electrostatic configuration; different values of $a$ label different
solutions of the same Stieltjes equations.  The external potential is
the same cubic potential that governs the Dyson--Riccati gas.  As shown
in Sec.~\ref{sec:SMquarticLargeN}, for any sequence of modes whose roots
condense onto a smooth limiting contour, the finite-$N$ Stieltjes
equations reduce to the planar Coulomb-gas saddle equation of the
nonconfining cubic Dyson model of Allez and Dumaz~\cite{AllezDumaz2015}.
Different sequences may correspond to different solutions or analytic
branches of this planar equation; the physical Dyson--Riccati problem
selects the appropriate large-$N$ branch separately.  Thus the exact
polynomial root configurations provide finite-$N$ algebraic
discretizations of the planar saddle equation, rather than a
one-to-one identification with its physical solution.

\subsection{Physical right partners and a regular construction}
\label{subsec:SMrightpartners}

For each polynomial mode $p_a$, the right equation
$\ell_Nv_a=\eps_av_a$ has two local solutions.  The physical one is fixed
by the projective boundary condition established in the preceding section.  With the unit-current
normalization of Eq.~\eqref{eq:concomitantmain},
\be
T\bigl(p_av_a'-p_a'v_a\bigr)
+(E+\lambda^2)p_av_a=1.
\label{eq:SMunitConcomitant}
\ee
Since $p_a$ is monic of degree $N-1$, this immediately gives
\be
p_a(\lambda)v_a(\lambda)
\sim\frac{1}{E+\lambda^2},
\qquad
v_a(\lambda)=O(\lambda^{-N-1}),
\qquad \lambda\to\pm\infty,
\label{eq:SMrightAsymptotics}
\ee
in agreement with the compactified regularity derived in the preceding section.
Dividing Eq.~\eqref{eq:SMunitConcomitant} by $Tp_a^2$ gives
\be
\frac{\dd}{\dd\lambda}
\left[
\frac{e^{U_E(\lambda)/T}v_a(\lambda)}{p_a(\lambda)}
\right]
=
\frac{1}{T}
\frac{e^{U_E(\lambda)/T}}{p_a(\lambda)^2}.
\label{eq:SMreductionFirstOrder}
\ee
Before integrating, note that the lower endpoint causes no ambiguity.
As $y\to-\infty$, one has $U_E(y)\sim y^3/3\to-\infty$ and
$p_a(y)\sim y^{N-1}$.  Hence
$e^{U_E(y)/T}/p_a(y)^2$ decays exponentially and is integrable at
$-\infty$.  Moreover, using Eq.~\eqref{eq:SMrightAsymptotics},
\[
\frac{e^{U_E(y)/T}v_a(y)}{p_a(y)}
\longrightarrow 0,
\qquad y\to-\infty .
\]
We may therefore integrate Eq.~\eqref{eq:SMreductionFirstOrder} from
$-\infty$ without an additional integration constant, obtaining the
compact form used in the main text,
\be
v_a(\lambda)
=
\frac1T e^{-U_E(\lambda)/T}p_a(\lambda)
\FP\!\int_{-\infty}^{\lambda}
\frac{e^{U_E(y)/T}}{p_a(y)^2}\,\dd y .
\label{eq:SMvFinitePart}
\ee

Here $\FP$ denotes the Hadamard finite part; it is needed only when
$p_a$ has a real zero on the integration interval.  To see explicitly
why such a zero does not produce a singularity in $v_a$, let $r$ be a
simple zero of $p_a$.  Expanding the integrand about $y=r$ gives
\be
\frac{e^{U_E(y)/T}}{p_a(y)^2}
=
\frac{e^{U_E(r)/T}}{p_a'(r)^2}
\left[
\frac{1}{(y-r)^2}
+
\left(
\frac{U_E'(r)}{T}
-\frac{p_a''(r)}{p_a'(r)}
\right)
\frac{1}{y-r}
+O(1)
\right].
\label{eq:SMnoSimplePole}
\ee
At a zero of $p_a$, the polynomial eigenvalue equation
$\ell_N^\dagger p_a=\eps_a p_a$ gives
$Tp_a''(r)=U_E'(r)p_a'(r)=(E+r^2)p_a'(r)$, equivalently the root
equation~\eqref{eq:SMStieltjesRoots}.  The coefficient of
$(y-r)^{-1}$ therefore vanishes.  Thus the integrand has only the
double-pole term,
\[
\frac{e^{U_E(y)/T}}{p_a(y)^2}
=
\frac{e^{U_E(r)/T}}{p_a'(r)^2}
\left[
\frac{1}{(y-r)^2}+O(1)
\right],
\]
and its finite-part primitive behaves as
\[
-\frac{e^{U_E(r)/T}}{p_a'(r)^2}\frac{1}{\lambda-r}
+O(1)
\qquad (\lambda\to r),
\]
with no logarithmic term.  The simple pole of the primitive is precisely
cancelled by
$p_a(\lambda)=p_a'(r)(\lambda-r)+\cdots$ in
Eq.~\eqref{eq:SMvFinitePart}.  Hence $v_a$ is regular at $r$, with
\be
v_a(r)=-\frac{1}{T p_a'(r)}.
\label{eq:SMvAtRoot}
\ee

Let us give the first two cases explicitly.  For $N=1$,
$p_1=1$, $\eps_1=0$, and
\be
v_1(\lambda)
=
\frac1T e^{-U_E(\lambda)/T}
\int_{-\infty}^{\lambda}e^{U_E(y)/T}\,\dd y ,
\label{eq:SMvN1}
\ee
which is Halperin's constant-current solution.  For $N=2$,
$\eps_a^2+E=0$ and $p_a(\lambda)=\lambda+\eps_a$, so
Eq.~\eqref{eq:SMvFinitePart} gives
\be
v_a(\lambda)
=
\frac1T e^{-U_E(\lambda)/T}(\lambda+\eps_a)
\FP\!\int_{-\infty}^{\lambda}
\frac{e^{U_E(y)/T}}{(y+\eps_a)^2}\,\dd y .
\label{eq:SMvN2FP}
\ee
Using $E=-\eps_a^2$, the integrand satisfies
\[
\frac{e^{U_E(y)/T}}{(y+\eps_a)^2}
=
\frac{y-\eps_a}{T}e^{U_E(y)/T}
-
\frac{\dd}{\dd y}
\left[
\frac{e^{U_E(y)/T}}{y+\eps_a}
\right],
\]
and hence the same regular solution can be written without a finite
part as
\be
v_a(\lambda)
=
-\frac1T
+
\frac{\lambda+\eps_a}{T^2}
e^{-U_E(\lambda)/T}
\int_{-\infty}^{\lambda}
(y-\eps_a)e^{U_E(y)/T}\,\dd y .
\label{eq:SMvN2Regular}
\ee
When $\eps_a$ is real, the first representation has an apparent pole at
$y=-\eps_a$, whereas the second involves only an ordinary integral and
makes the regular value $v_a(-\eps_a)=-1/T$ manifest. 
The same manipulation can be performed for any $N$, using a polynomial Hermite reduction which removes the finite-part prescription altogether, but for arbitrary $N$ the first-order form
\eqref{eq:SMnoPoleODE} is more economical.

For analytic work Eq.~\eqref{eq:SMvFinitePart} is often the shortest
representation.  For numerical work, however, it is preferable to avoid
finite parts altogether and integrate directly the first-order
concomitant equation
\be
T p_a v_a'
+
\bigl[(E+\lambda^2)p_a-Tp_a'\bigr]v_a
=1,
\label{eq:SMnoPoleODE}
\ee
using Eq.~\eqref{eq:SMvAtRoot} to continue the regular solution
smoothly across each real zero of $p_a$.
Equation~\eqref{eq:SMnoPoleODE} is exactly equivalent to the second-order
right eigenvalue equation together with the unit-concomitant
normalization, but contains no but contains no finite-part prescription.

\subsection{Biorthogonality, stationary kernel, and the IDOS}
\label{subsec:SMkernel}

We now use the left/right structure of the one-particle problem to
construct the stationary determinantal process.  The polynomial modes
$p_a$ and the physical right modes $v_a$ satisfy
$\ell_N^\dagger p_a=\eps_a p_a$ and
$\ell_N v_a=\eps_a v_a$, respectively.  Their natural pairing is the
bilinear, rather than Hermitian, pairing
\be
\langle p_a,v_b\rangle
:=
\int_{-\infty}^{\infty}
p_a(\lambda)v_b(\lambda)\,\dd\lambda .
\label{eq:SMbilinearPairing}
\ee
No complex conjugation is involved: $p_a$ and $v_b$ are left and right
eigenmodes of the non-self-adjoint one-body generator.  The integral is
convergent because the endpoint behavior established above gives
$p_a(\lambda)v_b(\lambda)=O(\lambda^{-2})$.

For arbitrary functions $p$ and $v$, define the Lagrange concomitant
\be
\mathcal B[p,v](\lambda)
:=
T\bigl(p(\lambda)v'(\lambda)-p'(\lambda)v(\lambda)\bigr)
+(E+\lambda^2)p(\lambda)v(\lambda).
\label{eq:SMconcomitant}
\ee
This is the boundary bilinear form associated with the adjoint pair
$\ell_N^\dagger,\ell_N$, and plays here the role analogous to the
Wronskian for a self-adjoint second-order problem.  A direct calculation
gives the Lagrange identity
\be
p\,\ell_N v-v\,\ell_N^\dagger p
=
\partial_\lambda\mathcal B[p,v].
\label{eq:SMLagrangeIdentity}
\ee
For a pair of eigenmodes this becomes
\be
\partial_\lambda\mathcal B[p_a,v_b]
=
(\eps_b-\eps_a)p_av_b .
\label{eq:SMconcomitantDerivative}
\ee
Integrating along the Riccati line therefore gives
\be
(\eps_b-\eps_a)
\int_{-\infty}^{\infty}p_a(\lambda)v_b(\lambda)\,\dd\lambda
=
\left.\mathcal B[p_a,v_b]\right|_{-\infty}^{+\infty}.
\label{eq:SMLagrangeIntegrated}
\ee
The compactified projective boundary condition identifies the two
endpoint values of the concomitant, so the right-hand side vanishes.
For distinct eigenvalues this gives the biorthogonality relation
\be
\int_{-\infty}^{\infty}p_a(\lambda)v_b(\lambda)\,\dd\lambda
=
g_a\delta_{ab},
\qquad
g_a:=
\int_{-\infty}^{\infty}p_a(\lambda)v_a(\lambda)\,\dd\lambda .
\label{eq:SMbiorthogonality}
\ee
Thus ``biorthogonal'' means that the two distinct families
$\{p_a\}$ and $\{v_a\}$ are mutually orthogonal under the bilinear
pairing~\eqref{eq:SMbilinearPairing}: $p_a$ pairs only with its
corresponding right partner $v_a$.

It is important that we do not normalize this pairing to unity.  As in
Eq.~\eqref{eq:concomitantmain} of the Main Text, the normalization of
each physical right mode has already been fixed by the unit-current
condition
\be
\mathcal B[p_a,v_a]=1.
\label{eq:SMunitCurrentNormalization}
\ee
Together with the chosen normalization of the polynomial $p_a$, this
fixes the relative normalization of the left/right pair.  Consequently
the biorthogonal norms $g_a$ in
Eq.~\eqref{eq:SMbiorthogonality} are in general not equal to one; they
are part of the physical finite-$N$ solution and, as shown below, their
inverse sum gives the IDOS.

More generally, without assuming a basis in which the pairing is
diagonal, define the Gram matrix
\be
G_{ab}
:=
\int_{-\infty}^{\infty}p_a(\lambda)v_b(\lambda)\,\dd\lambda .
\label{eq:SMGramMatrix}
\ee
Andr\'eief's identity then gives
\be
\int_{\mathbb R^N}
\det_{a,i}[p_a(\lambda_i)]
\det_{a,i}[v_a(\lambda_i)]\,\dd^N\lambda
=
N!\det G.
\label{eq:SMAndreiefNormalization}
\ee
Since the product of the two determinants is symmetric under
permutations of the $\lambda_i$, its integral over the ordered chamber
$\mathcal C_N$ is simply $\det G$.  Hence the normalization
$\cZ_N$ appearing in Eq.~\eqref{eq:stationaryJPD} of the Main Text is
\be
\cZ_N=\det G
=
\prod_{a=1}^N g_a
\label{eq:SMorderedNormalization}
\ee
in the nondegenerate biorthogonal basis.  
Equivalently, the normalized symmetric extension to $\mathbb R^N$ of
the stationary distribution is
\be
P_{\rm st}^{\rm sym}(\bm\lambda)
=
\frac{1}{N!\det G}
\det_{a,i}[p_a(\lambda_i)]
\det_{a,i}[v_a(\lambda_i)].
\label{eq:SMbiorthogonalEnsemble}
\ee
This is a biorthogonal ensemble and therefore defines a finite
determinantal point process for the $\lambda_i$'s on the line
\cite{Johansson2006Determinantal,Borodin2011DPP,Forrester2010}. 
More precisely, with
the standard convention, the correlation functions
\be
R_k(\lambda_1,\ldots,\lambda_k)
:=
\frac{N!}{(N-k)!}
\int_{\mathbb R^{N-k}}
P_{\rm st}^{\rm sym}(\lambda_1,\ldots,\lambda_N)
\prod_{j=k+1}^N\dd\lambda_j ,
\label{eq:SMRkDefinition}
\ee
are all determinants,
\be
R_k(\lambda_1,\ldots,\lambda_k)
=
\det_{1\leq i,j\leq k}
[K_N(\lambda_i,\lambda_j)].
\label{eq:SMdetCorrelations}
\ee
The corresponding finite-rank correlation kernel is
\be
K_N(\lambda,\mu)
=
\sum_{a,b=1}^N
p_a(\lambda)(G^{-1})_{ba}v_b(\mu).
\label{eq:SMGramKernel}
\ee
This Gram-matrix form is invariant under invertible changes of the
left and right bases and is therefore convenient also at degeneracies.
Although the biorthogonal kernel need not be Hermitian, the resulting
correlation functions are real and non-negative, as they are the
marginals of the physical stationary probability density
$P_{\rm st}^{\rm sym}$. At generic parameters, where
$G_{ab}=g_a\delta_{ab}$, Eq.~\eqref{eq:SMGramKernel} reduces to
\be
K_N(\lambda,\mu)
=
\sum_{a=1}^N
\frac{p_a(\lambda)v_a(\mu)}{g_a}.
\label{eq:SMdiagonalKernel}
\ee
In particular, the one-point density of the stationary Dyson--Riccati gas is
$n(\lambda)=R_1(\lambda)=K_N(\lambda,\lambda)$ and has total mass
$\int_{\mathbb R}n(\lambda)\,\dd\lambda=N$. With the
unit-current normalization, see \eqref{eq:SMunitConcomitant} and \eqref{eq:SMunitCurrentNormalization}, every diagonal
term satisfies $p_av_a\sim\lambda^{-2}$.  Therefore
\be 
K_N(\lambda,\lambda)
\sim
\frac{1}{\lambda^2}
\sum_{a=1}^N\frac{1}{g_a},
\qquad |\lambda|\to\infty.
\label{eq:SMkernelTail}
\ee
Comparing with the projective-tail identity
Eq.~\eqref{eq:SMprojectiveTail} gives the exact finite-$N$ current formula
\be
\nu_N(E)=\sum_{a=1}^N\frac{1}{g_a},
\qquad
g_a:=\int_{-\infty}^{\infty}p_a(\lambda)v_a(\lambda)\,\dd\lambda .
\label{eq:SMidosNormSum}
\ee
The same kernel also gives the total longitudinal volume growth,
\be
\Gamma_N(E)
=
\PV\!\int_{\mathbb R}
\lambda K_N(\lambda,\lambda)\,\dd\lambda.
\label{eq:SMgammaKernel}
\ee
Thus the finite matrix $M_N$, its polynomial eigenvectors, and their
physical right partners determine the full stationary determinantal
process, the IDOS, and the total Lyapunov growth.  
At exceptional parameter values where eigenvalues of $M_N$ coalesce, the
Gram-matrix representation, Eqs.~\eqref{eq:SMGramMatrix} and
\eqref{eq:SMGramKernel} provides the regular continuation.

A second, auxiliary determinantal structure will appear below in the
cubic-contour representation of $\mathcal Z_N^{(-)}$.  Its correlation
kernel should not be confused with the physical stationary kernel
$K_N$ above: the latter acts on the real Riccati eigenvalues and is built
from the left/right modes $(p_a,v_a)$, whereas the cubic kernel acts on
the complex contour $\Gamma_-$ and is built from complex orthogonal
polynomials.

\subsection{Fourier--Airy intertwining and the exact Hankel determinant}
\label{subsec:SMFourierAiry}

We now give the detailed derivation of
Eq.~\eqref{eq:characteristicTauRatio}.  The Main Text reduces the
$N$-body characteristic function, by Andr\'eief's identity, to
\be
\mathcal F_N(q;s)
=
\frac{\det I(q)}{\det I(0)},
\qquad
I_{ab}(q)
=
\int_{\mathbb R}
p_a(x)v_b(x)e^{-\ii qx}\,\dd x ,
\label{eq:SMoverlapStart}
\ee
and the purpose of this subsection is to evaluate the matrix $I(q)$
exactly.  The key observation is that, for $q>0$, the Fourier
transform of a physical right mode $v_b$ on the real line coincides,
up to a constant, with the Fourier transform of the polynomial left
mode $p_b$ multiplied by the weight $e^{-NV_E}$ of the outgoing
Kontsevich integral and integrated along the contour $\Gamma_-$.  The
overlap matrix then becomes the bilinear product of the orthogonal
polynomial problem at a complex-shifted energy, and its determinant is
the Hankel determinant $\tau_N$.

\paragraph{Dimensionless left and right equations.---}
Introduce
\be
x=\frac{\lambda}{T^{1/3}},
\qquad
s=-\frac{E}{T^{2/3}},
\qquad
e_a=\frac{\eps_a}{T^{1/3}} ,
\label{eq:SMFourierScaling}
\ee
and denote the rescaled functions $p_a(T^{1/3}x)$ and $v_a(T^{1/3}x)$
again by $p_a(x)$ and $v_a(x)$.  Since
$\ell_N=T^{1/3}\widehat\ell_N$, the adjoint
pair~\eqref{eq:ellpair} becomes
\be
\begin{aligned}
\widehat\ell_N^\dagger
&=
\partial_x^2-(x^2-s)\partial_x+(N-1)x,\\
\widehat\ell_N
&=
\partial_x^2+(x^2-s)\partial_x+(N+1)x,
\end{aligned}
\qquad
\widehat\ell_N^\dagger p_a=e_ap_a,
\qquad
\widehat\ell_Nv_a=e_av_a ,
\label{eq:SMDimensionlessAdjointPair}
\ee
and the weight of the outgoing Kontsevich
integral~\eqref{eq:overviewZ} takes the form
$e^{-NV_E(\lambda)}=e^{-x^3/3+sx}$.  The left mode belongs to the
invariant polynomial space $\mathcal P_{N-1}$,
\be
p_a(x)
=
\sum_{n=0}^{N-1}c_{an}x^n ,
\label{eq:SMPolynomialCoefficients}
\ee
on which the left operator acts as
\be
\widehat\ell_N^\dagger x^n
=
n(n-1)x^{n-2}
+ns\,x^{n-1}
+(N-1-n)x^{n+1},
\label{eq:SMDimensionlessMonomialAction}
\ee
the rescaled form of Eq.~\eqref{eq:monomialAction}.  The last
coefficient vanishes at $n=N-1$, which is why $\mathcal P_{N-1}$
closes exactly.

\paragraph{Fourier equation for a physical right mode.---}
For a fixed mode $b$ define
\be
\widetilde v_b(q)
:=
\int_{-\infty}^{\infty}
e^{-\ii qx}v_b(x)\,\dd x .
\label{eq:SMRightFourierDefinition}
\ee
The physical boundary condition of
Sec.~\ref{subsec:SMrightpartners} gives $v_b(x)=O(x^{-N-1})$ at both
ends, see Eq.~\eqref{eq:SMrightAsymptotics}.  Hence
$x^mv_b\in L^1(\mathbb R)$ for $0\leq m\leq N-1$, so that
$\widetilde v_b$ is an ordinary continuous Fourier transform and all
overlaps with polynomials of degree at most $N-1$ are obtained by
differentiating it,
\be
\int_{\mathbb R}
x^m v_b(x)e^{-\ii qx}\,\dd x
=
(\ii\partial_q)^m\widetilde v_b(q),
\qquad
0\leq m\leq N-1.
\label{eq:SMFourierMomentDerivative}
\ee
The right eigenvalue equation is
\be
\widehat\ell_Nv_b = v_b''+(x^2-s)v_b'+(N+1)xv_b=e_bv_b .
\label{eq:SMRightDimensionlessEquation}
\ee
Fourier transformation uses $\partial_x\mapsto\ii q$ and
$x\mapsto\ii\partial_q$; applied to
Eq.~\eqref{eq:SMRightDimensionlessEquation}, these rules hold for all
real $q$ in the sense of tempered distributions.  The point $q=0$
nevertheless requires care, for two reasons.  First, the algebraic
tails of $v_b$ limit the smoothness of $\widetilde v_b$ there: only
the moments $x^mv_b$ with $m\leq N-1$ are integrable, so
$\widetilde v_b$ is $N-1$ times continuously differentiable at $q=0$
but in general not analytic.  Second, the coefficient of the highest
derivative in the transformed equation is proportional to $q$, so
$q=0$ is a singular point of the resulting differential equation.  We
therefore treat it as an ordinary differential equation on the
half-line $q>0$; since its leading coefficient does not vanish there,
$\widetilde v_b$ is a classical solution.  One finds
\be
\left[
-q^2
-\ii q\,\partial_q^2
+\ii(N-1)\partial_q
-\ii s q
-e_b
\right]\widetilde v_b(q)
=
0,
\qquad q>0 .
\label{eq:SMFourierEquationQ}
\ee
In the complex variable
\be
z=s-\ii q,
\qquad
\ii\partial_q=\partial_z ,
\label{eq:SMzDefinition}
\ee
this equation becomes
\be
(\mathscr D_{s,N}+e_b)\,\widetilde v_b=0,
\qquad
\mathscr D_{s,N}
:=
(z-s)(\partial_z^2-z)
-(N-1)\partial_z .
\label{eq:SMFourierEquationZ}
\ee
For $q>0$ the variable $z$ runs vertically into the lower half-plane
from the point $z=s$.

\paragraph{Contour transform of polynomials.---}
The function $e^{-x^3/3}$ decays at infinity in three sectors of the
complex $x$-plane, centred on the directions $\arg x=0$ and
$\arg x=\pm2\pi/3$.  For any contour $\Gamma$ that starts and ends at
infinity in two different sectors, and for any polynomial $p$, define
\be
\mathcal T_\Gamma[p](z)
:=
\int_\Gamma p(x)\,e^{-x^3/3+zx}\,\dd x .
\label{eq:SMContourTransform}
\ee
The integral converges for all complex $z$, and three elementary
properties follow directly from the kernel
$K(x,z)=e^{-x^3/3+zx}$.

(i) \emph{Multiplication by $x$ becomes differentiation in $z$:}
$\mathcal T_\Gamma[xp]=\partial_z\mathcal T_\Gamma[p]$, since
$\partial_zK=xK$.  Consequently $\mathcal T_\Gamma[p]=p(\partial_z)F_\Gamma$
with $F_\Gamma:=\mathcal T_\Gamma[1]$.

(ii) \emph{$F_\Gamma$ solves the Airy equation:}
$(\partial_z^2-z)F_\Gamma=\int_\Gamma(x^2-z)K\,\dd x
=-\int_\Gamma\partial_xK\,\dd x=0$, because $K$ vanishes at both ends
of $\Gamma$.  Conversely, every solution of $F''=zF$ is of this form:
the three contours joining pairs of sectors add up to a contour that
can be deformed to zero, so two of them are independent, matching the
two-dimensional solution space of the Airy equation.  The two
contours relevant here are $\Gamma_-$, from $e^{-2\pi\ii/3}\infty$ to
$+\infty$, and its complex conjugate $\overline{\Gamma_-}$, from
$e^{2\pi\ii/3}\infty$ to $+\infty$.  Comparing with the
representation~\eqref{eq:outgoingAiryContourRepresentation} after the
substitution $w=-x$, which maps $\mathcal C_-$ onto $\Gamma_-$, gives
\be
F_{\Gamma_-}(z)=\ii\pi\,\mathcal A_-(z),
\qquad
F_{\overline{\Gamma_-}}(z)=-\ii\pi\,\mathcal A_+(z),
\qquad
\mathcal A_\mp=\Ai\mp\ii\Bi ,
\label{eq:SMAiryContours}
\ee
the second relation following from the first by complex conjugation
at real $z$.

(iii) \emph{Intertwining.}  Using $zK=\partial_xK+x^2K$ one finds
$\mathscr D_{s,N}K=-\partial_x\!\left[(\partial_x+x^2-s)K\right]-(N-1)xK$.
Inserting this under the integral and integrating by parts twice, with
no boundary terms, gives for every polynomial $p$
\be
\mathscr D_{s,N}\,\mathcal T_\Gamma[p]
=
-\mathcal T_\Gamma\!\left[\widehat\ell_N^\dagger p\right].
\label{eq:SMAiryIntertwiner}
\ee
The operator governing the Fourier transform of the right modes is
thus conjugate, through the contour transform, to minus the left
operator acting on polynomials.  In particular, if $p_b$ is a
polynomial eigenmode, $\widehat\ell_N^\dagger p_b=e_bp_b$, then
\be
(\mathscr D_{s,N}+e_b)\,
\mathcal T_\Gamma[p_b]
=
0
\label{eq:SMAiryMappedEigenmode}
\ee
for every admissible contour $\Gamma$, which is precisely the Fourier
equation~\eqref{eq:SMFourierEquationZ}.  In terms of the monomial
coefficients, using \eqref{eq:SMPolynomialCoefficients}
one has $\mathcal T_\Gamma[p_b](z)= \sum_nc_{bn} 
\mathcal T_\Gamma[x^n](z) =\sum_nc_{bn}F_\Gamma^{(n)}(z)$, 
where we denote $F_\Gamma^{(n)}(z)=\partial_z^n F_\Gamma(z)$.
Eq.~\eqref{eq:SMAiryIntertwiner} is in fact equivalent to the Airy-derivative
recursion $F^{(n+2)}=zF^{(n)}+nF^{(n-1)}$ compared term by term with
Eq.~\eqref{eq:SMDimensionlessMonomialAction}. Indeed, 
using the same recursion once more,
$zF^{(n-1)}
=F^{(n+1)}-(n-1)F^{(n-2)}$, one finds
\be
\left[
(z-s)(\partial_z^2-z)
-(N-1)\partial_z
\right]F^{(n)} =
-(N-1-n)F^{(n+1)}
-nsF^{(n-1)}
-n(n-1)F^{(n-2)} .
\label{eq:SMAiryIntertwiningMonomial}
\ee
Comparing with
Eq.~\eqref{eq:SMDimensionlessMonomialAction}, we see that the
coefficients are exactly the same, with an overall minus sign.

The functions $\mathcal T_{\Gamma_-}[p_b]$ and
$\mathcal T_{\overline{\Gamma_-}}[p_b]$ are linearly independent.
Indeed, a vanishing combination would give
$p_b(\partial_z)(\alpha F_{\Gamma_-}+\beta F_{\overline{\Gamma_-}})=0$,
i.e. an Airy solution annihilated by a constant-coefficient operator
of order at most $N-1$; such a function is a combination of
exponentials times polynomials, and no nonzero one solves the Airy
equation.  The two functions therefore span the complete solution
space of the second-order equation~\eqref{eq:SMFourierEquationZ} on
$q>0$.

\paragraph{Selection of the outgoing contour.---}
Since $v_b\in L^1(\mathbb R)$, the Riemann--Lebesgue lemma requires
\be
\widetilde v_b(q)\longrightarrow0,
\qquad q\to+\infty .
\label{eq:SMRiemannLebesgue}
\ee
In the contour representation the behaviour at large $q$ is
transparent.  At $z=s-\ii q$ the kernel contains the factor
$e^{-\ii qx}$, whose modulus is $e^{q\,\Imn x}$.  The contour
$\Gamma_-$ connects two sectors that both reach into the lower
half-plane, so it can be shifted rigidly to $\Gamma_--\ii c$ with any
$c>0$, keeping its ends in the same sectors.  On the shifted contour
$\Imn x\leq-c$, and therefore
$|\mathcal T_{\Gamma_-}[p](s-\ii q)|\leq C_c\,e^{-cq}$ for every $c>0$:
the transform along $\Gamma_-$ decays faster than any exponential.
This agrees with the asymptotics of
$\mathcal A_-(z)=2e^{-\ii\pi/3}\Ai(ze^{2\pi\ii/3})$, whose argument
approaches $\arg(ze^{2\pi\ii/3})=\pi/6$, inside the decaying sector of
$\Ai$~\cite{DLMF}.  By contrast, $\overline{\Gamma_-}$ necessarily
starts in the sector around $e^{2\pi\ii/3}\infty$, which lies in the
upper half-plane; correspondingly
$\mathcal A_+(z)=2e^{\ii\pi/3}\Ai(ze^{-2\pi\ii/3})$ is evaluated near
$\arg=5\pi/6$, where $\Ai$ grows exponentially, and so does
$p_b(\partial_z)\mathcal A_+$ for any nonzero $p_b$.  The growing solution is excluded by
Eq.~\eqref{eq:SMRiemannLebesgue}, and therefore
\be
\widetilde v_b(q)
=
\eta_b\,\mathcal T_{\Gamma_-}[p_b](s-\ii q)
=
\eta_b\int_{\Gamma_-}
p_b(x)\,
e^{-x^3/3+sx}\,
e^{-\ii qx}\,\dd x ,
\qquad q>0 .
\label{eq:SMPhysicalRightFourier}
\ee
where $\eta_b$ is independent of $q$, and $\eta_b\neq0$ since
$v_b\not\equiv0$.  Its value depends on the normalization of $v_b$
and cancels from the final ratio.

Equation~\eqref{eq:SMPhysicalRightFourier} has a simple meaning.  For
$q>0$ the physical right mode $v_b(x)\,\dd x$ on the real line and the
complex weight the complex weight
$p_b(x)e^{-x^3/3+sx}\,\dd x$
on the outgoing contour
$\Gamma_-$ produce the same Fourier transform: as far as the overlaps
$I_{ab}(q)$ are concerned, the right mode is the polynomial left mode
times the Kontsevich weight, transported from the real axis to
$\Gamma_-$.  This also explains why the result is one-sided in $q$.
For $q<0$ the factor $e^{-\ii qx}$ decays in the upper half-plane, and
the same argument selects $\overline{\Gamma_-}$, i.e. $\mathcal A_+$.
Since $\mathcal A_+(\bar z)=\overline{\mathcal A_-(z)}$ for real $s$,
this gives $\mathcal F_N(-q;s)=\overline{\mathcal F_N(q;s)}$, as
required for the characteristic function of a real variable.

\paragraph{Overlap matrix and the Hankel determinant.---}
Combining Eqs.~\eqref{eq:SMFourierMomentDerivative},
\eqref{eq:SMzDefinition} and~\eqref{eq:SMPhysicalRightFourier} with
property~(i) of the contour transform,
\be
I_{ab}(q)
=
\sum_{m=0}^{N-1}c_{am}\,\partial_z^m\widetilde v_b(q)
=
\eta_b\,\mathcal T_{\Gamma_-}[p_ap_b](z)
=
\eta_b\int_{\Gamma_-}p_a(x)p_b(x)\,e^{-x^3/3+zx}\,\dd x,
\qquad z=s-\ii q .
\label{eq:SMOverlapExpanded}
\ee
The overlap matrix is therefore the symmetric bilinear
product~\eqref{eq:SMbilinear} of the orthogonal-polynomial problem,
evaluated at the complex-shifted value $z=s-\ii q$ of $s$, i.e. at
the energy $E+\ii qT^{2/3}$.  Expanding in monomials,
\be
I(q)
=
C\,M(z)\,C^{\mathsf T}\mathsf E,
\qquad
M_{mn}(z)=\mu_{m+n}(z),
\qquad
\mu_k(z):=\int_{\Gamma_-}x^ke^{-x^3/3+zx}\,\dd x
=\ii\pi\,\mathcal A_-^{(k)}(z),
\label{eq:SMOverlapFactorization}
\ee
with $C_{an}=c_{an}$ and
$\mathsf E=\operatorname{diag}(\eta_1,\ldots,\eta_N)$.  The Hankel
structure is explicit: the moment depends only on $m+n$.  For generic
parameters the $N$ polynomial eigenmodes form a basis of
$\mathcal P_{N-1}$, so $C$ is invertible, and
\be
\det I(q)
=
(\det C)^2\det\mathsf E\,
\det_{0\leq m,n\leq N-1}\left[\mu_{m+n}(z)\right]
=
(\ii\pi)^N(\det C)^2\det\mathsf E\;\tau_N(s-\ii q),
\label{eq:SMOverlapDeterminant}
\ee
with $\tau_N$ the Airy Hankel determinant~\eqref{eq:AiryTauDefinition}.
Equivalently, by Andr\'eief's identity, the Hankel determinant of the
moments is the $N$-fold contour integral
$\frac{1}{N!}\int_{\Gamma_-^N}V_N(\bm x)^2\prod_je^{-x_j^3/3+zx_j}\,\dd^Nx$,
i.e. the outgoing Kontsevich integral~\eqref{eq:overviewZ} at the
shifted energy.  The matrices $C$ and $\mathsf E$ carry all dependence
on the choice and normalization of the bases but are independent of
$q$.  Taking $q\to0_+$,
\be
\det I(0)
=
(\ii\pi)^N(\det C)^2\det\mathsf E\;\tau_N(s),
\label{eq:SMOverlapZero}
\ee
and all basis-dependent factors cancel in the ratio,
\be
\mathcal F_N(q;s)
=
\frac{\tau_N(s-\ii q)}{\tau_N(s)}
=
\frac{\mathcal Z_N^{(-)}(E+\ii qT^{2/3})}{\mathcal Z_N^{(-)}(E)},
\qquad
q>0.
\label{eq:SMCharacteristicTauDerivation}
\ee
This proves Eq.~\eqref{eq:characteristicTauRatio}.

Equation~\eqref{eq:SMOverlapZero} has a useful corollary.  Up to the
factor $T^{N/3}$ coming from the change of variables, $\det I(0)$ is
the determinant of the Gram matrix~\eqref{eq:SMGramMatrix}, i.e. the
normalization of the stationary density on the ordered chamber,
Eq.~\eqref{eq:SMbiorthogonalEnsemble}.  Since this density is a
probability density, $\det G\neq0$, and therefore
\be
\tau_N(s)\neq0\qquad\text{for all real } s .
\label{eq:SMtauNonvanishing}
\ee
Equivalently, the outgoing Kontsevich integral
$\mathcal Z_N^{(-)}\propto\tau_N(-E/T^{2/3})$ has no zeros on the real
energy axis, so that $\log\mathcal Z_N^{(-)}$, and with it
$\Xi_N=\Gamma_N-\ii\pi\nu_N$, is well defined for every real $E$.

The derivation was written for generic parameters, where the
polynomial eigenvalues are nondegenerate and the eigenvectors form an
ordinary basis.  The determinant ratio itself is basis independent,
however, and the physical stationary process varies continuously with
$s$.  Equation~\eqref{eq:SMCharacteristicTauDerivation}, as well as
Eq.~\eqref{eq:SMtauNonvanishing}, therefore extend through isolated
degeneracies by continuity; equivalently, one may work there with a
Jordan (confluent) polynomial basis, in which the same determinant
identity is obtained directly.

\paragraph{Check at $N=1$.---}
For one channel the polynomial sector contains only $p_1=1$, with
$e_1=0$, and Eq.~\eqref{eq:SMFourierEquationZ} reduces to the Airy
equation itself.  Equation~\eqref{eq:SMPhysicalRightFourier} gives
$\widetilde v_1(q)=\ii\pi\eta_1\mathcal A_-(s-\ii q)$, and
\be
\mathcal F_1(q;s)
=
\frac{\mathcal A_-(s-\ii q)}
{\mathcal A_-(s)},
\label{eq:SMNOneFourierCheck}
\ee
which is Halperin's result~\cite{Halperin1965}, since
$\tau_1=\mathcal A_-$.  For arbitrary $N$ the single Airy function is
replaced by the Hankel determinant generated by the polynomial sector.

\paragraph{Explicit illustration for $N=2$.---}
For $N=2$ the polynomial sector is
$\mathcal P_1=\operatorname{span}\{1,x\}$, and
Eq.~\eqref{eq:SMDimensionlessMonomialAction} gives
$\widehat\ell_2^\dagger1=x$ and $\widehat\ell_2^\dagger x=s$.  For
$p=c_0+c_1x$ the eigenvalue equation
$\widehat\ell_2^\dagger p=ep$ becomes $sc_1=ec_0$, $c_0=ec_1$, hence
$e^2=s$.  Away from the degeneracy $s=0$ we fix a branch of
$\sqrt s$ and take
\be
e_\pm=\pm\sqrt s,
\qquad
p_\pm(x)=x\pm\sqrt s ,
\qquad
C=
\begin{pmatrix}
\sqrt s&1\\
-\sqrt s&1
\end{pmatrix},
\qquad
(\det C)^2=4s .
\label{eq:SMN2PolynomialModes}
\ee
The final result depends only on $s$, not on the branch.  By
Eq.~\eqref{eq:SMPhysicalRightFourier} the Fourier transforms of the two
physical right modes are
$\widetilde v_\pm=\eta_\pm(\mu_1\pm\sqrt s\,\mu_0)$, and
Eq.~\eqref{eq:SMOverlapExpanded} gives
$I_{ab}=\eta_b\left[\mu_2+(e_a+e_b)\mu_1+e_ae_b\mu_0\right]$.  Using
$\mu_2=z\mu_0$, which is property~(ii),
\be
I(q)
=
\begin{pmatrix}
(z+s)\mu_0+2\sqrt s\,\mu_1 &
(z-s)\mu_0\\[1mm]
(z-s)\mu_0 &
(z+s)\mu_0-2\sqrt s\,\mu_1
\end{pmatrix}
\begin{pmatrix}
\eta_+&0\\
0&\eta_-
\end{pmatrix},
\label{eq:SMN2OverlapMatrix}
\ee
with all moments evaluated at $z=s-\ii q$.  Its determinant is
\be
\det I(q)
=
4s\,\eta_+\eta_-\left(z\mu_0^2-\mu_1^2\right)
=
4s\,\eta_+\eta_-
\det\begin{pmatrix}\mu_0&\mu_1\\ \mu_1&\mu_2\end{pmatrix}
=
(\ii\pi)^2\,4s\,\eta_+\eta_-\,\tau_2(s-\ii q),
\label{eq:SMN2OverlapDeterminant}
\ee
where $\tau_2(z)=z\mathcal A_-^2(z)-[\mathcal A_-'(z)]^2$.  This is the
$N=2$ instance of Eq.~\eqref{eq:SMOverlapDeterminant}, with
$(\det C)^2=4s$.  The prefactor cancels between $q$ and $q=0$, giving
\be
\mathcal F_2(q;s)
=
\frac{
(s-\ii q)\mathcal A_-^2(s-\ii q)
-[\mathcal A_-'(s-\ii q)]^2
}{
s\mathcal A_-^2(s)-[\mathcal A_-'(s)]^2
},
\qquad q>0 .
\label{eq:SMN2Characteristic}
\ee
At $s=0$ the two polynomial eigenvalues coalesce and the basis
$p_\pm$ degenerates, as signalled by $(\det C)^2=4s\to0$, but the
Hankel expression has a regular confluent continuation and remains
valid there.

\section{Quartic Schr\"odinger representation and the large-$N$ saddle}
\label{sec:SMquarticLargeN}

\subsection{Schr\"odinger gauge and projective boundary conditions}
\label{subsec:SMquarticGauge}

The same one-body problem admits a useful Schr\"odinger representation.
Set
\be
g(\lambda)=\frac{U_E(\lambda)}{2T},
\qquad
\phi_a(\lambda)=e^{-g(\lambda)}p_a(\lambda),
\qquad
\chi_a(\lambda)=e^{g(\lambda)}v_a(\lambda).
\label{eq:SMgaugeModes}
\ee
The left and right equations then become the same second-order
differential equation,
\be
H_N\phi_a=-\eps_a\phi_a,
\qquad
H_N\chi_a=-\eps_a\chi_a,
\label{eq:SMquarticEigenproblem}
\ee
with quartic differential expression
\be
H_N
=
-T\partial_\lambda^2
+
\frac{(E+\lambda^2)^2}{4T}
-N\lambda.
\label{eq:SMquarticHamiltonian}
\ee
Moreover the unit concomitant becomes the constant Wronskian
\be
T\bigl(\phi_a\chi_a'-\phi_a'\chi_a\bigr)=1.
\label{eq:SMquarticWronskian}
\ee

It is important that Eq.~\eqref{eq:SMquarticHamiltonian} specifies a
differential expression, not by itself an $L^2(\mathbb R)$ spectral
problem.  The physical stationary sector inherits the projective
boundary condition.  In particular,
\be
\phi_a(\lambda)
\sim
\lambda^{N-1}e^{-U_E(\lambda)/(2T)},
\qquad
\chi_a(\lambda)
\sim
\lambda^{-N-1}e^{U_E(\lambda)/(2T)},
\qquad |\lambda|\to\infty,
\label{eq:SMquarticBC}
\ee
with the two functions decaying at opposite ends of the real axis.
Thus the algebraic levels $-\eps_a$ belong to the projective/current
realization selected by the original Riccati problem; they should not be
identified with the ordinary bound-state spectrum of the quartic
potential.  This distinction will be kept throughout the large-$N$
interpretation below.

\subsection{Semiclassical filled Fermi sea}
\label{subsec:SMFermiSea}

Now take $N\to\infty$ at fixed $J$, so that $T=J^2/(2N)$.  Then
Eq.~\eqref{eq:SMquarticHamiltonian} can be written
\be
H_N
=
N\left[
-\frac{J^2}{2N^2}\partial_\lambda^2
+W_E(\lambda)
\right],
\qquad
W_E(\lambda)
=
\frac{(E+\lambda^2)^2}{2J^2}-\lambda .
\label{eq:SMsemiclassicalHamiltonian}
\ee

The effective semiclassical parameter is therefore
$\hbar_{\rm eff}=J/N$.  The $N$ algebraic one-body levels fill a Fermi
sea of the classical Hamiltonian $p^2/2+W_E(\lambda)$.  If $\mu_F$
denotes its leading Fermi level, the semiclassical Weyl count gives
\be
N
=
\frac{1}{2\pi\hbar_{\rm eff}}
\iint_{p^2/2+W_E(\lambda)\leq\mu_F}
\dd p\,\dd\lambda
=
\frac{N}{2\pi J}
\iint_{p^2/2+W_E(\lambda)\leq\mu_F}
\dd p\,\dd\lambda
\quad\Longrightarrow\quad
1
=
\frac{1}{2\pi J}
\iint_{p^2/2+W_E(\lambda)\leq\mu_F}
\dd p\,\dd\lambda .
\label{eq:SMThomasFermi}
\ee
The exact trace identity~\eqref{eq:SMsumEpsZero} says that the sum of the
filled algebraic one-body energies $-\eps_a$ is zero.  Semiclassically,
this becomes the zero-mean-energy condition
\[
\frac{1}{2\pi J}
\iint_{p^2/2+W_E(\lambda)\leq\mu_F}
\left[\frac{p^2}{2}+W_E(\lambda)\right]
\dd p\,\dd\lambda
=0 .
\]

In the confined regime $E<E^*$ these conditions have a particularly
simple solution.  The quartic potential $W_E$ then has a tilted
double-well structure, with a deeper minimum at
$\lambda_+(E)>0$ and a secondary, higher minimum at
$\lambda_-(E)<0$, separated by a local maximum, see the figure above.

\begin{figure}[t]
\centering
\resizebox{\linewidth}{!}{%
\begin{tikzpicture}[x=1.18cm,y=1.05cm,font=\small,line cap=round,line join=round]
\begin{scope}[xshift=0.0cm]
\node at (0.175,2.72) {(a) $E<E^*$};
\node at (0.175,2.34) {$\epsilon=-5/4$};
\begin{scope}
\clip (-1.55,-1.4) rectangle (1.9,2.0);
\path[fill=blue!12,draw=none] (0.292893218813,1.0625) plot[domain=0.292893218813:1.707106781187,samples=200] (\x,{(\x*\x-1.25)^2-\x}) -- (1.707106781187,1.0625) -- cycle;
\draw[black!65,densely dashed] (-1.55,1.0625) -- (1.9,1.0625);
\draw[black,thick] plot[domain=-1.55:1.9,samples=240] (\x,{(\x*\x-1.25)^2-\x});
\fill (-1,1.0625) circle[radius=1.9pt];
\draw[fill=white,thick] (-0.20710678118654757,1.664213562373095) circle[radius=1.9pt];
\end{scope}
\draw[->,black!65] (-1.55,-1.48) -- (1.98,-1.48);
\node[anchor=west] at (2.01,-1.48) {$x$};
\node[anchor=south] at (-1.55,2.02) {$w_\epsilon$};
\draw[->,black!65] (-1.55,-1.48) -- (-1.55,1.99);
\foreach \v in {-1,0,1} {\draw[black!65] (\v,-1.48) -- (\v,-1.55);\node[anchor=north] at (\v,-1.60) {$\v$};}
\node[anchor=west,inner sep=0pt] at (2.0,1.0625) {$\mu_F/s_J$};
\end{scope}
\begin{scope}[xshift=5.5cm]
\node at (0.175,2.72) {(b) $E=E^*$};
\node at (0.175,2.34) {$\epsilon=-3/4$};
\begin{scope}
\clip (-1.55,-1.4) rectangle (1.9,2.0);
\path[fill=blue!12,draw=none] (-0.500000000000,0.75) plot[domain=-0.500000000000:1.500000000000,samples=200] (\x,{(\x*\x-0.75)^2-\x}) -- (1.500000000000,0.75) -- cycle;
\draw[black!65,densely dashed] (-1.55,0.75) -- (1.9,0.75);
\draw[black,thick] plot[domain=-1.55:1.9,samples=240] (\x,{(\x*\x-0.75)^2-\x});
\fill (-0.5,0.75) circle[radius=1.9pt];
\end{scope}
\draw[->,black!65] (-1.55,-1.48) -- (1.98,-1.48);
\node[anchor=west] at (2.01,-1.48) {$x$};
\node[anchor=south] at (-1.55,2.02) {$w_\epsilon$};
\draw[->,black!65] (-1.55,-1.48) -- (-1.55,1.99);
\foreach \v in {-1,0,1} {\draw[black!65] (\v,-1.48) -- (\v,-1.55);\node[anchor=north] at (\v,-1.60) {$\v$};}
\node[anchor=west,inner sep=0pt] at (2.0,0.75) {$3/4$};
\end{scope}
\begin{scope}[xshift=11.0cm]
\node at (0.175,2.72) {(c) $E>E^*$};
\node at (0.175,2.34) {$\epsilon=-1/2$};
\begin{scope}
\clip (-1.55,-1.4) rectangle (1.9,2.0);
\draw[black,thick] plot[domain=-1.55:1.9,samples=240] (\x,{(\x*\x-0.5)^2-\x});
\end{scope}
\draw[->,black!65] (-1.55,-1.48) -- (1.98,-1.48);
\node[anchor=west] at (2.01,-1.48) {$x$};
\node[anchor=south] at (-1.55,2.02) {$w_\epsilon$};
\draw[->,black!65] (-1.55,-1.48) -- (-1.55,1.99);
\foreach \v in {-1,0,1} {\draw[black!65] (\v,-1.48) -- (\v,-1.55);\node[anchor=north] at (\v,-1.60) {$\v$};}
\end{scope}
\end{tikzpicture}%
}
\caption{Evolution of the tilted quartic potential
$w_\epsilon(x)=W_E(\lambda)/s_J=(x^2+\epsilon)^2-x$,
where $x=\lambda/s_J$ and $\epsilon=E/s_J^2$.
All panels use the same scales; the horizontal axes are below the curves.
(a) Below the edge, the higher minimum (filled dot) and the intervening
maximum (open dot) are distinct. The dashed line is the Fermi level;
shading highlights the classically allowed interval in the deeper
right-hand well. The isolated contact at the higher minimum has zero
phase-space area.
(b) At the edge, the two stationary points merge at $x_*=-1/2$
(filled dot), where $w'=w''=0$ and $w'''\ne0$.
The Fermi level is $3/4$ and the occupied interval is $[-1/2,3/2]$.
(c) Above the edge, the secondary minimum and maximum have disappeared.
No continuation of the real secondary-minimum Fermi level is drawn.}
\label{fig:SMquarticMerger}
\end{figure}
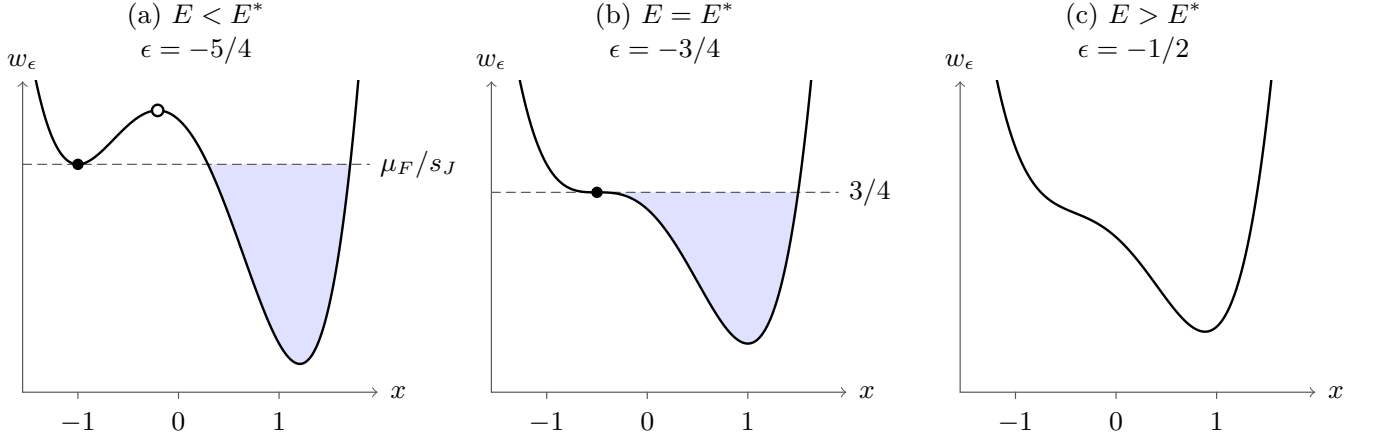

Let
$\lambda_-(E)$ denote this secondary minimum, characterized by
\[
W_E'(\lambda_-)=0,
\qquad
2\lambda_-(E+\lambda_-^2)=J^2 .
\]
The Fermi level coincides exactly with the value of the potential at
this higher minimum,
\be
\mu_F
=
W_E(\lambda_-)
=
\frac{J^2}{8\lambda_-^2}-\lambda_- .
\label{eq:SMFermiLevel}
\ee
Indeed, setting $a^2:=-J^2/\lambda_->0$, one finds the factorization
\be
2J^2\bigl[W_E(\lambda)-\mu_F\bigr]
=
(\lambda-\lambda_-)^2
\left[(\lambda+\lambda_-)^2-a^2\right].
\label{eq:SMFermiFactorization}
\ee
The occupied component of the Fermi sea is therefore the compact
interval
\[
-\lambda_- -a
\leq \lambda
\leq
-\lambda_- +a ,
\]
lying in the deeper right-hand well of $W_E$.
Integrating explicitly over the occupied phase-space region, first over
$p$, and using Eq.~\eqref{eq:SMFermiFactorization}, gives
\be
\iint_{\rm occ}\dd p\,\dd\lambda
=
2\int_{-\lambda_--a}^{-\lambda_-+a}
\sqrt{2\bigl[\mu_F-W_E(\lambda)\bigr]}\,\dd\lambda
=
2\pi J .
\label{eq:SMFermiArea}
\ee
so Eq.~\eqref{eq:SMThomasFermi} is satisfied exactly; the same
factorization also gives the zero mean energy above.  Thus the
state-counting and trace conditions select the same Fermi level.

 Thus, at leading semiclassical order the particles occupy
the deeper right-hand well, although the exact finite-$N$
modes obey projective, rather than $L^2$, boundary conditions.
At the critical merger discussed below,
$\lambda_-\to\lambda_*=-s_J/2$ and
$\mu_F\to\mu_F^*=3s_J/4$.

\subsection{Stieltjes roots and the Allez--Dumaz planar solution}
\label{subsec:SMADSaddle}

The same planar limit is visible directly in the polynomial roots.  For
a sequence of modes, introduce the normalized empirical root measure
\be
\mu_a^{(N)}
:=
\frac{1}{N-1}
\sum_{j=1}^{N-1}\delta_{r_j^{(a)}} .
\label{eq:SMrootDensityFiniteN}
\ee
The roots need not be real at finite $N$.  If, as $N\to\infty$, the
relevant roots condense on a smooth contour $\Sigma$ with limiting
density $\varrho(r)\,\dd r$, then using $2T=J^2/N$ in
Eq.~\eqref{eq:SMStieltjesRoots} gives
\be
\lambda^2+E
=
J^2\,\PV\!\int_{\Sigma}
\frac{\varrho(r)}{\lambda-r}\,\dd r,
\qquad
\int_{\Sigma}\varrho(r)\,\dd r=1 .
\label{eq:SMADCoulombSaddle}
\ee
This is precisely the planar Coulomb-gas saddle equation underlying the
nonconfining cubic Dyson model of Allez and Dumaz~\cite{AllezDumaz2015}.
The exact roots of the finite-$N$ polynomial modes therefore provide an
algebraic discretization of this saddle.

It is useful to recall the corresponding planar solution more
explicitly.  
To avoid confusion with the spectral DOS $\rho_N(E)$, let
$\sigma_E(\lambda)$ denote the normalized large-$N$ Riccati eigenvalue
density.  In terms of the exact finite-$N$ one-point density
$n(\lambda)$ used in the Main Text, which is normalized to $N$, this
means $\sigma_E(\lambda)=\lim_{N\to\infty}N^{-1}n(\lambda)$ whenever
the limit exists. The saddle is encoded by a resolvent $G(z)$, normalized by
$G(z)\sim z^{-1}$ as $z\to\infty$.  Its one-cut parametrization can be
written
\be
\begin{aligned}
G(z)
&=
\frac{1}{J^2}
\left[
z^2+E
-(z+d)\sqrt{(z-\Lambda_-)(z-\Lambda_+)}
\right],
\\
\Lambda_\pm
&=
d\pm\frac{J}{\sqrt d},
\qquad
d^3+Ed+\frac{J^2}{2}=0 ,
\end{aligned}
\label{eq:SMADResolvent}
\ee
where the square-root branch is fixed by $G(z)\sim z^{-1}$.  The
physical branch of $d(E)$ is the one obtained by analytic continuation
from $d(E)\sim-\ii\sqrt E$ at large positive $E$.

In the confined phase $E<E^*$, this branch becomes the largest positive
real solution of Eq.~\eqref{eq:SMADResolvent}; the cut is the real
interval $[\Lambda_-,\Lambda_+]$ and the stationary current vanishes.
Taking the discontinuity of the resolvent gives the compact density
\be
\sigma_E^{\rm conf}(\lambda)
=
\frac{\lambda+d}{\pi J^2}
\sqrt{(\Lambda_+-\lambda)(\lambda-\Lambda_-)}\,
\mathbf 1_{[\Lambda_-,\Lambda_+]}(\lambda) .
\label{eq:SMADConfinedDensity}
\ee
Thus in this phase the planar solution is an ordinary equilibrium
Coulomb gas on a compact real interval.  Notice that this is the same
confined regime in which the Thomas--Fermi sea of
Sec.~\ref{subsec:SMFermiSea} is trapped in the deeper well of $W_E$.

For $E>E^*$, $d(E)$ moves off the real axis and the real compact cut is
lost.  The physical Allez--Dumaz solution is then obtained by continuing
the same resolvent onto the current-carrying branch,
\be
\sigma_E^{\rm flow}(\lambda)
=
-\frac{1}{\pi}\Imn G(\lambda+\ii0)
=
\frac{1}{\pi J^2}
\Imn\!\left[
(\lambda+d)
\sqrt{(\lambda-d)^2-\frac{J^2}{d}}
\right],
\qquad \lambda\in\mathbb R ,
\label{eq:SMADFlowingDensity}
\ee

 The algebraic tails follow by expanding the square root
on the inherited branch:
\[
(\lambda+d)\sqrt{(\lambda-d)^2-\frac{J^2}{d}}
=
\lambda^2+E-\frac{J^2}{\lambda}
-\frac{J^2}{\lambda^2}
\left(d+\frac{J^2}{8d^2}\right)
+O(\lambda^{-3}).
\]
The first three terms are real, hence
\[
\sigma_E^{\rm flow}(\lambda)\sim\frac{j_E}{\lambda^2},
\qquad
j_E=-\frac{1}{\pi}\Imn
\left(d+\frac{J^2}{8d^2}\right)>0,
\qquad \lambda\to\pm\infty.
\]
Here $j_E=\lim_{N\to\infty}\nu_N(E)/N$ is the
stationary winding rate per channel.

 For the matrix Schr\"odinger problem, the confined and flowing
branches correspond respectively to the region below the limiting
spectral edge and to the spectral bulk above it
\cite{FyodorovLeDoussal2020,FyodorovLeDoussalOssipov2025}.

The Thomas--Fermi and electrostatic pictures are therefore two
representations of the same planar algebraic sector:
Eq.~\eqref{eq:SMADCoulombSaddle} uses the zeros of the exact polynomial
modes, whereas Eq.~\eqref{eq:SMThomasFermi} uses the semiclassical phase
space of their quartic-gauge representatives.  In the confined phase
both reduce to a compact real saddle; beyond the transition the physical
Riccati problem selects its current-carrying analytic continuation.

\subsection{Critical point, critical density, and semiclassical scaling}
\label{subsec:SMcriticalScaling}

The collective edge is particularly transparent in the quartic
Schr\"odinger representation.  From
Eq.~\eqref{eq:SMsemiclassicalHamiltonian},
\be
W_E'(\lambda)
=
\frac{2\lambda(E+\lambda^2)}{J^2}-1,
\qquad
W_E''(\lambda)
=
\frac{2(E+3\lambda^2)}{J^2},
\qquad
W_E'(\lambda)=0
\ \Longleftrightarrow\
2\lambda^3+2E\lambda-J^2=0 .
\label{eq:SMquarticDerivatives}
\ee
The transition occurs when the secondary minimum introduced in
Sec.~\ref{subsec:SMFermiSea} merges with the intervening maximum, so
$W_E'(\lambda_*)=W_E''(\lambda_*)=0$.  Writing
$s_J=(2J^2)^{1/3}$ gives
\be
\lambda_*=-\frac{s_J}{2},
\qquad
E^*=-\frac{3s_J^2}{4},
\qquad
\mu_F^*=W_{E^*}(\lambda_*)=\frac{3s_J}{4}.
\label{eq:SMquarticCriticalPoint}
\ee
The last value is the critical limit of the Fermi level found in
Eq.~\eqref{eq:SMFermiLevel}.

Comparing the extremum equation in
Eq.~\eqref{eq:SMquarticDerivatives} with the cubic equation for $d$ in
Eq.~\eqref{eq:SMADResolvent} identifies $d=-\lambda$.  In the confined
phase, $d=-\lambda_-(E)$, where $\lambda_-(E)$ is the higher local
minimum of $W_E$ used in Sec.~\ref{subsec:SMFermiSea}.  At criticality
$d_*=s_J/2$, and the endpoints in Eq.~\eqref{eq:SMADResolvent} become
$\Lambda_-^*=-s_J/2=\lambda_*$ and $\Lambda_+^*=3s_J/2$.  Taking the
critical limit of Eq.~\eqref{eq:SMADConfinedDensity} therefore gives
\be
\sigma_*(\lambda)
=
\frac{1}{\pi J^2}
\left(\lambda+\frac{s_J}{2}\right)^{3/2}
\left(\frac{3s_J}{2}-\lambda\right)^{1/2}
\mathbf 1_{[-s_J/2,\,3s_J/2]}(\lambda) .
\label{eq:SMADCriticalDensity}
\ee
Thus the generic square-root left edge becomes a $3/2$-order zero at
the transition.  Equivalently, the double zero of the planar spectral
curve collides with the left endpoint of the cut.

The same quartic potential also determines the $N$-dependent width of
the critical region.  Set
$\lambda=\lambda_*+\delta\lambda$ and $E=E^*+\delta E$.  Expanding
around Eq.~\eqref{eq:SMquarticCriticalPoint} gives
\be
W_E(\lambda)
=
\mu_F^*
-\frac{\delta E}{s_J}
-\frac{2}{s_J^2}
\left(
\delta E\,\delta\lambda+\delta\lambda^3
\right)
+
O\!\left(
\delta E^2,\,
\delta E\,\delta\lambda^2,\,
\delta\lambda^4
\right).
\label{eq:SMquarticCriticalExpansion}
\ee
The $\lambda$-independent term proportional to $\delta E$ merely shifts
the local Fermi level.  In $H_N/N$, the kinetic term is of order
$N^{-2}\delta\lambda^{-2}$, while the first spatial variation of the
potential is cubic, of order $\delta\lambda^3$.  Balancing these two
terms, and then balancing the mixed term
$\delta E\,\delta\lambda$ with the cubic one, gives
\be
\delta\lambda=O(N^{-2/5}),
\qquad
\delta E=O(N^{-4/5}) .
\label{eq:SMquarticCriticalScales}
\ee
 Thus the $N^{-4/5}$ collective-edge window follows directly
from balancing the kinetic term against the local cubic
variation of the quartic Schr\"odinger potential, without
using either the Thouless argument of Sec.(\ref{sec:SMThouless}) or the string equations
derived below.

\subsection{Relation to the planar string equation}
\label{subsec:SMplanarBridge}

The orthogonal-polynomial analysis of the outgoing cubic matrix
integral, developed in the next section, reaches exactly the same planar
saddle in recurrence variables.  At the endpoint $n=N$, the planar
string equations give Eq.~\eqref{eq:SMendpointcubic},
$d^3+Ed+J^2/2=0$, which is precisely the cubic equation already obtained
for the Allez--Dumaz resolvent in Eq.~\eqref{eq:SMADResolvent}.  The
identification $d=-\lambda$ established above simultaneously maps it
onto the extremum equation of the quartic potential,
Eq.~\eqref{eq:SMquarticDerivatives}.

 At the critical point this gives $d_*=s_J/2=-\lambda_*$,
while the recurrence analysis also gives $r_*=s_J^2/4$.
Thus the continuum limit of the finite-$N$ Stieltjes roots,
the semiclassical filled Fermi sea, the Allez--Dumaz planar
gas, and the planar string equations describe the same
large-$N$ solution.
The Thouless argument developed later provides a complementary physical
interpretation of its breakdown near the edge: the
$N^{-4/5}$ window is reached when the length explored
on the relevant energy scale becomes comparable to the
localization length.
Both this argument and the local quartic-potential balance
recover the critical scale, but not the crossover profile.
To determine that profile, we retain the discrete corrections
to the string equations, as developed next.

\section{String equations and critical regime}

\subsection{Complex orthogonal polynomials, string equations, and Toda identity}

We give the details leading to Eq.~\eqref{eq:string}.  For fixed $N$ define
the symmetric bilinear product
\be
\langle f,g\rangle_-:=
\int_{\Gamma_-}f(\lambda)g(\lambda)e^{-NV_E(\lambda)}\,\dd\lambda,
\qquad
V_E(\lambda)=\frac{2}{J^2}
\left(\frac{\lambda^3}{3}+E\lambda\right).
\label{eq:SMbilinear}
\ee
There is no complex conjugation in~\eqref{eq:SMbilinear}; the product is
therefore not positive, and the quantities below are in general complex.
Let $\pi_n(\lambda)=\lambda^n+\cdots$ be the monic polynomials satisfying
\be
\langle\pi_n,\pi_m\rangle_-=h_n\delta_{nm}.
\label{eq:SMorthogonality}
\ee
We suppress for the moment the $N$-dependence of
$\pi_{n,N}$, $\beta_{n,N}$ and $r_{n,N}$, which will be restored in the
large-$N$ analysis.

It is useful to distinguish these polynomials from the polynomial
eigenmodes $p_a$ of Sec.~\ref{subsec:SMfinitepoly}.  For fixed $N$, the
$p_a$, $a=1,\ldots,N$, form, at generic parameters, an eigenbasis of
$\ell_N^\dagger|_{\mathcal P_{N-1}}$, whereas the
$\pi_n$, $n=0,\ldots,N-1$, form the degree-graded monic basis obtained
by orthogonalizing
$1,\lambda,\ldots,\lambda^{N-1}$ with respect to the contour bilinear
form~\eqref{eq:SMbilinear}.  Thus both families span the same polynomial
space and are related by an invertible change of basis,
\be
p_a(\lambda)
=
\sum_{n=0}^{N-1}A_{an}\pi_n(\lambda),
\qquad
A\in GL(N,\mathbb C),
\label{eq:SMpPiChangeBasis}
\ee
but they are adapted to different structures: the $p_a$ diagonalize
the one-body Riccati operator, while the $\pi_n$ are adapted to the
cubic-contour moment functional.

For the following low-$N$ examples only, let us display explicitly the
dependence on the physical matrix size.  Set
\be
T_N:=\frac{J^2}{2N},
\qquad
s_N:=-\frac{E}{T_N^{2/3}},
\qquad
\mu_k^{(N)}(E)
:=
\int_{\Gamma_-}
\lambda^k e^{-NV_E(\lambda)}\,\dd\lambda .
\label{eq:SMcubicMoments}
\ee
Using $\lambda=-T_N^{1/3}w$ in
Eq.~\eqref{eq:outgoingAiryContourRepresentation} gives the exact moment
formula
\be
\mu_k^{(N)}(E)
=
\ii\pi\,T_N^{(k+1)/3}\,
\mathcal A_-^{(k)}(s_N).
\label{eq:SMmomentAiry}
\ee
Hence the first orthogonal polynomials are
\be
\begin{aligned}
N=1:\qquad&
\pi_0^{(1)}(\lambda)=1,
\\[1mm]
N=2:\qquad&
\pi_0^{(2)}(\lambda)=1,
\qquad
\pi_1^{(2)}(\lambda)
=
\lambda-\frac{\mu_1^{(2)}}{\mu_0^{(2)}},
\\[1mm]
N=3:\qquad&
\pi_0^{(3)}(\lambda)=1,
\qquad
\pi_1^{(3)}(\lambda)
=
\lambda-\frac{\mu_1^{(3)}}{\mu_0^{(3)}},
\\
&
\pi_2^{(3)}(\lambda)
=
\lambda^2
+
\frac{
\mu_1^{(3)}\mu_2^{(3)}
-\mu_0^{(3)}\mu_3^{(3)}
}{
\mathfrak d_2^{(3)}
}\lambda
+
\frac{
\mu_1^{(3)}\mu_3^{(3)}
-\bigl(\mu_2^{(3)}\bigr)^2
}{
\mathfrak d_2^{(3)}
},
\\[-1mm]
&
\hspace{18mm}
\mathfrak d_2^{(3)}
:=
\mu_0^{(3)}\mu_2^{(3)}
-\bigl(\mu_1^{(3)}\bigr)^2 .
\end{aligned}
\label{eq:SMlowNpi}
\ee
In particular, even $\pi_1^{(2)}$ and $\pi_1^{(3)}$ are different
polynomials, because the contour weight $e^{-NV_E}$ itself depends on
the physical matrix size $N$.  We again suppress this $N$-dependence
below.

To see why multiplication by $\lambda$ gives a three-term recurrence,
expand the degree-$n+1$ polynomial $\lambda\pi_n$ in the orthogonal
basis.  Monicity fixes the coefficient of $\pi_{n+1}$ to one.  
Since $\{\pi_0,\ldots,\pi_{n-1}\}$ spans all polynomials of degree
at most $n-1$, orthogonality implies
$\langle\pi_n,q\rangle_-=0$ for every $\deg q<n$.
Thus, for $k\leq n-2$, $\deg(\lambda\pi_k)\leq n-1$ and, using the
symmetry of the bilinear product,
$\langle\lambda\pi_n,\pi_k\rangle_-
=\langle\pi_n,\lambda\pi_k\rangle_-=0$. Hence only three terms remain,
\be
\lambda\pi_n
=
\pi_{n+1}+\beta_n\pi_n+r_n\pi_{n-1},
\qquad
\beta_n=\frac{\langle\lambda\pi_n,\pi_n\rangle_-}{h_n},
\qquad
r_n=\frac{h_n}{h_{n-1}},
\label{eq:SMrecurrence}
\ee
with $r_0=0$.  The last identity follows by taking the bilinear product of the recurrence with $\pi_{n-1}$, which gives
$\langle\lambda\pi_n,\pi_{n-1}\rangle_-=r_n h_{n-1}$.
By symmetry this equals
$\langle\pi_n,\lambda\pi_{n-1}\rangle_-=h_n$, using the recurrence for
$\pi_{n-1}$ and orthogonality. Hence $r_n=h_n/h_{n-1}$.
The same orthogonal polynomials also factorize the $N$-fold integral
defining $\mathcal Z_N^{(-)}$ in Eq.~\eqref{eq:overviewZ}. Since the
$\pi_n$ are monic,
\be 
V_N(\bm\lambda)
=\det_{1\leq i,j\leq N}[\pi_{j-1}(\lambda_i)].
\ee 
Applying Andr\'eief's identity to the two Vandermonde determinants gives
\be 
\mathcal Z_N^{(-)}
=
\det_{0\leq j,k<N}\langle\pi_j,\pi_k\rangle_-
=
\prod_{n=0}^{N-1}h_n .
\ee 

The norms themselves are ratios of successive Hankel minors.  Using
Eq.~\eqref{eq:SMmomentAiry}, and defining
$\tau_0(s):=1$ and
\[
\tau_n(s)
:=
\det_{0\leq j,k<n}
\left[\mathcal A_-^{(j+k)}(s)\right],
\]
one obtains
\be
h_n
=
\frac{
\det_{0\leq j,k\leq n}[\mu_{j+k}]
}{
\det_{0\leq j,k<n}[\mu_{j+k}]
}
=
\ii\pi\,T^{(2n+1)/3}
\frac{\tau_{n+1}(s)}{\tau_n(s)},
\qquad
s=-\frac{E}{T^{2/3}},
\label{eq:SMhnTau}
\ee
where here and below $T=J^2/(2N)$ refers to the fixed physical matrix
size.  For example, writing $A:=\mathcal A_-(s)$,
\be
\begin{aligned}
h_0
&=
\ii\pi T^{1/3}A,
\\
h_1
&=
\ii\pi T\,
\frac{sA^2-(A')^2}{A},
\\
h_2
&=
\ii\pi T^{5/3}
\frac{
-A^3+2sA^2A'-2(A')^3
}{
sA^2-(A')^2
}.
\end{aligned}
\label{eq:SMfirstHn}
\ee
Consequently the complete product telescopes to
\be
\mathcal Z_N^{(-)}(E)
=
\prod_{n=0}^{N-1}h_n
=
(\ii\pi)^N T^{N^2/3}\tau_N(s).
\label{eq:SMZTauExact}
\ee
Thus, with the contour orientation used here,
the $E$-independent constant in Eq.~\eqref{eq:overviewZ} is
$C_N=(\ii\pi)^N$.

The same orthogonal-polynomial construction defines an auxiliary
finite-rank kernel on the cubic contour,
\be
K_N^{\rm cub}(\lambda,\mu)
=
e^{-NV_E(\mu)}
\sum_{n=0}^{N-1}
\frac{\pi_n(\lambda)\pi_n(\mu)}{h_n},
\qquad
\lambda,\mu\in\Gamma_- .
\label{eq:SMcubicKernel}
\ee
Equivalently, the Christoffel--Darboux identity gives
\be
K_N^{\rm cub}(\lambda,\mu)
=
\frac{e^{-NV_E(\mu)}}{h_{N-1}}
\frac{
\pi_N(\lambda)\pi_{N-1}(\mu)
-
\pi_{N-1}(\lambda)\pi_N(\mu)
}{
\lambda-\mu
}.
\label{eq:SMcubicCD}
\ee
Indeed, the normalized cubic-contour weight
\be
\frac{1}{N!\mathcal Z_N^{(-)}}
V_N(\bm\lambda)^2
\prod_{j=1}^N
e^{-NV_E(\lambda_j)}\,\dd^N\lambda,
\qquad
\lambda_j\in\Gamma_-,
\label{eq:SMcubicEnsemble}
\ee
has algebraic $k$-point marginals
\be
R_{k,N}^{\rm cub}(\lambda_1,\ldots,\lambda_k)
=
\det_{1\leq i,j\leq k}
\left[
K_N^{\rm cub}(\lambda_i,\lambda_j)
\right].
\label{eq:SMcubicCorrelations}
\ee
Since $\Gamma_-$ is a complex contour and
$e^{-NV_E(\lambda)}\,\dd\lambda$ is not a positive measure,
Eq.~\eqref{eq:SMcubicEnsemble} is in general a normalized complex
weight rather than a probability distribution.  We therefore refer to
it as a \emph{determinantal contour ensemble}, not as a determinantal
point process; its determinantal identities are algebraic consequences
of Andr\'eief's formula, but its correlation functions need not be real
or non-negative.

The contour kernel~\eqref{eq:SMcubicKernel} is therefore distinct from
the physical stationary Riccati kernel~\eqref{eq:SMGramKernel}.
The latter acts on $\mathbb R$ and uses the left/right pair $(p_a,v_a)$,
whereas $K_N^{\rm cub}$ acts on $\Gamma_-$ and uses the dual contour
pair
$\bigl(\pi_n,e^{-NV_E}\pi_n/h_n\bigr)$.  Their left polynomial sectors
are the same finite-dimensional space $\mathcal P_{N-1}$, related by
Eq.~\eqref{eq:SMpPiChangeBasis}, while their right sectors are different.
The connection between them is the Fourier--Airy transformation derived
in End Matter: the Fourier transforms of the physical modes $v_a$
become linear combinations of the Airy derivatives
$\mathcal A_-^{(n)}$, whose moment matrix is precisely the Hankel matrix
underlying $\mathcal Z_N^{(-)}$.

The string equations follow from two integrations by parts. Since the
weight decays at both ends of $\Gamma_-$, boundary terms vanish. First,
\be
0=\int_{\Gamma_-}
\partial_\lambda\!
\left[\pi_n(\lambda)^2e^{-NV_E(\lambda)}\right]\dd\lambda .
\label{eq:SMIBP1}
\ee
Expanding the derivative gives
\[
0=
2\langle\pi_n',\pi_n\rangle_-
-N\langle V_E'\pi_n,\pi_n\rangle_- .
\]
Because $\pi_n'$ has degree $n-1$, orthogonality gives
$\langle\pi_n',\pi_n\rangle_-=0$, and hence
$\langle V_E'\pi_n,\pi_n\rangle_-=0$.
Using $V_E'=2(\lambda^2+E)/J^2$ and 
using the recurrence once more,
\[
\lambda^2\pi_n
=
\lambda\pi_{n+1}
+\beta_n\lambda\pi_n
+r_n\lambda\pi_{n-1}.
\]
The coefficient of $\pi_n$ in the three terms on the right is,
respectively, $r_{n+1}$, $\beta_n^2$, and $r_n$; all other terms are
orthogonal to $\pi_n$. Therefore
\be 
\langle\lambda^2\pi_n,\pi_n\rangle_-
=
(\beta_n^2+r_n+r_{n+1})h_n.
\ee 
leading to the first string equation in Eq.~\eqref{eq:string}. 
Similarly,
\be
0=\int_{\Gamma_-}
\partial_\lambda\!
\left[\pi_n(\lambda)\pi_{n-1}(\lambda)
e^{-NV_E(\lambda)}\right]\dd\lambda .
\label{eq:SMIBP2}
\ee

Expanding Eq.~\eqref{eq:SMIBP2} gives
\[
0=
\langle\pi_n',\pi_{n-1}\rangle_-
+\langle\pi_n,\pi_{n-1}'\rangle_-
-N\langle V_E'\pi_n,\pi_{n-1}\rangle_- .
\]
Since $\pi_{n-1}'$ has degree $n-2$,
$\langle\pi_n,\pi_{n-1}'\rangle_-=0$. Moreover, monicity implies
$\pi_n'=n\pi_{n-1}+\text{(lower-degree terms)}$, hence
$\langle\pi_n',\pi_{n-1}\rangle_-=nh_{n-1}$.
Using $V_E'=2(\lambda^2+E)/J^2$, the $E$ term vanishes by
orthogonality, so that
\[
\langle\lambda^2\pi_n,\pi_{n-1}\rangle_-
=
\frac{nJ^2}{2N}h_{n-1}.
\]
On the other hand, multiplying
$\lambda\pi_n=\pi_{n+1}+\beta_n\pi_n+r_n\pi_{n-1}$
by $\lambda$ and taking the product with $\pi_{n-1}$ gives
\[
\begin{aligned}
\langle\lambda^2\pi_n,\pi_{n-1}\rangle_-
={}&
\langle\lambda\pi_{n+1},\pi_{n-1}\rangle_-
+\beta_n\langle\lambda\pi_n,\pi_{n-1}\rangle_- \\
&+r_n\langle\lambda\pi_{n-1},\pi_{n-1}\rangle_- .
\end{aligned}
\]
The first term vanishes by orthogonality, while the recurrence gives
$\langle\lambda\pi_n,\pi_{n-1}\rangle_-=r_nh_{n-1}$ and
$\langle\lambda\pi_{n-1},\pi_{n-1}\rangle_-=\beta_{n-1}h_{n-1}$.
Therefore
\[
\langle\lambda^2\pi_n,\pi_{n-1}\rangle_-
=
r_n(\beta_n+\beta_{n-1})h_{n-1},
\]
and comparison with the previous expression yields
the second string equation in Eq.~\eqref{eq:string}. 
Thus the string equations are exact finite-$N$ identities obeyed by the
recurrence coefficients selected by orthogonality on the contour
$\Gamma_-$.

We finally record the Toda identity used below. Introduce
$\theta=-2NE/J^2$, so that the $E$-dependent part of the weight is
$e^{\theta\lambda}$. The polynomials $\pi_n$ depend on $\theta$, but
since they are monic, $\partial_\theta\pi_n$ has degree at most $n-1$.
Differentiating
$h_n=\langle\pi_n,\pi_n\rangle_-$ therefore gives
\[
\partial_\theta h_n
=
2\langle\partial_\theta\pi_n,\pi_n\rangle_-
+\langle\lambda\pi_n,\pi_n\rangle_- .
\]
The first term vanishes by orthogonality, while the recurrence relation
gives
$\langle\lambda\pi_n,\pi_n\rangle_-=\beta_n h_n$. Hence
\be
\partial_\theta\log h_n=\beta_n.
\label{eq:SMToda1}
\ee
To determine $\partial_\theta\pi_n$ more precisely, differentiate
$\langle\pi_n,\pi_k\rangle_-=0$ for $k<n$:
\[
0=
\langle\partial_\theta\pi_n,\pi_k\rangle_-
+\langle\pi_n,\partial_\theta\pi_k\rangle_-
+\langle\lambda\pi_n,\pi_k\rangle_- .
\]
The second term vanishes since $\partial_\theta\pi_k$ has degree below
$n$. For $k\leq n-2$ the last term also vanishes, while for $k=n-1$
it equals $r_nh_{n-1}=h_n$. Since
$\partial_\theta\pi_n$ has degree at most $n-1$, this proves
\[
\partial_\theta\pi_n=-r_n\pi_{n-1}.
\]
Writing $\pi_n(\lambda)=\lambda^n+c_n\lambda^{n-1}+\cdots$, the
three-term recurrence~\eqref{eq:SMrecurrence} gives
$\beta_n=c_n-c_{n+1}$, whereas the last identity gives
$\partial_\theta c_n=-r_n$. Hence
\be
\partial_\theta\beta_n=r_{n+1}-r_n.
\label{eq:SMToda2}
\ee
Using
$\mathcal Z_N^{(-)}=\prod_{n=0}^{N-1}h_n$, Eq.~\eqref{eq:SMToda1}
gives
\be 
\partial_\theta\log\mathcal Z_N^{(-)}
=
\sum_{n=0}^{N-1}\beta_n .
\ee 
Differentiating once more and using Eq.~\eqref{eq:SMToda2},
\be 
\partial_\theta^2\log\mathcal Z_N^{(-)}
=
\sum_{n=0}^{N-1}(r_{n+1}-r_n) = r_N \quad \Rightarrow \quad \partial_E^2\log\mathcal Z_N^{(-)}
=
\left(\frac{2N}{J^2}\right)^2 r_N .
\label{eq:SMToda}
\ee 
since the sum telescopes, and since $r_0=0$, recalling 
that $\theta=-2NE/J^2$. 

\subsection{Planar $N=\infty$ limit and the critical edge}

We now restore explicitly the $N$-dependence of the recurrence
coefficients, writing $\beta_{n,N}$ and $r_{n,N}$.  The exact string
equations~\eqref{eq:string} hold at every finite $N$.  Their planar
limit is obtained by sending $N\to\infty$ while keeping
\be
t=\frac{n}{N}
\label{eq:SMtdef}
\ee
fixed.  Away from a critical point, the recurrence coefficients are
assumed to approach smooth functions of $t$,
\be
\beta_{n,N}\longrightarrow d(t),
\qquad
r_{n,N}\longrightarrow r(t),
\qquad N\to\infty .
\label{eq:SMplanaransatz}
\ee
This is the $N=\infty$ or planar limit.  Since changing $n$ by one
changes $t$ only by $1/N$, neighbouring coefficients have the same
leading limit, for instance
\be
r_{n+1,N}=r(t)+O(N^{-1}),
\qquad
\beta_{n-1,N}=d(t)+O(N^{-1}).
\label{eq:SMneighbors}
\ee
Substituting these leading expressions into the exact string equations
gives
\be
d(t)^2+2r(t)+E=0,
\qquad
2r(t)d(t)=\frac{tJ^2}{2}.
\label{eq:SMplanarstring}
\ee
Eliminating $r(t)$ yields the planar cubic equation
\be
F(d;t,E):=
d^3+Ed+\frac{tJ^2}{2}=0.
\label{eq:SMplanarcubic}
\ee

The physical $N\times N$ Hankel determinant probes the endpoint
$n=N$, hence $t=1$.  
In particular, the Toda relation
\eqref{eq:SMToda} involves precisely the endpoint coefficient $r_{N,N}$.
At this endpoint,
\be
d^3+Ed+\frac{J^2}{2}=0.
\label{eq:SMendpointcubic}
\ee
This is precisely the algebraic equation governing the large-$N$
transition. 
As already found from the Allez--Dumaz planar resolvent in
Sec.~\ref{subsec:SMADSaddle}, Eq.~\eqref{eq:SMendpointcubic} is
identical to the cubic equation for $d$ in
Eq.~\eqref{eq:SMADResolvent}.
The critical point occurs when two roots of the cubic
merge,
\be
F(d_*;1,E^*)=0,
\qquad
\partial_dF(d_*;1,E^*)=3d_*^2+E^*=0.
\label{eq:SMdoubleRoot}
\ee
With $s_J=(2J^2)^{1/3}$ this gives
\be
d_*=\frac{s_J}{2},
\qquad
r_*=\frac{s_J^2}{4},
\qquad
E^*=-\frac{3s_J^2}{4}.
\label{eq:SMcriticalValues}
\ee
This is the same large-$N$ edge obtained in the matrix Schr\"odinger
analyses of
Refs.~\cite{FyodorovLeDoussal2020,FyodorovLeDoussalOssipov2025},
and corresponds to the confined-to-flowing transition of the
nonconfining Dyson gas found by Allez and Dumaz~\cite{AllezDumaz2015}.

It is useful to distinguish this strict $N=\infty$ result from the
ordinary large-$N$ corrections around it.  Away from the critical
point, smoothness in $t$ allows one to expand neighbouring coefficients,
for example
\be
r_{n+1,N}
=
r\left(t+\frac1N\right)
=
r(t)+\frac1N r'(t)+\cdots ,
\label{eq:SMregularExpansion}
\ee
and similarly for $\beta_{n\pm1,N}$.  Such an expansion gives the
regular finite-$N$ corrections to the planar solution.

At the edge this regular expansion becomes singular.  Indeed,
differentiating~\eqref{eq:SMplanarcubic} with respect to $t$ gives
\be
\bigl(3d^2+E\bigr)\,\partial_t d+\frac{J^2}{2}=0,
\qquad
\partial_t d=
-\frac{J^2/2}{3d^2+E}.
\label{eq:SMdtdivergence}
\ee
The denominator vanishes at $(d_*,E^*)$, so the smooth dependence on
$t$ assumed in the regular large-$N$ expansion breaks down there.

The same singularity is visible directly in the dependence on $E$.
Writing $d=d_*+\delta d$ and $E=E^*+\delta E$ in
Eq.~\eqref{eq:SMendpointcubic}, the term linear in $\delta d$ vanishes
at the double root, giving at leading order
\be
d_*\,\delta E+3d_*\,(\delta d)^2+\cdots=0,
\qquad\text{i.e.}\qquad
E-E^*\simeq-3(d-d_*)^2 .
\label{eq:SMdoubleRootExpansion}
\ee
Thus the two planar branches separate with a square-root singularity,
$d-d_*\sim(E^*-E)^{1/2}$.  Near this point the $1/N$ corrections from
the discrete shifts $n\to n\pm1$ can no longer be treated
perturbatively.  One must instead scale simultaneously the distance
$n-N$, the deviations of $\beta_{n,N}$ and $r_{n,N}$ from their
critical values, and $E-E^*$.  This is the double-scaling regime
described next.

\subsection{Double scaling and the Painlev\'e-I equation}

We now resolve the breakdown of the planar approximation at the double
root.  Let $\ell$ denote the number of neighbouring polynomial degrees
participating in the critical region, and let $A$ denote the typical
deviation of $\beta_{n,N}$ and $r_{n,N}$ from their critical values.
Three effects enter the exact string equations at the first nontrivial
order.  The degeneracy of the planar cubic makes the leading
nonlinearity quadratic, of order $A^2$.  The discreteness in $n$ produces
second differences of order $A/\ell^2$, while moving a distance $\ell$
from $n=N$ changes $t=n/N$ by $O(\ell/N)$.  Hence the critical regime
requires
\be
A^2\sim\frac{A}{\ell^2}\sim\frac{\ell}{N}.
\label{eq:SMbalance}
\ee
The first equality gives $A\sim\ell^{-2}$, and the second then gives
$\ell^5\sim N$.  Thus
\be
\ell\sim N^{1/5},
\qquad
A\sim N^{-2/5}.
\label{eq:SMcriticalScales1}
\ee
Since near the double root $E-E^*=O(A^2)$, the corresponding energy
window is
\be
E-E^*=O(N^{-4/5}).
\label{eq:SMcriticalScales2}
\ee
This explains the three scales quoted in the main text.

It is convenient to introduce
\be
\delta=N^{-1/5},
\qquad
\xi_n=(n-N)\delta,
\qquad
E=E^*+s_J^2\delta^4 X.
\label{eq:SMDSvariables}
\ee
Then
\be
\frac{n}{N}=1+\delta^4\xi_n,
\qquad
n\to n\pm1
\quad\Longleftrightarrow\quad
\xi_n\to\xi_n\pm\delta .
\label{eq:SMlatticeShift}
\ee
Thus the $O(N^{1/5})$ lattice window around $n=N$ becomes a continuum
coordinate $\xi=O(1)$.

To expand the string equations it is useful to remember that
$\beta_{n,N}$ is naturally associated with the site $n$, whereas
$r_{n,N}$ multiplies $\pi_{n-1}$ in the recurrence and is naturally
associated with the bond $n-\frac12$.  We therefore write
\be
\begin{aligned}
\beta_{n,N}
&=
s_J\left[
\frac12+\delta^2 b(\xi_n)
+\delta^4 b_1(\xi_n)+\cdots
\right],\\
r_{n,N}
&=
s_J^2\left[
\frac14+
\delta^2 a\!\left(\xi_n-\frac{\delta}{2}\right)
+\delta^4 a_1\!\left(\xi_n-\frac{\delta}{2}\right)
+\cdots
\right].
\end{aligned}
\label{eq:SMDSansatz}
\ee
The midpoint convention makes the combinations entering the string
equations symmetric.  For example,
\be
a\!\left(\xi-\frac{\delta}{2}\right)
+a\!\left(\xi+\frac{\delta}{2}\right)
=
2a(\xi)+\frac{\delta^2}{4}a''(\xi)+O(\delta^4).
\label{eq:SMmidpointa}
\ee

Consider first
$\beta_{n,N}^2+r_{n,N}+r_{n+1,N}+E=0$.
The critical values
$d_*=s_J/2$, $r_*=s_J^2/4$ and
$E^*=-3s_J^2/4$ cancel the order-one terms.  At order $\delta^2$ one
finds
\be
b+2a=0.
\label{eq:SMleadingRelation}
\ee
At the next order, $\delta^4$, the same equation gives
\be
b^2+b_1+2a_1+\frac14a''+X=0.
\label{eq:SMDSfirst}
\ee

We next expand the second string equation,
\be
r_{n,N}\bigl(\beta_{n,N}+\beta_{n-1,N}\bigr)
=
\frac{nJ^2}{2N}.
\label{eq:SMsecondStringRepeat}
\ee
It is natural here to expand about the bond $n-\frac12$.  The two
neighbouring site values occur symmetrically,
\be
b\!\left(\xi+\frac{\delta}{2}\right)
+b\!\left(\xi-\frac{\delta}{2}\right)
=
2b(\xi)+\frac{\delta^2}{4}b''(\xi)+O(\delta^4).
\label{eq:SMmidpointb}
\ee
Using $s_J^3=2J^2$, the right-hand side becomes
\be
\frac{nJ^2}{2N}
=
\frac{s_J^3}{4}\left(1+\delta^4\xi+\cdots\right).
\label{eq:SMdegreeExpansion}
\ee
At order $\delta^2$ the second string equation again gives
\be
a+\frac{b}{2}=0,
\label{eq:SMleadingRelation2}
\ee
which is identical to Eq.~\eqref{eq:SMleadingRelation}.  At order
$\delta^4$ it gives
\be
a_1+\frac{b_1}{2}
+\frac1{16}b''
+2ab
=
\frac{\xi}{4}.
\label{eq:SMDSsecond}
\ee

The two $O(\delta^4)$ equations contain the subleading functions
$a_1$ and $b_1$, but they can be eliminated.  From
Eq.~\eqref{eq:SMleadingRelation}, $b=-2a$, while
Eqs.~\eqref{eq:SMDSfirst} and~\eqref{eq:SMDSsecond} give respectively
\be
b_1+2a_1
=
-b^2-\frac14a''-X
\label{eq:SMelim1}
\ee
and
\be
b_1+2a_1
=
\frac{\xi}{2}-\frac18b''-4ab.
\label{eq:SMelim2}
\ee
Equating these two expressions and using $b=-2a$ yields the closed
continuum equation
\be
a''(\xi)+24a(\xi)^2+\xi+2X=0.
\label{eq:SMcontinuuma}
\ee

To bring it to canonical form, define
\be
u(\xi)=-4a(\xi).
\label{eq:SMudef}
\ee
Then
\be
u''=6u^2+4(\xi+2X).
\label{eq:SMuEquation}
\ee
Finally introduce
\be
\mathfrak t=4^{1/5}(\xi+2X),
\qquad
u(\xi)=4^{2/5}y(\mathfrak t).
\label{eq:SMcanonicalRescaling}
\ee
Equation~\eqref{eq:SMuEquation} becomes
\be
y''(\mathfrak t)=6y(\mathfrak t)^2+\mathfrak t,
\label{eq:SMPainleveI}
\ee
which is the canonical Painlev\'e-I equation.

Using $b=-2a$ and the rescaling above, the leading recurrence
coefficients may equivalently be written
\be
\begin{aligned}
\beta_{n,N}
&=
\frac{s_J}{2}
+\frac{s_J}{2^{1/5}N^{2/5}}\,
y(\mathfrak t)+\cdots,\\
r_{n,N}
&=
\frac{s_J^2}{4}
-\frac{s_J^2}{2^{6/5}N^{2/5}}\,
y(\mathfrak t)+\cdots,
\end{aligned}
\label{eq:SMfinalDSansatz}
\ee
with
\be
\mathfrak t
=
2^{7/5}X
+4^{1/5}(n-N)N^{-1/5}.
\label{eq:SMmathfrakt}
\ee
This is the double-scaling ansatz quoted in the End Matter.  At this
stage the discrete string equations determine the Painlev\'e-I
\emph{equation}; the choice of the particular solution requires the
additional information carried by the contour $\Gamma_-$ and is
discussed separately below.

\paragraph{From the recurrence coefficient to the density.}
The Toda identity~\eqref{eq:SMToda} directly relates the endpoint
recurrence coefficient to the spectral observables. Since
\[
\Xi_N(E)=-T\partial_E\log\cZ_N^{(-)}(E),
\qquad
T=\frac{J^2}{2N},
\]
one finds
\be
\partial_E\Xi_N(E)
=
-T\partial_E^2\log\cZ_N^{(-)}
=
-\frac{r_{N,N}}{T}.
\label{eq:SMXiToda}
\ee
Writing $\Xi_N=\Gamma_N-\ii\pi\nu_N$ and
$\rho_N=\partial_E\nu_N$, the imaginary part of
Eq.~\eqref{eq:SMXiToda} gives the exact relation
\be
\rho_N(E)=\frac{1}{\pi T}\Imn r_{N,N}.
\label{eq:SMrhoFromr}
\ee
Thus only the imaginary part of the critical correction to $r_{N,N}$
contributes to the density; the critical planar value
$r_*=s_J^2/4$ is real.

At $n=N$ one has $\xi=0$ and therefore
$\mathfrak t=2^{7/5}X$. Equation~\eqref{eq:SMfinalDSansatz} gives
\be
r_{N,N}
=
\frac{s_J^2}{4}
-\frac{s_J^2}{2^{6/5}N^{2/5}}
y(2^{7/5}X)+\cdots .
\label{eq:SMrEndpoint}
\ee
Since $s_J^3=2J^2$, one has
$T=s_J^3/(4N)$. Substitution into
Eq.~\eqref{eq:SMrhoFromr} therefore yields
\be
\rho_N(E)
\sim
-\frac{2^{4/5}N^{3/5}}{\pi s_J}
\Imn y(2^{7/5}X).
\label{eq:SMrhoScaling}
\ee
Defining
\be
\Phi'(X):=
-\frac{2^{4/5}}{\pi}\Imn y(2^{7/5}X),
\label{eq:SMPhiPrime}
\ee
one recovers
\be
\rho_N(E)\sim\frac{N^{3/5}}{s_J}\Phi'(X).
\label{eq:SMrhoPhi}
\ee
Finally, since
$E=E^*+s_J^2N^{-4/5}X$ and
$\rho_N=\partial_E\nu_N$, integration gives the corresponding form
\be
\nu_N(E)=s_JN^{-1/5}\Phi(X),
\label{eq:SMnuPhi}
\ee
up to an $X$-independent constant. Its determination, together with
the choice of the relevant Painlev\'e-I solution, requires the global
contour information and is addressed separately.

\subsection{Selection of the Painlev\'e-I solution}

The double-scaling analysis above determines the differential equation
\[
y''=6y^2+\mathfrak t,
\]
but not its solution.  This distinction is essential here.  Different
solutions can have the same algebraic asymptotic expansion while differing
by exponentially small terms, and on the $E<E^*$ side these terms produce
the spectral density.  The required solution is fixed by the contour
$\Gamma_-$ inherited from the finite-$N$ cubic integral.

To make this identification we use the complex cubic matrix model studied
by Bleher and Dea\~no~\cite{BleherDeano2013,BleherDeano2016}.  Their
potential can be written
\be
V_{\rm BD}(z;u)=\frac{z^2}{2}-u z^3 .
\label{eq:SMBDpotential}
\ee
For $E<0$, our potential $V_E(\lambda)$ is brought exactly to this form
by the affine transformation
\be
\lambda=B_E+A_Ez,\qquad
B_E=\sqrt{-E},\qquad
A_E=-\frac{J}{2(-E)^{1/4}} .
\label{eq:SMBDmap}
\ee
Indeed, using $B_E^2+E=0$ and
$A_E^2=J^2/(4B_E)$, one obtains
\be
V_E(B_E+A_Ez)
=
{\rm const}+\frac{z^2}{2}-u(E)z^3,
\qquad
u(E)=\frac{J}{12(-E)^{3/4}} .
\label{eq:SMBDcoupling}
\ee
Thus our outgoing cubic integral is, apart from an $E$-dependent
translation and elementary prefactors, a member of the cubic family
analyzed in Refs.~\cite{BleherDeano2013,BleherDeano2016}.

Bleher and Dea\~no found a critical coupling
\be
u_c=\frac{3^{1/4}}{18}.
\label{eq:SMBDuc}
\ee
At our edge
$E^*=-3s_J^2/4$, with $s_J=(2J^2)^{1/3}$,
Eq.~\eqref{eq:SMBDcoupling} gives precisely
\be
u(E^*)=u_c .
\label{eq:SMBDcriticalmatch}
\ee
Hence the critical point found from our planar string equations is
exactly the critical point of their cubic model.

The agreement also fixes the normalization of the double-scaling
coordinate.  In the notation of Ref.~\cite{BleherDeano2016}, the
Painlev\'e variable $\lambda_{\rm BD}$ is defined by
\be
N^{4/5}(u-u_c)=c_1\lambda_{\rm BD},
\qquad
c_1=2^{-12/5}3^{-7/4}.
\label{eq:SMBDscaling}
\ee
Expanding Eq.~\eqref{eq:SMBDcoupling} for
$E=E^*+s_J^2N^{-4/5}X$ gives
\be
N^{4/5}(u-u_c)=u_c X+o(1).
\label{eq:SMBDexpansion}
\ee
Since $u_c/c_1=2^{7/5}$, one obtains
\be
\lambda_{\rm BD}=2^{7/5}X .
\label{eq:SMBDvariablematch}
\ee
This is exactly the value
$\mathfrak t=2^{7/5}X$ obtained from our string equations at the physical
endpoint $n=N$.

It remains to identify the contour.  Bleher and Dea\~no define the cubic
model on admissible complex contours connecting decay sectors of the
cubic exponential.  Their family may be parametrized by a parameter
$\alpha$; the endpoint $\alpha=1$ corresponds to a particular contour
$\Gamma_0$ and selects the Painlev\'e-I tritronqu\'ee solution denoted
$y_1$ in Ref.~\cite{BleherDeano2016}.  Under
Eq.~\eqref{eq:SMBDmap}, since $A_E<0$, our outgoing contour
\be
\Gamma_-:\ e^{-2\pi\ii/3}\infty\longrightarrow+\infty
\label{eq:SMoutgoingAgain}
\ee
is mapped to
\be
e^{\ii\pi/3}\infty\longrightarrow-\infty .
\label{eq:SMmappedContour}
\ee
The ray $\arg z=\pi/3$ lies in the same cubic decay sector as the
corresponding ray used for $\Gamma_0$ and may be deformed to it without
crossing a growth sector.  The orientation is reversed; this only
multiplies each one-dimensional moment by $-1$, and hence the
$N$-particle integral by the $E$-independent factor $(-1)^N$.  It does
not affect logarithmic derivatives or normalized observables.  Therefore
our contour corresponds precisely to the Bleher--Dea\~no choice
$\alpha=1$.

Consequently the Painlev\'e solution entering the physical
$E+\ii0$ boundary value is
\be
y(\mathfrak t)=y_1(\mathfrak t),
\label{eq:SMphysicalPI}
\ee
the tritronqu\'ee solution selected by that contour.  The conjugate
contour $\overline{\Gamma_-}$, corresponding to the opposite
$E-\ii0$ boundary value, selects the conjugate solution.  Thus the
finite-$N$ choice between the two outgoing Airy combinations is carried
in the double-scaling limit into the corresponding choice of
Painlev\'e-I Stokes data.

The important point is that Eq.~\eqref{eq:SMPainleveI} determines only
the local critical equation, whereas the original Riccati boundary
condition fixes the global contour and therefore the particular solution
of that equation.  The asymptotic consequences of this choice are
discussed in the main text.

\paragraph{Painlev\'e-I tau-function interpretation.}
The same result has a useful partition-function formulation.  Let
$\mathcal F(X)$ denote the $O(1)$ singular part of
$\log\mathcal Z_N^{(-)}$ in the double-scaling limit, after subtraction
of its regular large-$N$ contribution.  Combining the exact Toda
identity~\eqref{eq:SMToda}, the endpoint scaling
\eqref{eq:SMrEndpoint}, and the physical selection
$y=y_1$ in Eq.~\eqref{eq:SMphysicalPI}, one may define the associated
Painlev\'e-I tau function by
\be
\frac{\dd^2}{\dd\mathfrak t^2}
\log\tau_{\rm PI}^{(1)}(\mathfrak t)
=
-y_1(\mathfrak t),
\qquad
\mathcal F(X)
=
\log\tau_{\rm PI}^{(1)}(2^{7/5}X).
\label{eq:SMPITau}
\ee
The second relation is understood up to an affine function of
$\mathfrak t$, which is absorbed into the regular part of the free
energy.  This is the partition-function form of the
Bleher--Dea\~no double-scaling result.  Since
$\Xi_N=-T\partial_E\log\mathcal Z_N^{(-)}$ and
$\nu_N=-\pi^{-1}\Imn\Xi_N$, the critical IDOS scaling function may
equivalently be written
\be
\Phi(X)
=
\frac{2^{-3/5}}{\pi}
\Imn\frac{\dd}{\dd\mathfrak t}
\log\tau_{\rm PI}^{(1)}(\mathfrak t),
\qquad
\mathfrak t=2^{7/5}X .
\label{eq:SMPhiTau}
\ee
Differentiating Eq.~\eqref{eq:SMPhiTau} reproduces
Eq.~\eqref{eq:SMPhiPrime}; the tau-function formulation therefore
encodes the same critical result directly at the level of the
partition function.

\section{Thouless heuristics for critical scaling identifications}
\label{sec:SMThouless}

The exponents of the critical regime, $E-E^*\sim N^{-4/5}$, $\nu_N\sim N^{-1/5}$, $\rho_N\sim N^{3/5}$ and the forbidden-side exponent $5/4$, follow exactly from the string equations. It is nevertheless instructive to see how they arise from elementary transport considerations, which also explains why they differ from the familiar $N^{-2/3}$ edge scaling of Gaussian random matrices although the large-$N$ density of states vanishes as a square root in both cases. The argument below is a heuristic in the spirit of Thouless: it uses only mean-field ($N=\infty$) quantities, computed away from the critical window, and determines the window as the scale at which these quantities stop being self-consistent. It fixes powers of $N$, not scaling functions. Throughout we measure energies in units of $s_J^2$ and lengths in units of $s_J^{-1}$, and write $\delta=E-E^*$.

\subsection{Why the Gaussian counting argument has to be modified}

For an $N\times N$ GUE matrix with semicircular density $\rho(\lambda)\simeq\pi^{-1}\sqrt{2-\lambda}$ one estimates the width $\delta$ of the edge window by requiring that it contains of order one eigenvalue, $N\int_{2-\delta}^2\rho\,d\lambda\sim N\delta^{3/2}\sim 1$, whence $\delta\sim N^{-2/3}$. Implicit in this count is the assumption that an eigenvector at the edge extends over the whole matrix, so that all $N$ levels compete for the window. In a quasi-one-dimensional system this is not the case: an eigenstate near the edge occupies only a finite portion of the wire, and the count must be performed in that portion. The size of this portion depends itself on the distance $\delta$ to the edge, and this is what changes the exponents. The same mechanism operates in random banded matrices (see the last subsection), where the number of eigenvalues is finite and the density is exactly semicircular, and yet the edge window scales as $W^{-4/5}$ rather than $M^{-2/3}$.

\subsection{Mean-field Green function and the planar cubic}

At $N=\infty$ the disorder-averaged Green function of \eqref{eq:Hmain} is obtained in the self-consistent Born approximation. Since the variance of each matrix element of $W(x)$ is $2T=J^2/N$, diagrams with crossing impurity lines are suppressed by powers of $1/N$, exactly as in the planar limit of the Gaussian ensembles, and the self-consistent Born approximation is exact in this limit. Because the noise is $\delta$-correlated in $x$, the self-energy carries no momentum dependence and is proportional to the local Green function $g=N^{-1}\Tr\,\bar G(x,x)$,
\be
\Sigma(E)=2T\,N\,g=J^2g,
\qquad
\bar G(k)=\frac{1}{E-\Sigma(E)-k^2}.
\label{eq:SMThSigma}
\ee
Note that the factor $1/N$ in the variance is compensated by the summation over $N$ intermediate channels, so that $\Sigma$ and all per-channel quantities below are independent of $N$. The averaged propagator has the clean functional form with $E$ replaced by $E-\Sigma$. Writing $E-\Sigma=-\kappa^2$ with $\operatorname{Re}\kappa>0$, one has $\bar G(x,x')=-e^{-\kappa|x-x'|}/(2\kappa)$ and $g=-1/(2\kappa)$, and the self-consistency condition $-\kappa^2=E+J^2/(2\kappa)$ becomes
\be
\kappa^3+E\kappa+\frac{J^2}{2}=0 ,
\label{eq:SMThcubic}
\ee
which is the planar cubic~\eqref{eq:SMplanarcubic} with $d\equiv\kappa$. The identification is not only formal: with $r=J^2/(4d)$ the exact relation $\rho_N=(\pi T)^{-1}\Imn r_{N,N}$ of Eq.~\eqref{eq:SMrhoFromr} gives in the planar limit $\rho_N=N\rho_1$, where $\rho_1=-\pi^{-1}\Imn g=|\Imn\kappa|/(2\pi|\kappa|^2)$ is the density of states per channel and unit length of the self-consistent Born approximation. The planar recurrence coefficient $d$ is therefore the inverse decay length of the averaged Green function.

The edge is the double root of~\eqref{eq:SMThcubic}, $\kappa_*=s_J/2$, $E^*=-3\kappa_*^2$. At this point $E-\operatorname{Re}\Sigma=-\kappa_*^2<0$: the edge lies below the bottom of the clean band and is created entirely by the disorder. There is no propagating clean velocity; the averaged wave decays on the length $\kappa_*^{-1}$, which plays the role of the mean free path. For $\delta>0$ the double root splits into a complex pair, $\kappa-\kappa_*\simeq-\ii\sqrt{\delta/3}$ (retarded branch), so that
\be
\rho_1(E)\simeq\frac{1}{2\pi\kappa_*^2}\sqrt{\frac{\delta}{3}} ,
\qquad
\rho_N=N\rho_1 .
\label{eq:SMThrho}
\ee
In the scaling variable $\delta=s_J^2N^{-4/5}X$ this reproduces exactly the large-$X$ behaviour~\eqref{eq:PhiPlusAsym} of the critical density~\eqref{eq:criticalDensity}.

\subsection{Scattering rate and diffusion along the wire}

The imaginary part of the self-energy is the golden-rule rate of scattering out of a given channel,
\be
\frac{1}{\tau}=-2\,\Imn\Sigma=2\pi J^2\rho_1 .
\label{eq:SMThtau}
\ee
A particle scatters into each of the $N$ channels with squared amplitude $J^2/N$ and final density of states $\rho_1$, so the rate is independent of $N$. Near the edge it vanishes as $\sqrt\delta$, simply because there are few states to scatter into.

The diffusion constant along the wire follows from the Kubo formula. The current vertex $2k$ is odd in $k$ while the impurity lines carry no momentum, so the ladder corrections to the current vertex vanish and the bare bubble built from $\bar G$ gives the full $N=\infty$ (Drude) conductivity. Using $\Imn\bar G(k)\simeq 2\kappa_*\Imn\kappa/(\kappa_*^2+k^2)^2$ near the edge, the conductivity per channel is $\sigma_1=\pi^{-1}\int\frac{dk}{2\pi}(2k)^2[\Imn\bar G(k)]^2=(\Imn\kappa)^2/(2\pi\kappa_*^3)$. The Einstein relation $\sigma_1=\rho_1D$ then gives
\be
D=\frac{|\Imn\kappa|}{\kappa_*}=2\pi\kappa_*\,\rho_1\simeq\frac{1}{\kappa_*}\sqrt{\frac{\delta}{3}} .
\label{eq:SMThD}
\ee
Two features are essential. First, $D$ is independent of $N$: the number of channels enters the problem only through the total density of states, not through the speed of spreading along the wire. Second, $D$ is proportional to the density of states and vanishes as $\sqrt\delta$ at the edge. The diffusive description applies on times longer than $\tau\sim\delta^{-1/2}$, i.e. on lengths longer than $\sqrt{D\tau}\sim\kappa_*^{-1}$.

\subsection{Localization length}

In a quasi-one-dimensional wire the localization length is the length on which the dimensionless Thouless conductance $g(L)=E_{\rm Th}(L)/\Delta(L)$ drops to one~\cite{Thouless1972,Dorokhov1982,EfetovLarkin1983}. Here $E_{\rm Th}(L)=D/L^2$ is the inverse time needed to diffuse across a segment of length $L$, and $\Delta(L)=1/(\rho_N L)$ is the mean level spacing in that segment, with $\rho_N=N\rho_1$ the density of states per unit length summed over all channels. Hence $g(L)=\rho_ND/L$ and, up to a numerical factor of order one,
\be
L_{\rm loc}(E)\sim\rho_N D=N\rho_1D\sim N\delta .
\label{eq:SMThxi}
\ee
Deep in the band the same formula gives the familiar $L_{\rm loc}\sim N\ell$, with $\ell$ the mean free path. Near the edge the localization length shrinks linearly in $\delta$, because both the density of states and the diffusion constant vanish there as $\sqrt\delta$.

\subsection{The critical window}

To resolve the structure of the spectrum on the energy scale $\delta$, a state has to evolve for a time $t_\delta\sim1/\delta$. During this time it spreads diffusively over the length
\be
L_\delta=\sqrt{D\,t_\delta}\sim\delta^{-1/4},
\label{eq:SMThLdelta}
\ee
again independent of $N$, and longer than the mean free path for $\delta\ll1$. The relevant volume for the counting argument is therefore $N$ channels times the length $L_\delta$. The number of levels within $\delta$ of the edge in this volume is
\be
n(\delta)=L_\delta\,N\!\int_{E^*}^{E^*+\delta}\!\rho_1(E)\,dE\sim\delta^{-1/4}\,N\delta^{3/2}=N\delta^{5/4},
\label{eq:SMThcount}
\ee
in place of $N\delta^{3/2}$ for the Gaussian ensembles. The extra factor $\delta^{-1/4}$ is the $\delta$-dependence of the explored volume. The edge window is the range of energies where this number is of order one,
\be
n(\delta)\sim1
\quad\Longleftrightarrow\quad
E-E^*\sim s_J^2N^{-4/5},
\label{eq:SMThwindow}
\ee
in agreement with the double-scaling analysis. In terms of the scaling variable, $n(\delta)\sim|X|^{5/4}$.

The same condition has a second, equivalent reading. Setting $L_\delta\sim L_{\rm loc}(\delta)$ with~\eqref{eq:SMThxi} and~\eqref{eq:SMThLdelta} gives $\delta^{-1/4}\sim N\delta$, which is again $N\delta^{5/4}\sim1$. This is not a coincidence: by construction $E_{\rm Th}(L_\delta)=\delta$, and the counting condition states $\Delta(L_\delta)=\delta$, so the Thouless conductance of the explored segment is of order one. The critical window is thus the energy range in which a single localization volume contains of order one level. Outside it, for $\delta\gg N^{-4/5}$, each localization volume contains many levels and the smooth mean-field density applies. Inside it, levels are resolved individually and the mean-field description fails. For the Gaussian ensembles there is no spatial structure, the localization volume is the whole matrix, and one recovers $N\delta^{3/2}\sim1$.

The remaining critical exponents follow at once. At the window the integrated density of states is $\nu_N\sim N\delta^{3/2}\sim N^{-1/5}$ and the density is $\rho_N\sim N\delta^{1/2}\sim N^{3/5}$, as in Eq.~\eqref{eq:criticalDensity}. The edge states extend over $L_\delta\sim L_{\rm loc}\sim N^{1/5}$ mean free paths, i.e. over $N\cdot N^{1/5}=N^{6/5}$ channel lengths. The deviation of the planar recurrence coefficient, $d-d_*=\kappa-\kappa_*\sim\sqrt\delta\sim N^{-2/5}$, reproduces the amplitude $A\sim N^{-2/5}$ of the double-scaling ansatz. The number $N^{1/5}$ of polynomial degrees that participate in the critical region of the string equations has the same magnitude as $L_\delta$; we note this correspondence without attempting to derive it.

On the forbidden side the quantity $n(|\delta|)$ also controls the exponent of the density. For the Gaussian ensembles the right tail of the Tracy--Widom law is $\exp[-\frac43N\delta^{3/2}]$, i.e. minus the logarithm of the probability of a level at distance $\delta$ outside the edge is of the order of the number of levels the mean field would place in a window of the same size. The same rule applied to~\eqref{eq:SMThcount} gives $-\log\rho_N\sim N|\delta|^{5/4}$, which is the exponent of Eq.~\eqref{eq:Phiasym} and of the single-eigenvalue activation barrier of Ref.~\cite{FyodorovLeDoussalOssipov2025}. We stress that this last step is an observation about the structure of the result rather than an independent derivation.

The status of the ingredients is as follows. The planar Green function, the cubic~\eqref{eq:SMThcubic} and the edge density~\eqref{eq:SMThrho} are exact at $N=\infty$. The diffusion constant~\eqref{eq:SMThD} is the exact $N=\infty$ Drude value. The localization length~\eqref{eq:SMThxi} and the counting~\eqref{eq:SMThcount} are heuristic: they extrapolate mean-field quantities to the scale where, by construction, the $1/N$ (weak-localization) corrections become of order one. Such corrections modify numerical factors but not the powers of $N$, which are confirmed independently by the exact double-scaling analysis.

\subsection{Random banded matrices}

The same reasoning applies to $M\times M$ random banded matrices of bandwidth $W\gg1$, normalized so that the semicircle has its edge at $2$, with matrix elements of variance $\sim1/W$ within the band. The density of states per site is semicircular, $\rho\propto\sqrt\delta$ with $\delta=2-\lambda$. A state on a given site is coupled to $\sim W$ others, each with squared amplitude $\sim1/W$, so the golden-rule hopping rate is $\sim\rho$ and each hop displaces the particle by $\sim W$ sites. Hence $D\sim W^2\rho\propto W^2\sqrt\delta$ in units of sites squared per unit time, and the localization length in sites is $L_{\rm loc}\sim\rho D\sim W^2\rho^2$, which in the bulk reduces to the known $L_{\rm loc}\propto W^2$~\cite{FyodorovMirlin1991}. A state resolving the window $\delta$ explores
\be
L_\delta=\sqrt{D/\delta}\sim W\delta^{-1/4}
\ee
sites, and the number of levels within $\delta$ of the edge in this volume is $L_\delta\,\delta^{3/2}\sim W\delta^{5/4}$. Setting it to one gives the edge scale $\delta\sim W^{-4/5}$ of Silvestrov and Sodin~\cite{Silvestrov1997,Sodin2010}, and, as before, the equivalent condition $L_\delta\sim L_{\rm loc}(\delta)$. The edge states then extend over $L_\delta\sim W\cdot W^{1/5}=W^{6/5}$ sites, which is precisely the normalization $W^{6/5}M^{-1}$ in the definition of $\sigma_2(\xi)$ in the main text: $\sigma_2$ counts edge levels per localization volume. If $M\ll W^{6/5}$ the edge state fills the whole matrix, the volume in the counting saturates at $M$, and one recovers $M\delta^{3/2}\sim1$, i.e. the Gaussian edge scale $M^{-2/3}$. The dictionary between the two problems is that a block of $W$ consecutive sites corresponds to one mean free path of the wire carrying $N$ channels, so that $W$ plays the role of $N$.

\section{Fixed-$N$ formulas and large-$|E|$ asymptotics}
\label{sec:SMfixedNasymptotics}

In this section we obtain series for the IDOS and the total Lyapunov exponent for large $|E|$ for any $N$. For $E>0$ we use the cubic matrix integral, while for $E<0$ we use the equivalent Hankel determinant representation. Before turning to asymptotics, we give for concreteness the explicit formula for $N=1,2$ from the exact result 
Eq.~\eqref{eq:KontsevichHalperinMain} in the main text.

\subsection{Explicit results for $N=1$ and $N=2$}
\label{subsec:SMlowNobservables}

Set $s=-E/T^{2/3}$ and, in this subsection only,
$A=\Ai(s)$, $B=\Bi(s)$, with primes denoting derivatives with respect
to $s$.  Recall that $T=J^2/(2N)$.

For $N=1$, Eq.~\eqref{eq:AiryTauDefinition} gives
$\tau_1=A-\ii B$.  Using the Airy Wronskian
$AB'-A'B=1/\pi$, Eq.~\eqref{eq:KontsevichHalperinMain} yields
\be
\Gamma_1(E)=T^{1/3}\frac{AA'+BB'}{A^2+B^2},
\qquad
\nu_1(E)=\frac{T^{1/3}}{\pi^2(A^2+B^2)} .
\label{eq:SMN1GammaNu}
\ee
These are Halperin's exact scalar IDOS and Lyapunov exponent
~\cite{Halperin1965,ComtetTourigny2017}.

For $N=2$, the Airy equation gives
$\tau_2=s(A-\ii B)^2-(A'-\ii B')^2$ and
$\tau_2'=(A-\ii B)^2$, hence
$\Xi_2=T^{1/3}(A-\ii B)^2/\tau_2$.  Define
\be
\mathcal D_2(s):=|\tau_2|^2
=\left[s(A^2-B^2)-(A'^2-B'^2)\right]^2
+4\left(sAB-A'B'\right)^2 .
\label{eq:SMN2Denominator}
\ee
Taking real and imaginary parts gives
\be
\Gamma_2(E)=T^{1/3}
\frac{s(A^2+B^2)^2-(AA'+BB')^2+\pi^{-2}}{\mathcal D_2(s)},
\qquad
\nu_2(E)=\frac{2T^{1/3}}{\pi^2}
\frac{AA'+BB'}{\mathcal D_2(s)} .
\label{eq:SMN2GammaNu}
\ee
As an independent algebraic check, multiplying
$\Xi_2/T^{1/3}$ by $\overline{\tau_2}/\overline{\tau_2}$
and using the Airy Wronskian gives
\be
(A-\ii B)^2\overline{\tau_2}
=s(A^2+B^2)^2-(AA'+BB')^2+\pi^{-2}
-\frac{2\ii}{\pi}(AA'+BB') ,
\label{eq:SMN2check}
\ee
whose real and imaginary parts reproduce
Eq.~\eqref{eq:SMN2GammaNu}.  Thus already at $N=2$ both spectral
counting and total Lyapunov growth are explicit rational combinations
of ordinary Airy functions and their derivatives.

\subsection{Perturbative expansion for $E>0$ using the cubic matrix integral}
\label{sec:positive-E-perturbation}

The cubic matrix integral provides a particularly efficient starting
point for the high-energy, or weak-disorder, expansion.  We use
\be
\Xi_N(E)=-T\,\partial_E\log \mathcal Z_N^{(-)}(E),
\label{eq:pertXiZN}
\ee
with
\be
\mathcal Z_N^{(-)}(E)
=\frac{1}{N!}\int_{\Gamma_-^N}\Delta(\boldsymbol\lambda)^2
\prod_{j=1}^N
\exp\!\left[-\frac{1}{T}
\left(\frac{\lambda_j^3}{3}+E\lambda_j\right)\right]
\dd^N\lambda .
\label{eq:pertZN}
\ee
For $E>0$ set $k=\sqrt E$ and
$\eps=T/E^{3/2}=T/k^3\ll1$.  The contour $\Gamma_-$ selects the
saddle $\lambda_*=-\ii k$, and we write
\be
\lambda_j=-\ii k+e^{\ii\pi/4}
\sqrt{\frac{T}{2k}}\,x_j .
\label{eq:pertShift}
\ee
Since the potential is exactly cubic, the expansion terminates,
\be
-\frac1T\left(\frac{\lambda_j^3}{3}+E\lambda_j\right)
=\frac{2\ii k^3}{3T}-\frac{x_j^2}{2}+g x_j^3,
\qquad
g=-\frac{e^{3\ii\pi/4}}{3\,2^{3/2}}\sqrt{\eps},
\quad
g^2=-\frac{\ii}{72}\eps .
\label{eq:pertCubicExpansion}
\ee
Including the Jacobian and the Vandermonde scaling gives
\be
\mathcal Z_N^{(-)}
=C_N e^{2\ii Nk^3/(3T)}
\left(\frac{T}{2k}\right)^{N^2/2}
\left\langle e^{gA}\right\rangle_0,
\qquad
A:=\Tr X^3,
\label{eq:pertGaussianReduction}
\ee
where $C_N$ is independent of $E$ and
$\langle\cdots\rangle_0$ denotes the normalized GUE average with
weight $e^{-\Tr X^2/2}$.

The remaining perturbation is therefore the connected-cumulant
expansion of the single cubic observable $A$.  Since the Gaussian
measure is invariant under $X\to-X$, all odd cumulants vanish and
\be
\log\left\langle e^{gA}\right\rangle_0
=\sum_{m\ge1}\frac{g^{2m}}{(2m)!}\,\kappa_{2m}(A) .
\label{eq:pertEvenCumulants}
\ee
Because $g^2\propto-\ii\eps$, successive terms generate integer powers
of $\eps$: odd powers in $\log\mathcal Z_N^{(-)}$ are imaginary and
even powers are real.  Consequently $\Gamma_N$ contains
$T,T^3,T^5,\ldots$, whereas $\nu_N$ contains
$T^0,T^2,T^4,\ldots$.

The weak-disorder expansion has a topological
organization~\cite{LandoZvonkin2004}.
A connected Wick contraction in $\kappa_{2m}$ gives a
trivalent ribbon graph with $2m$ vertices and $3m$ edges.
At genus $h$, it has $F=m+2-2h$ index loops and contributes
$N^F$. Thus $\kappa_{2m}$ is a finite polynomial in
$N^{m+2},N^m,\ldots$.
Since $g^{2m}\propto N^{-m}$ at fixed $E$ and $J$,
the contribution $g^{2m}\kappa_{2m}/(2m)!$ to the logarithm
of the matrix integral carries powers $N^{2-2h}$.
This is precisely the 't~Hooft genus expansion of the
cubic one-matrix model, with a complex coupling fixed
by $E$ and $J$. The first three nonzero cumulants are
\be
\begin{aligned}
\kappa_2(A)&=3N(4N^2+1),\\
\kappa_4(A)&=648N^2(8N^2+7),\\
\kappa_6(A)&=29160N(336N^4+664N^2+105).
\end{aligned}
\label{eq:pertCumulants}
\ee
Substitution into Eq.~\eqref{eq:pertEvenCumulants} gives
\be
\begin{aligned}
\log\left\langle e^{gA}\right\rangle_0
={}&-\frac{\ii N(4N^2+1)}{48}\eps
-\frac{N^2(8N^2+7)}{192}\eps^2\\
&+\frac{\ii N(336N^4+664N^2+105)}{9216}\eps^3
+O(\eps^4).
\end{aligned}
\label{eq:pertLogAverage}
\ee
Hence, up to an $E$-independent constant,
\be
\begin{aligned}
\log\mathcal Z_N^{(-)}
={}&\frac{2\ii N E^{3/2}}{3T}
+\frac{N^2}{2}\log\!\left(\frac{T}{2\sqrt E}\right)
-\frac{\ii N(4N^2+1)}{48}\eps\\
&-\frac{N^2(8N^2+7)}{192}\eps^2
+\frac{\ii N(336N^4+664N^2+105)}{9216}\eps^3
+O(\eps^4).
\end{aligned}
\label{eq:pertLogZN}
\ee
Using $\partial_E\eps^m=-3m\eps^m/(2E)$ in
Eq.~\eqref{eq:pertXiZN}, we obtain
\be
\begin{aligned}
\Xi_N(E)
={}&-\ii N\sqrt E+\frac{N^2T}{4E}
-\frac{\ii N(4N^2+1)T^2}{32E^{5/2}}
-\frac{N^2(8N^2+7)T^3}{64E^4}\\
&+\frac{\ii N(336N^4+664N^2+105)T^4}
{2048E^{11/2}}
+O\!\left(\frac{T^5}{E^7}\right).
\end{aligned}
\label{eq:pertXiExpansion}
\ee
Since $\Xi_N=\Gamma_N-\ii\pi\nu_N$, the two components are
\bea
\Gamma_N(E)&=&
\frac{N^2T}{4E}
\left[
1-\frac{8N^2+7}{16}
\left(\frac{T}{E^{3/2}}\right)^2
+O\!\left(\frac{T^4}{E^6}\right)
\right],
\label{eq:pertGammaFinal}\\
\nu_N(E)&=&
\frac{N\sqrt E}{\pi}
\left[
1+\frac{4N^2+1}{32}
\left(\frac{T}{E^{3/2}}\right)^2
-\frac{336N^4+664N^2+105}{2048}
\left(\frac{T}{E^{3/2}}\right)^4
+O\!\left(\frac{T^6}{E^9}\right)
\right].
\label{eq:pertNuFinal}
\eea
For $N=1$ the latter reduces to
\be
\nu_1(E)=\frac{\sqrt E}{\pi}
\left[
1+\frac{5}{32}\left(\frac{T}{E^{3/2}}\right)^2
-\frac{1105}{2048}\left(\frac{T}{E^{3/2}}\right)^4
+\cdots
\right],
\label{eq:pertN1Check}
\ee
in agreement with the large-negative-$s$ expansion of the exact scalar
Airy formula.  The advantage of the cubic-integral representation is
that the entire perturbative series is reduced to connected Gaussian
cumulants of $\Tr X^3$; no additional interaction vertices arise at
higher orders.

\subsection{Negative-energy expansion from the Airy determinant}
\label{sec:negative-E-expansion}

For $E<0$, set $z=|E|/T^{2/3}$ and
$\eps=z^{-3/2}=T/|E|^{3/2}\ll1$.  The corresponding instanton action
is $S(z)=4z^{3/2}/3=4/(3\eps)=4|E|^{3/2}/(3T)$.  Recall that
\be
\tau_N(z)=
\det_{0\le j,k\le N-1}
\left[
\Ai^{(j+k)}(z)-\ii\Bi^{(j+k)}(z)
\right].
\label{eq:negEtau}
\ee
Introduce the real Hankel matrices
\be
\mathsf A_N(z)=
\left[\Ai^{(j+k)}(z)\right]_{j,k=0}^{N-1},
\qquad
\mathsf B_N(z)=
\left[\Bi^{(j+k)}(z)\right]_{j,k=0}^{N-1}.
\label{eq:negEAB}
\ee
Then
\be
\tau_N(z)=(-\ii)^N\det\mathsf B_N(z)\,
\det\!\left[\mathsf I+\ii\mathsf X_N(z)\right],
\qquad
\mathsf X_N:=\mathsf B_N^{-1}\mathsf A_N .
\label{eq:negEFactorization}
\ee
For $z\gg1$,
\be
\Bi(z)\sim\frac{z^{-1/4}}{\sqrt\pi}e^{2z^{3/2}/3},
\qquad
\Ai(z)\sim\frac{z^{-1/4}}{2\sqrt\pi}e^{-2z^{3/2}/3},
\label{eq:negEAiryAsymptotics}
\ee
and similarly for their derivatives, so that
$\mathsf X_N=O(e^{-S})$.  The logarithm therefore separates into a
dominant algebraic sector and exponentially small instanton sectors,
\be
\begin{aligned}
\log\tau_N
={}&-\frac{\ii\pi N}{2}+\log\det\mathsf B_N
+\Tr\log(\mathsf I+\ii\mathsf X_N)\\
={}&-\frac{\ii\pi N}{2}+\log\det\mathsf B_N
+\ii\Tr\mathsf X_N+\frac12\Tr\mathsf X_N^2
-\frac{\ii}{3}\Tr\mathsf X_N^3+\cdots .
\end{aligned}
\label{eq:negElogtau}
\ee
The real all-$\Bi$ determinant gives the algebraic expansion of the
total Lyapunov exponent.  Expanding it to the same order as above gives, recalling
that $\eps=T/|E|^{3/2}\ll1$
\be
\Gamma_N(E)
\sim N\sqrt{|E|}
\left[
1-\frac{N}{4}\eps
-\frac{4N^2+1}{32}\eps^2
-\frac{N(8N^2+7)}{64}\eps^3
-\frac{336N^4+664N^2+105}{2048}\eps^4
+O(\eps^5)
\right],
\label{eq:negEGammaFinal}
\ee
up to exponentially small corrections beginning at order $e^{-2S}$.
For $N=1$ this agrees with the large-$z$ expansion of
$T^{1/3}\Bi'(z)/\Bi(z)$, while for $N=2$ it agrees with the
large-$s$ expansion of the exact formula~\eqref{eq:SMN2GammaNu}.

By contrast, the IDOS is nonperturbative in $T$ and comes from the
imaginary part of Eq.~\eqref{eq:negElogtau}.  At leading exponential
order only the one-$\Ai$ sector contributes.  Defining
$R_N(z):=\Tr(\mathsf B_N^{-1}\mathsf A_N)$, one has
\be
\nu_N(E)
=-\frac{T^{1/3}}{\pi}\,\partial_zR_N(z)+O(e^{-3S}) .
\label{eq:negEnuRN}
\ee
There is no $e^{-2S}$ contribution to the phase because even powers of
$\mathsf X_N$ in Eq.~\eqref{eq:negElogtau} are real, whereas odd
powers are imaginary.

The one-instanton term can be evaluated systematically from the
large-$z$ Airy expansions, or equivalently as the ratio between the
sum of Hankel determinants containing one $\Ai$ column and the
dominant all-$\Bi$ determinant.  To the required order,
\bea
R_N(z)&=&
\frac{2^{3N-4}}{(N-1)!}\,
\eps^{-(N-1)}e^{-4/(3\eps)}
\left[
1+r_1(N)\eps+r_2(N)\eps^2+O(\eps^3)
\right],
\label{eq:negERNexpansion}\\
r_1(N)&=&
-\frac{21N^2-51N+35}{24},
\qquad
r_2(N)=
\frac{
441N^4-2880N^3+7059N^2-8340N+3745
}{1152}.
\label{eq:negErCoefficients}
\eea
The leading prefactor has a simple saddle interpretation: the scalar
subdominant-to-dominant Airy ratio contributes $\frac12e^{-S}$,
moving one eigenvalue to the subdominant saddle gives the
cross-Vandermonde factor $(8/\eps)^{N-1}$, and the Gaussian
fluctuations contribute $1/(N-1)!$.

Differentiating Eq.~\eqref{eq:negERNexpansion} according to
Eq.~\eqref{eq:negEnuRN} gives
\be
\begin{aligned}
\nu_N(E)\simeq{}&
\frac{\sqrt{|E|}}{\pi}\frac{8^{N-1}}{(N-1)!}
\left(\frac{|E|^{3/2}}{T}\right)^{N-1}
\exp\!\left[-\frac{4|E|^{3/2}}{3T}\right]\\
&\times
\left[
1-\frac{21N^2-33N+17}{24}
\frac{T}{|E|^{3/2}}
+\frac{
441N^4-2124N^3+3711N^2-3408N+1225
}{1152}
\left(\frac{T}{|E|^{3/2}}\right)^2
+O\!\left(\frac{T^3}{|E|^{9/2}}\right)
\right].
\end{aligned}
\label{eq:negEnuFinal}
\ee
For $N=1$ this reduces to
\be
\nu_1(E)\simeq
\frac{\sqrt{|E|}}{\pi}
\exp\!\left[-\frac{4|E|^{3/2}}{3T}\right]
\left[
1-\frac{5}{24}\frac{T}{|E|^{3/2}}
-\frac{155}{1152}
\left(\frac{T}{|E|^{3/2}}\right)^2+\cdots
\right],
\label{eq:negEN1check}
\ee
in agreement with the direct large-$z$ expansion of the exact scalar
Airy formula.

The determinant representation therefore provides a natural
trans-series organization on the $E<0$ side: the all-$\Bi$ determinant
is the perturbative Lyapunov sector, one $\Ai$ insertion gives the
leading Lifshitz tail of the IDOS, higher algebraic corrections dress
the same one-instanton sector, and higher odd numbers of $\Ai$
insertions generate the next exponentially smaller contributions to
the IDOS.

\section{Asymptotics of the critical scaling functions}
\label{sec:SMcriticalAsymptotics}

We now derive the asymptotic expansions of the critical IDOS scaling
function $\Phi(X)$ and of its Lyapunov counterpart.  The Painlev\'e-I
equation itself was obtained in Eq.~\eqref{eq:SMPainleveI}, and the
outgoing Riccati contour was shown in
Eq.~\eqref{eq:SMphysicalPI} to select the tritronqu\'ee solution
$y_1(\mathfrak t)$.  Here we extract the asymptotic information carried
by that particular solution and recover, in particular, the two limiting
behaviors quoted in Eqs.~\eqref{eq:PhiPlusAsym} and
\eqref{eq:Phiasym} of the Main Text.

It is useful first to keep the real and imaginary critical observables
together.  The exact Toda relation~\eqref{eq:SMXiToda} determines the
variation of $\Xi_N$ across the critical window through the endpoint
recurrence coefficient $r_{N,N}$.  Its critical expansion contains the
regular planar value $r_*=s_J^2/4$, together with the singular
Painlev\'e correction in Eq.~\eqref{eq:SMrEndpoint}.  Upon integration
with respect to $E$, the former contributes only to a background that
is smooth through $E=E^*$, whereas the latter gives the universal
double-scaling contribution.  We therefore write
\be
\Xi_N(E)
=
\Xi_{N,\mathrm{reg}}(E)
+
s_JN^{-1/5}\mathcal X(X)
+
o(N^{-1/5}),
\qquad
\mathcal X(X):=\Psi(X)-\ii\pi\Phi(X),
\label{eq:SMcriticalComplexScaling}
\ee
where $\Xi_{N,\mathrm{reg}}(E)$ denotes the part of the characteristic
exponent that is smooth through the critical point.
Using Eq.~\eqref{eq:SMXiToda}, the endpoint expansion
\eqref{eq:SMrEndpoint}, and
$E=E^*+s_J^2N^{-4/5}X$, one obtains
\be
\mathcal X'(X)
=
2^{4/5}y_1(2^{7/5}X),
\qquad
\Psi'(X)
=
2^{4/5}\Re y_1(2^{7/5}X),
\qquad
\Phi'(X)
=
-\frac{2^{4/5}}{\pi}\Imn y_1(2^{7/5}X).
\label{eq:SMPsiPhiFromPI}
\ee
The last relation is precisely the scaling formula
Eq.~\eqref{eq:criticalDensity} of the Main Text.  Thus the two critical
observables are respectively the real and imaginary parts of the same
Painlev\'e-I solution.

\subsection{Allowed side: $X\to+\infty$}
\label{subsec:SMcriticalAllowed}

For positive $\mathfrak t$, the physical tritronqu\'ee approaches the
complex algebraic branch
$y_1(\mathfrak t)\sim-\ii\sqrt{\mathfrak t/6}$.  Its complete formal
large-$\mathfrak t$ expansion is obtained by inserting
\be
y_1(\mathfrak t)
=
-\frac{\ii}{\sqrt6}\,
\mathfrak t^{1/2}
\left[
1+\sum_{n\ge1}c_n\,\mathfrak t^{-5n/2}
\right],
\qquad c_0=1,
\label{eq:SMPIAllowedExpansion}
\ee
into the Painlev\'e-I equation~\eqref{eq:SMPainleveI} and matching
successive powers of $\mathfrak t^{-5/2}$.  This gives the recursion
\be
c_n
=
\frac{\ii[25(n-1)^2-1]}{8\sqrt6}\,c_{n-1}
-\frac12\sum_{j=1}^{n-1}c_jc_{n-j},
\qquad n\ge1.
\label{eq:SMPIcnRecursion}
\ee
The first coefficients are
\be
c_1=-\frac{\ii\sqrt6}{48},
\qquad
c_2=\frac{49}{768},
\qquad
c_3=\frac{1225\,\ii\sqrt6}{9216},
\qquad
c_4=-\frac{4412401}{1179648}.
\label{eq:SMPIfirstcn}
\ee

The recursion alternates between purely imaginary odd coefficients and
real even coefficients.  Through Eq.~\eqref{eq:SMPsiPhiFromPI}, the
even coefficients therefore generate the algebraic expansion of
$\Phi$, while the odd coefficients generate that of $\Psi$.
Substituting $\mathfrak t=2^{7/5}X$ and integrating term by term gives
\be
\begin{aligned}
\Phi(X)
&\sim
\frac{4}{3\pi\sqrt3}X^{3/2}
+C_\Phi
+\sum_{m\ge1}A_mX^{3/2-5m},
&
A_m
&=
\frac{2^{\,2-7m}}
{\pi\sqrt3\,(3-10m)}\,c_{2m},
\\
\Psi(X)
&\sim
\sum_{m\ge0}B_mX^{-1-5m},
&
B_m
&=
-\frac{2^{-2-7m}}
{(1+5m)\sqrt6}\,
\Imn c_{2m+1}.
\end{aligned}
\label{eq:SMAllowedPhiPsiGeneral}
\ee
The local Painlev\'e expansion determines $\Psi$ only up to an additive
real constant.  Such a constant changes only the smooth Lyapunov
background and is therefore fixed by global matching rather than by
the local double-scaling analysis.  Likewise, the integration constant
$C_\Phi$ is not determined by the local large-$X$ expansion; its value
is fixed by the global normalization of the IDOS, equivalently by
matching to the physical solution selected by the outgoing contour.
The first terms are
\be
\begin{aligned}
\Phi(X)
&\sim
\frac{4}{3\pi\sqrt3}X^{3/2}
+C_\Phi
-\frac{7}{24576\,\pi\sqrt3}X^{-7/2}
+\frac{259553}{4831838208\,\pi\sqrt3}X^{-17/2}
+\cdots,
\\
\Psi(X)
&\sim
\frac{1}{192X}
-\frac{1225}{28311552\,X^6}
+\cdots,
\qquad X\to+\infty.
\end{aligned}
\label{eq:SMAllowedPhiPsiFirstTerms}
\ee
Differentiating the leading term of $\Phi$ gives
$\Phi'(X)\sim2X^{1/2}/(\pi\sqrt3)$, reproducing
Eq.~\eqref{eq:PhiPlusAsym} of the Main Text.  The subleading terms in
Eq.~\eqref{eq:SMAllowedPhiPsiFirstTerms} are therefore the systematic
Painlev\'e corrections to the sharp large-$N$ square-root edge.

\subsection{Forbidden side: $X\to-\infty$}
\label{subsec:SMcriticalForbidden}

The opposite limit has a qualitatively different structure.  Put
$x=-\mathfrak t=2^{7/5}|X|>0$.  Equation~\eqref{eq:SMPainleveI}
then reads
$\dd^2y/\dd x^2=6y^2-x$, and the physical solution approaches the
positive real branch $y_1(-x)\sim\sqrt{x/6}$.  Its complete algebraic
expansion is therefore real.  The Lyapunov scaling function is visible
already in this perturbative sector, whereas the IDOS arises only from
exponentially small Stokes corrections.

\paragraph{Algebraic, or zero-instanton, sector.---}
Write
\be
y_1^{(0)}(-x)
=
\sqrt{\frac{x}{6}}
\left[
1+\sum_{n\ge1}a_nx^{-5n/2}
\right],
\qquad a_0=1.
\label{eq:PIForbiddenZeroInstanton}
\ee
Substitution into Painlev\'e I and matching powers of $x^{-5/2}$ gives
\be
a_n
=
\frac{25(n-1)^2-1}{8\sqrt6}\,a_{n-1}
-\frac12\sum_{j=1}^{n-1}a_ja_{n-j},
\qquad n\ge1.
\label{eq:PIForbiddenanRecursion}
\ee
The first coefficients are
\be
a_1=-\frac{\sqrt6}{48},
\qquad
a_2=-\frac{49}{768},
\qquad
a_3=-\frac{1225\sqrt6}{9216},
\qquad
a_4=-\frac{4412401}{1179648},
\label{eq:PIForbiddenFirstan}
\ee
and hence
\be
y_1^{(0)}(-x)
=
\sqrt{\frac{x}{6}}
-\frac{1}{48}x^{-2}
-\frac{49}{768\sqrt6}x^{-9/2}
+\cdots .
\label{eq:PIForbiddenZeroFirstTerms}
\ee
Every term in this expansion is real.  Equation
\eqref{eq:SMPsiPhiFromPI} therefore implies that
$\Phi(X)$ vanishes to all algebraic orders as $X\to-\infty$.
The forbidden-side IDOS is consequently a genuinely
nonperturbative Stokes effect.

\paragraph{One-instanton fluctuation.---}
To determine the leading exponentially small contribution, linearize
around the algebraic solution by writing
$y_1(-x)=y_1^{(0)}(-x)+\delta y(x)$.  At leading exponential order,
the fluctuation obeys
$\delta y''=12y_1^{(0)}(-x)\delta y$.

We first determine the exponential decay of this solution by a WKB
ansatz $\delta y(x)\sim A(x)e^{-S(x)}$, where $A(x)$ varies only
algebraically with $x$.  At leading WKB order,
$[S'(x)]^2\simeq12y_1^{(0)}(-x)$.  Since
$12y_1^{(0)}(-x)\sim2\sqrt6\,x^{1/2}$, the decaying branch satisfies
$S'(x)\sim\kappa x^{1/4}$, with
$\kappa=\sqrt{2\sqrt6}=2^{3/4}3^{1/4}$.  Hence
\be
S(x)
=
\int^x\kappa u^{1/4}\,\dd u
=
\frac{4\kappa}{5}x^{5/4}
=
\frac{2^{11/4}3^{1/4}}{5}x^{5/4}.
\label{eq:PIForbiddenAction}
\ee
The irrelevant additive constant in $S$ is absorbed into the overall
amplitude of $\delta y$.  We refer to $S(x)$ as the WKB, or
one-instanton, action: it is the exponent governing the leading
nonperturbative correction, $\delta y\propto e^{-S(x)}$.

Including the algebraic WKB prefactor and its systematic corrections,
the decaying one-instanton solution may be written
\be
y^{(1)}(x)
=
x^{-1/8}e^{-S(x)}
\sum_{n\ge0}b_nx^{-5n/4},
\qquad
b_0=1.
\label{eq:PIForbiddenOneInstanton}
\ee
The powers now occur in steps of $x^{-5/4}$, half the spacing of the
zero-instanton expansion.  Writing
$\rho_n=-1/8-5n/4$, substitution of
Eq.~\eqref{eq:PIForbiddenOneInstanton} into the linearized equation
gives
\be
b_n
=
\frac{2}{5\kappa n}
\left[
\kappa^2
\sum_{k=1}^{\lfloor(n+1)/2\rfloor}
a_kb_{n+1-2k}
-
\rho_{n-1}(\rho_{n-1}-1)b_{n-1}
\right],
\qquad
n\ge1.
\label{eq:PIForbiddenbnRecursion}
\ee
The first coefficients are
\be
b_1=-\frac{5}{32\kappa},
\qquad
b_2=\frac{75\sqrt6}{8192}.
\label{eq:PIForbiddenFirstbn}
\ee
Thus, once the perturbative coefficients $a_n$ are known, the
one-instanton fluctuation and all of its algebraic corrections are
generated recursively.

\paragraph{Kapaev Stokes multiplier and the physical branch.---}
The linearized analysis above determines the shape of the
one-instanton sector but not its overall normalization.  This
normalization is fixed by the Stokes data of the particular
tritronqu\'ee selected by the outgoing contour.

We use here the quasi-linear Stokes formula of
Kapaev~\cite{Kapaev2004}.  His Painlev\'e-I equation has exactly our
canonical normalization,
$y''(\zeta)=6y(\zeta)^2+\zeta$.  To avoid confusion with our physical
solution $y_1$, let us denote Kapaev's two adjacent tritronqu\'ee
solutions by $y_0^{\rm K}$ and $y_1^{\rm K}$.  They are the solutions
specified by the Stokes data in Eqs.~(2.65) and~(2.66) of
Ref.~\cite{Kapaev2004}.  Corollary~2.4, Eq.~(2.67) of that reference,
gives their exponentially small difference in the sector containing
the negative real axis:
\be
y_1^{\rm K}(\zeta)-y_0^{\rm K}(\zeta)
\sim
\frac{\ii}{\sqrt\pi}\,
2^{-11/8}3^{-1/8}
(e^{-\ii\pi}\zeta)^{-1/8}
\exp\left[
-\frac{2^{11/4}3^{1/4}}{5}
(e^{-\ii\pi}\zeta)^{5/4}
\right].
\label{eq:PIForbiddenKapaevJump}
\ee
In Eq.~\eqref{eq:PIForbiddenKapaevJump} we use only the leading
exponentially small term of Kapaev's Stokes formula.  This term fixes
the overall normalization of the one-instanton sector; the algebraic
corrections multiplying the same exponential are generated
independently by the recursion~\eqref{eq:PIForbiddenbnRecursion}.

The dictionary with our variables is simply
$\zeta=\mathfrak t=-x$, with $x>0$.  Hence
$e^{-\ii\pi}\zeta=x$ on the negative Painlev\'e axis, and the
exponential in Eq.~\eqref{eq:PIForbiddenKapaevJump} is precisely
$e^{-S(x)}$, with $S(x)$ defined in
Eq.~\eqref{eq:PIForbiddenAction}.  Thus Kapaev's result becomes
\be
y_1^{\rm K}(-x)-y_0^{\rm K}(-x)
\sim
\frac{\ii}{\sqrt\pi}\,
2^{-11/8}3^{-1/8}
x^{-1/8}e^{-S(x)} .
\label{eq:PIForbiddenKapaevJumpReal}
\ee

On this Stokes ray the two lateral tritronqu\'ee solutions are complex
conjugates and have the same real algebraic expansion.  Their
imaginary parts are therefore one half of the jump in
Eq.~\eqref{eq:PIForbiddenKapaevJumpReal}, with opposite signs.  The
physical $E+\ii0$ solution selected in
Eq.~\eqref{eq:SMphysicalPI} is the branch with negative imaginary
part: indeed Eq.~\eqref{eq:SMPsiPhiFromPI} requires
$\Imn y_1<0$ for the positive IDOS.  Consequently
\be
\Imn y_1(-x)
\sim
-\frac{2^{-19/8}3^{-1/8}}{\sqrt\pi}\,
x^{-1/8}e^{-S(x)}
\left[
1+\sum_{n\ge1}b_nx^{-5n/4}
\right].
\label{eq:PIForbiddenImy1}
\ee
The overall factor in this equation is fixed by Kapaev's Stokes
formula, while the coefficients $b_n$ are the algebraic
one-instanton corrections generated by the recursion
\eqref{eq:PIForbiddenbnRecursion}.

\paragraph{Critical IDOS tail.---}
Since $x=2^{7/5}|X|$, the action becomes
\be
S(x)
=
B|X|^{5/4},
\qquad
B:=\frac{16}{5}\sqrt2\,3^{1/4}.
\label{eq:PIForbiddenActionX}
\ee
Using Eq.~\eqref{eq:SMPsiPhiFromPI}, one obtains
\be
\Phi'(X)
\sim
K|X|^{-1/8}e^{-B|X|^{5/4}}
\left[
1+\sum_{n\ge1}q_n|X|^{-5n/4}
\right],
\qquad
K=\frac{2^{-7/4}3^{-1/8}}{\pi^{3/2}},
\qquad
q_n=2^{-7n/4}b_n .
\label{eq:PIForbiddenPhiPrimeExpansion}
\ee
Together with the physical normalization $\Phi(-\infty)=0$, integration
gives
\be
\Phi(X)
\sim
C_\Phi^{(-)}
|X|^{-3/8}
e^{-B|X|^{5/4}}
\left[
1+\sum_{n\ge1}p_n|X|^{-5n/4}
\right],
\qquad
C_\Phi^{(-)}
=
\frac{1}{2^{17/4}3^{3/8}\pi^{3/2}} .
\label{eq:PIForbiddenPhiExpansion}
\ee
The coefficients $p_n$ follow directly by differentiating this
expression and comparing with
Eq.~\eqref{eq:PIForbiddenPhiPrimeExpansion}.  Defining
$\beta=4\sqrt2\,3^{1/4}$ and $p_0=1$ gives
\be
p_n
=
q_n-\frac{10n-7}{8\beta}\,p_{n-1},
\qquad n\ge1,
\qquad
p_1=-\frac{17\sqrt2\,3^{3/4}}{768}.
\label{eq:PIForbiddenpnRecursion}
\ee
The first refined forbidden-side asymptotics is therefore
\be
\Phi(X)
\sim
\frac{|X|^{-3/8}}
{2^{17/4}3^{3/8}\pi^{3/2}}
\exp\!\left[
-\frac{16}{5}\sqrt2\,3^{1/4}|X|^{5/4}
\right]
\left[
1-\frac{17\sqrt2\,3^{3/4}}{768}|X|^{-5/4}
+\cdots
\right].
\label{eq:PIForbiddenPhiRefined}
\ee
Taking the logarithm of
Eq.~\eqref{eq:PIForbiddenPhiPrimeExpansion} gives at leading order
\be
\log\Phi'(X)
\sim
-\frac{16}{5}\sqrt2\,3^{1/4}|X|^{5/4},
\qquad X\to-\infty,
\label{eq:SMForbiddenLogPhiPrime}
\ee
which is Eq.~\eqref{eq:Phiasym} of the Main Text.

In physical variables,
$|X|=N^{4/5}(E^*-E)/s_J^2$, and therefore the same exponent is
$\mathcal B/T$, with
$\mathcal B=\frac45\sqrt2\,3^{1/4}s_J^{1/2}(E^*-E)^{5/4}$.
This is the single-eigenvalue activation barrier of
Ref.~\cite{FyodorovLeDoussalOssipov2025}.  Its deep forbidden-tail
limit gives $4|E|^{3/2}/(3T)$, continuously matching the finite-$N$
Lifshitz exponent in Eq.~\eqref{eq:Lifmain} of the Main Text and in
Sec.~\ref{sec:negative-E-expansion} above.

\paragraph{Multi-instanton structure.---}
The complete forbidden-side solution is naturally organized as a
Painlev\'e trans-series,
\be
y_1(-x)
=
y_1^{(0)}(-x)
+
\sum_{k\ge1}\sigma^k y^{(k)}(x),
\qquad
y^{(k)}(x)
\sim
x^{(4-5k)/8}e^{-kS(x)}
\left[
\text{series in }x^{-5/4}
\right].
\label{eq:PIForbiddenTransSeries}
\ee
For the physical tritronqu\'ee the parameter $\sigma$ is not arbitrary:
it is fixed by the outgoing Stokes data.  Once the sectors
$y^{(1)},\ldots,y^{(k-1)}$ are known, the equation determining
$y^{(k)}$ is linear, with a source constructed from lower instanton
sectors.

On the physical Stokes line $\sigma$ is purely imaginary.  Hence even
instanton sectors contribute to the real part of $y_1$ and odd sectors
to its imaginary part.  It follows from
Eq.~\eqref{eq:SMPsiPhiFromPI} that the IDOS contains the sectors
$e^{-S},e^{-3S},e^{-5S},\ldots$, whereas the exponentially small
corrections to the Lyapunov sector start with
$e^{-2S},e^{-4S},\ldots$.  This is the critical counterpart of the
odd/even instanton structure of the finite-$N$ Airy determinant in
Sec.~\ref{sec:negative-E-expansion}.

\paragraph{Forbidden-side Lyapunov expansion.---}
The algebraic part of $\Psi$ follows directly from the zero-instanton
solution.  Substituting Eq.~\eqref{eq:PIForbiddenZeroInstanton} into
Eq.~\eqref{eq:SMPsiPhiFromPI} gives
\be
\Psi'(X)
\sim
\frac{2}{\sqrt3}|X|^{1/2}
\left[
1+\sum_{n\ge1}
a_n2^{-7n/2}|X|^{-5n/2}
\right].
\label{eq:PIForbiddenPsiPrimeExpansion}
\ee
Term-by-term integration yields
\be
\Psi(X)
\sim
C_\Psi^{(-)}
-\frac{4}{3\sqrt3}|X|^{3/2}
+\sum_{n\ge1}D_n|X|^{(3-5n)/2},
\qquad
D_n
=
-\frac{4\,2^{-7n/2}}
{\sqrt3(3-5n)}\,a_n .
\label{eq:PIForbiddenPsiExpansion}
\ee
The first terms are
\be
\Psi(X)
\sim
C_\Psi^{(-)}
-\frac{4}{3\sqrt3}|X|^{3/2}
-\frac{1}{192|X|}
-\frac{7\sqrt3}{73728}|X|^{-7/2}
-\frac{1225}{28311552}|X|^{-6}
+\cdots .
\label{eq:PIForbiddenPsiFirstTerms}
\ee
The additive constant $C_\Psi^{(-)}$ is global matching information and
cannot be fixed by the local Painlev\'e expansion alone.

The resulting hierarchy is therefore completely systematic.  The
zero-instanton sector generates the algebraic expansion of the Lyapunov
scaling function; the one-instanton sector, together with the Kapaev
Stokes multiplier, gives the full leading Lifshitz-tail expansion of
the IDOS; the two-instanton sector gives the first exponentially small
correction to $\Psi$; and the three-instanton sector gives the next
exponential correction to $\Phi$.  All these sectors are different
components of the same tritronqu\'ee solution selected by the original
Riccati boundary condition.

\section{Global normalization and numerical evaluation of the critical scaling function}
\label{sec:phi-numerics}

The local Painlev\'e asymptotics of
Sec.~\ref{sec:SMcriticalAsymptotics} determine $\Phi'(X)$ and all its
algebraic and exponentially small corrections, but they leave the
additive constant $C_\Phi$ in the allowed-side expansion
\eqref{eq:SMAllowedPhiPsiFirstTerms} undetermined.  This constant is
global information: it depends on the unique tritronqu\'ee solution
selected by the outgoing contour and on the physical normalization
$\Phi(-\infty)=0$.  We determine it here by solving the corresponding
Painlev\'e-I boundary-value problem across the full critical crossover.

Using Eq.~\eqref{eq:SMPsiPhiFromPI}, the physical normalization gives
\be
\Phi(X)
=
-\frac{2^{-3/5}}{\pi}
\int_{-\infty}^{\,2^{7/5}X}
\Imn y_1(\mathfrak t)\,\dd\mathfrak t .
\label{eq:PhiNumIntegral}
\ee
Thus the problem is genuinely global: the same solution $y_1$ must
match the exponentially small Stokes sector as
$\mathfrak t\to-\infty$ and the complex algebraic branch as
$\mathfrak t\to+\infty$.  The main numerical task is to determine the physical solution
$y_1(\mathfrak t)$ throughout the crossover; once it is known,
the integral in Eq.~\eqref{eq:PhiNumIntegral} can be evaluated
by numerical quadrature.

\paragraph{Global boundary-value problem.---}
A shooting method would integrate the Painlev\'e-I equation
from initial values of $y_1$ and $y_1'$ at a large negative
$\mathfrak t$, adjusting these values to recover the required
behaviour at large positive $\mathfrak t$.  This procedure is
numerically sensitive: the initial imaginary part is
exponentially small, and small errors can excite a growing
perturbation that drives the solution away from the physical
branch.  We therefore solve a two-point boundary-value problem
by adaptive collocation, imposing the asymptotic conditions
at both endpoints simultaneously.

We restrict the numerical computation to the finite interval to
$-L\leq\mathfrak t\leq R$, with $L=10$ and $R=30$.
At $\mathfrak t=-L$ we impose the forbidden-side expansion obtained by
combining Eqs.~\eqref{eq:PIForbiddenZeroInstanton} and
\eqref{eq:PIForbiddenImy1}, retaining the terms through $a_6$ in the
algebraic sector and through $b_7$ in the one-instanton sector.
At $\mathfrak t=R$ we impose the allowed-side expansion
\eqref{eq:SMPIAllowedExpansion}, retaining the terms through $c_7$.

  In the adaptive collocation method, $y_1(\mathfrak t)$ is represented locally by piecewise
polynomials on an adaptive mesh, and the Painlev\'e-I equation is
enforced at selected collocation points within each mesh interval.
The maximum residual obtained by substituting the numerical solution
back into the Painlev\'e-I equation is below $5\times10^{-11}$, and the
result is stable under changes of $L$, $R$, and of the truncation
orders.

As a representative value,
\be
y_1(0)
=
0.1517346228157
-
0.1102416564465\,\ii .
\label{eq:PhiNumy1zero}
\ee

\paragraph{Global normalization and determination of $C_\Phi$.---}
Once $y_1$ is known, Eq.~\eqref{eq:PhiNumIntegral} determines $\Phi$
without an arbitrary additive constant.  The exponentially small
contribution from $(-\infty,-L)$ is evaluated analytically from the
forbidden-side trans-series, so the normalization does not depend on
the numerical cutoff.  In particular,
\be
\Phi(0)=0.01191950433362 .
\label{eq:PhiNumZero}
\ee

The global constant is extracted by subtracting the known allowed-side
asymptotic terms.  Define
\be
\begin{aligned}
C_\Phi(X)
:={}&
\Phi(X)
-\frac{4}{3\pi\sqrt3}X^{3/2}
+\frac{7}{24576\pi\sqrt3}X^{-7/2}
\\
&-
\frac{259553}{4831838208\pi\sqrt3}X^{-17/2}.
\end{aligned}
\label{eq:PhiNumCphiEstimator}
\ee
According to Eq.~\eqref{eq:SMAllowedPhiPsiFirstTerms},
$C_\Phi=\lim_{X\to+\infty}C_\Phi(X)$, up to the still higher algebraic
corrections displayed there.  The numerical solution gives
\be
C_\Phi^{\rm num}=O(10^{-13}),
\qquad
\text{i.e.}\qquad
C_\Phi=0
\quad\text{within numerical accuracy}.
\label{eq:PhiNumCphi}
\ee
This is the main global result of the calculation.  Neither local
asymptotic expansion determines $C_\Phi$: its numerical vanishing is
obtained only after matching the physical normalization
$\Phi(-\infty)=0$ to the selected allowed-side tritronqu\'ee branch
through the full crossover.

\begin{figure}[t]
\centering
\includegraphics[width=0.95\linewidth]{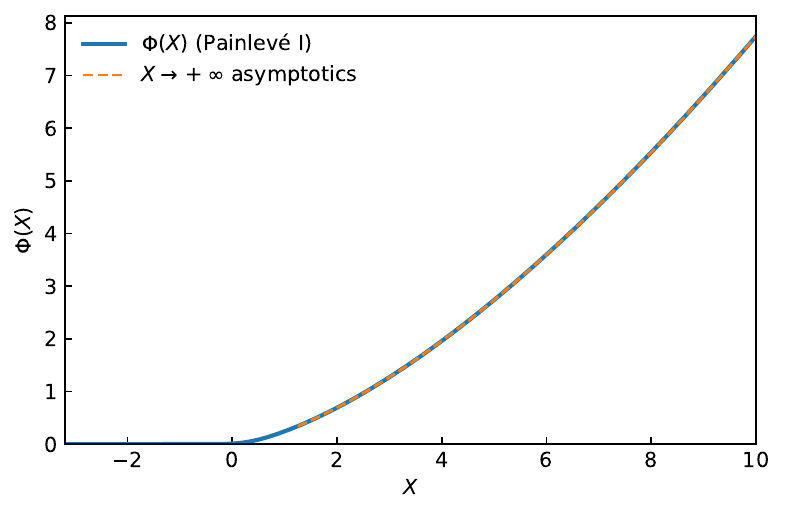}
\caption{
Critical IDOS scaling function $\Phi(X)$ obtained from the physical
Painlev\'e-I tritronqu\'ee $y_1$.  The solid line is the global
boundary-value solution and the dashed line is the allowed-side
expansion~\eqref{eq:SMAllowedPhiPsiFirstTerms} with $C_\Phi=0$.
The two become indistinguishable already for moderately positive $X$.
}
\label{fig:phi-main}
\end{figure}

The resulting scaling function is shown in
Fig.~\ref{fig:phi-main}.  It is exponentially small on the forbidden
side and rapidly approaches the planar $X^{3/2}$ law on the allowed
side.

\paragraph{Allowed-side asymptotic check.---}
With the globally determined value $C_\Phi=0$, the allowed-side
expansion~\eqref{eq:SMAllowedPhiPsiFirstTerms} provides a stringent
check of both the numerical solution and the metric
$\mathfrak t=2^{7/5}X$.  Keeping the two displayed algebraic
corrections, the relative errors are
\be
\begin{array}{c|ccccc}
X
&1&1.5&2&3&4
\\ \hline
\text{relative error}
&6.0\times10^{-4}
&-2.0\times10^{-5}
&6.0\times10^{-7}
&3.4\times10^{-10}
&1.8\times10^{-13}
\end{array}.
\label{eq:PhiNumAllowedErrors}
\ee
Thus the large-$X$ expansion is already extremely accurate for
$X\simeq2$--$3$.

\begin{figure}[t]
\centering
\includegraphics[width=0.95\linewidth]{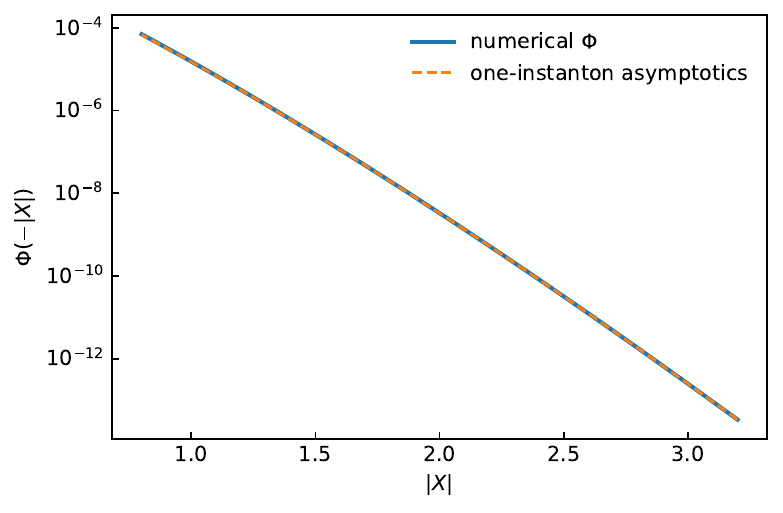}
\caption{
Forbidden-side critical IDOS.  The solid line is the global
Painlev\'e-I result and the dashed line is the one-instanton expansion
\eqref{eq:PIForbiddenPhiRefined}, including its first algebraic
correction.  The agreement rapidly improves as $X\to-\infty$.
}
\label{fig:phi-forbidden}
\end{figure}

\paragraph{Forbidden-side asymptotic check.---}
On the opposite side the IDOS is the one-instanton Stokes contribution
derived in Eq.~\eqref{eq:PIForbiddenPhiRefined}.  Its relative errors
against the global numerical solution are
\be
\begin{array}{c|ccccc}
X
&-1&-1.5&-2&-2.5&-3
\\ \hline
\text{relative error}
&-1.33\times10^{-2}
&-5.31\times10^{-3}
&-2.72\times10^{-3}
&-1.60\times10^{-3}
&-1.04\times10^{-3}
\end{array}.
\label{eq:PhiNumForbiddenErrors}
\ee
The numerical and one-instanton curves are essentially
indistinguishable on a logarithmic scale for $X\lesssim-1.5$, as shown
in Fig.~\ref{fig:phi-forbidden}.

The two asymptotic comparisons probe complementary parts of the
construction.  The allowed side tests the ordinary Painlev\'e-I power
series and the metric factor $\mathfrak t=2^{7/5}X$, whereas the
forbidden side also tests the WKB action, the Kapaev Stokes multiplier,
and the absolute normalization inherited from $\Phi(-\infty)=0$.

For reference, numerical values of the globally normalized scaling
function are collected in Table~\ref{tab:phi-numerics}.

\begin{table}[t]
\centering
\caption{Numerical values of the globally normalized critical IDOS
scaling function $\Phi(X)$.}
\label{tab:phi-numerics}
\begin{tabular}{r@{\qquad}l}
\hline\hline
$X$ & $\Phi(X)$ \\
\hline
-3    & $2.4952095307\times10^{-13}$ \\
-2.5  & $3.2082982880\times10^{-11}$ \\
-2    & $3.3039434568\times10^{-9}$  \\
-1.5  & $2.6263692409\times10^{-7}$  \\
-1.25 & $2.0907292163\times10^{-6}$  \\
-1    & $1.5241800141\times10^{-5}$  \\
-0.75 & $0.000100383183$              \\
-0.5  & $0.000585322298$              \\
-0.25 & $0.002925733007$              \\
0     & $0.011919504334$              \\
0.25  & $0.037255518753$              \\
0.5   & $0.086959414022$              \\
0.75  & $0.158491204849$              \\
1     & $0.244845490214$              \\
1.5   & $0.450154937067$              \\
2     & $0.693058807100$              \\
2.5   & $0.968584038735$              \\
3     & $1.273238425852$              \\
4     & $1.960280108184$              \\
5     & $2.739575119676$              \\
6     & $3.601265165696$              \\
8     & $5.544510550405$              \\
10    & $7.748689091881$              \\
\hline\hline
\end{tabular}
\end{table}

\section{Comparison with the localized random-band edge}
\label{sec:SMRBM}

We now compare the critical function $\Phi(X)$
with the localized spectral edge of one-dimensional unitary random band
matrices.  We conjecture
equality of the \emph{one-point cumulative edge intensities}, up to
horizontal and vertical metric factors.  Equality of the full point
processes is neither required nor asserted.

\subsection{Historical context and Sodin's localized edge}
\label{subsec:SMRBMSodin}

Random band matrices interpolate between mean-field random matrices and
quasi-one-dimensional localization.  The early works
Refs.~\cite{CasatiMolinariIzrailev1990,FyodorovMirlin1991,FyodorovMirlin1994}
established the characteristic bandwidth-dependent localization-volume
scaling and its nonlinear-$\sigma$-model interpretation.

The spectral edge was analyzed diagrammatically by
Silvestrov~\cite{Silvestrov1997}, who used the topological organization
of band-matrix graphs and identified the non-mean-field edge crossover.
Sodin~\cite{Sodin2010} subsequently gave a rigorous treatment for
$M\times M$ random band matrices of bandwidth $W$, where $M$ is the
matrix size.  The crossover to the ordinary random-matrix edge occurs
at $W\sim M^{5/6}$; Sodin explicitly notes before Theorem~1.1 that this
scale had been obtained previously at the physical level by Silvestrov.
In the localized regime $1\ll W\ll M^{5/6}$, Sodin's Theorem~1.2 shows
that the spectral edge is resolved on an energy window of width
$W^{-4/5}$, with a cumulative edge intensity of order $M/W^{6/5}$.
Equivalently, the natural edge coordinate is
$\xi=(2-\lambda)W^{4/5}$ when the limiting spectral edge is normalized
to $2$.

We use the conventional normalization in which the limiting semicircle
has upper edge $2$.  If $\lambda_\alpha$ are the RBM eigenvalues, define
the localized-edge coordinate
$\xi_\alpha=(2-\lambda_\alpha)W^{4/5}$ and the cumulative intensity
\be
\sigma_2(\xi)
:=
\lim_{\substack{M,W\to\infty\\1\ll W\ll M^{5/6}}}
\frac{W^{6/5}}{M}\,
\aver{\#\{\alpha:\xi_\alpha<\xi\}} .
\label{eq:SMRBMsigmaDef}
\ee
This is the unitary, or random-phase, measure $\sigma_2$ of
Ref.~\cite{Sodin2010}, after rescaling Sodin's semicircle edge from
$1$ to $2$.

Sodin's Theorem~1.2 proves, in particular,
\be
\sigma_2(\xi)
=
\frac{2}{3\pi}\xi^{3/2}+O(\xi),
\qquad
\xi\to+\infty,
\qquad
\sigma_2(\xi)
\leq
C\exp[-c_0|\xi|^{5/4}],
\qquad
\xi\to-\infty ,
\label{eq:SMRBMSodinAsymptotics}
\ee
for positive constants $C,c_0$.  Thus the theorem fixes the
$W^{-4/5}$ edge scale and the $W^{6/5}$ cumulative normalization, and
establishes a stretched-exponential upper bound on the
$|\xi|^{5/4}$ scale, but does not determine its sharp exponential
coefficient.  For the random-phase case Sodin further conjectures a
sharper constant-level allowed-side remainder in his Eq.~(8.1); that
refinement is not part of Theorem~1.2.

Liu and Zou~\cite{LiuZou2024} subsequently developed the edge problem
for more general band profiles and dimensions.  Their
Theorem~1.3, Eqs.~(1.11)--(1.12), describes the subcritical regime,
while Theorem~1.4, Eqs.~(1.13)--(1.14), describes the critical regime.
Let $H$ denote the $M\times M$ complex-Hermitian random band matrix.
Their method probes its edge spectrum through disorder averages of
normalized traces of high-degree Chebyshev-type polynomials of $H$.
In the complex Gaussian case relevant here, the one-point observable
reduces to $\aver{n^{-1}\Tr U_n(H/2)}$, where $U_n$ is the
Chebyshev polynomial of the second kind and $n$ is its degree; we
return to this quantity explicitly in
Sec.~\ref{subsec:SMRBMDiagrams}.  Near the upper edge, writing
$\lambda=2-\delta$, the corresponding Chebyshev edge asymptotics
[Eq.~(B.69) of Ref.~\cite{LiuZou2024}] depends on the combination
$n\sqrt{\delta}$.  Hence a polynomial of degree $n$ resolves spectral
distances $\delta\sim n^{-2}$.  In the one-dimensional localized
regime their diagrammatic power counting selects $n\sim W^{2/5}$, and
therefore $\delta\sim W^{-4/5}$, reproducing Sodin's edge scale.  Their
diagram expansion is naturally trivalent and has the ultraviolet
structure of a $\phi^3$ theory.  More recently
Ref.~\cite{LiuZou2026} established Markov-chain comparison principles,
notably Theorems~1.3 and~1.4 there, which extend critical and
subcritical edge statistics across broad classes of inhomogeneous
variance profiles.  These results strengthen the universality picture
on the RBM side, but do not imply the correspondence with the Riccati
scaling function proposed below.  The recent works
Refs.~\cite{ErdosRiabov2025,Yau2026} concern broader progress on RBM
delocalization and universality and are not used in the specific
coefficient matching that follows.

\subsection{Continuum input and a regularized moment of $\Phi$}
\label{subsec:SMRBMPhiInput}

The comparison requires three pieces of information already derived
above.  From Eq.~\eqref{eq:SMAllowedPhiPsiFirstTerms},
\be
\Phi(X)
\sim
A_+X^{3/2}
+C_\Phi
+a_\Phi X^{-7/2}
+O(X^{-17/2}),
\qquad
A_+=\frac{4}{3\pi\sqrt3},
\qquad
a_\Phi=-\frac{7}{24576\pi\sqrt3},
\label{eq:SMRBMPhiAllowed}
\ee
as $X\to+\infty$.  The global matching in
Sec.~\ref{sec:phi-numerics} gives $C_\Phi=0$ within numerical accuracy,
Eq.~\eqref{eq:PhiNumCphi}; we use this physical normalization in the
regularized quantities below.  On the forbidden side,
Eq.~\eqref{eq:PIForbiddenPhiRefined} gives
\be
\log\Phi(X)
\sim
-A_-|X|^{5/4},
\qquad
A_-=
\frac{16}{5}\sqrt2\,3^{1/4},
\qquad
X\to-\infty .
\label{eq:SMRBMPhiForbidden}
\ee

A second independent characteristic of the crossover is obtained by
subtracting the sharp planar edge.  Set
$X_+=\max(X,0)$,
$\Phi_0(X)=A_+X_+^{3/2}$ and
$\delta\Phi(X)=\Phi(X)-\Phi_0(X)$.  With the global normalization
$C_\Phi=0$, $\delta\Phi$ tends to zero at both ends and has a finite
integral.  Define
\be
b_\Phi
:=
\frac23
\int_{-\infty}^{\infty}\delta\Phi(X)\,\dd X .
\label{eq:SMRBMbPhiDef}
\ee
With this normalization the moment can be obtained analytically,
without numerical quadrature.  After subtracting the same planar
background, the critical form of the Thouless relation gives
\be
\delta\Psi(X)
=
\int_{-\infty}^{\infty}
\log|X-Y|\,\delta\Phi'(Y)\,\dd Y
=
\int_{-\infty}^{\infty}
\frac{\delta\Phi(Y)}{X-Y}\,\dd Y .
\label{eq:SMRBMcriticalThouless}
\ee
Here polynomial and $X$-independent pieces belong to the smooth
background already separated in Eq.~\eqref{eq:SMcriticalComplexScaling}.
Expanding
Eq.~\eqref{eq:SMRBMcriticalThouless} for $X\to+\infty$ gives
\be
\delta\Psi(X)
=
\frac{1}{X}
\int_{-\infty}^{\infty}\delta\Phi(Y)\,\dd Y
+O(X^{-2}).
\label{eq:SMRBMThoulessMoment}
\ee
On the other hand, the direct Painlev\'e-I expansion
\eqref{eq:SMAllowedPhiPsiFirstTerms}
gives
$\Psi(X)=1/(192X)+O(X^{-6})$.  Therefore
\be
\int_{-\infty}^{\infty}\delta\Phi(X)\,\dd X
=
\frac1{192},
\qquad
b_\Phi=\frac1{288}.
\label{eq:SMRBMbPhiExact}
\ee
Thus, once the global normalization $C_\Phi=0$ is imposed, the
Painlev\'e expansion and the Thouless relation fix the shape
coefficient needed below to $b_\Phi=1/288$.

\subsection{Liu--Zou ribbon diagrams and the RBM shape coefficient}
\label{subsec:SMRBMDiagrams}

We now extract from the Liu--Zou expansion the RBM quantity that will
fix the second metric factor in the conjecture.  The detailed graph
construction and the explicit evaluation of the relevant integrals are
given in Sec.~\ref{subsec:SMRBMdiagramDetails}; here we summarize the
main mechanism and the resulting coefficient.

As explained above, Liu and Zou~\cite{LiuZou2024} probe the spectrum
through disorder averages of normalized traces of high-degree
Chebyshev-type polynomials of the random band matrix $H$.  Expanding
such a trace produces long closed band walks.  After disorder
averaging in the complex-Hermitian class, the surviving paired walks
reduce to orientable ribbon graphs.  Long chains of degree-two vertices
become diffusive propagators, while the remaining critical vertices are
generically trivalent, which is the origin of the associated
$\phi^3$ structure.

For a connected ribbon graph $G=(V,E)$, with vertex set $V$ and edge
set $E$, let $\alpha_e>0$ denote the continuum length associated with
an edge $e\in E$.  The Gaussian integrations over the internal
longitudinal coordinates are governed by the first Symanzik, or
Kirchhoff, polynomial
\be
\mathcal U_G(\{\alpha_e\})
=
\sum_{T\ {\rm spanning\ tree}}
\prod_{e\notin T}\alpha_e ,
\label{eq:SMRBMSymanzik}
\ee
which is Eq.~(3.5) of Ref.~\cite{LiuZou2024}.  In one spatial
dimension the corresponding graph weight contains
$\mathcal U_G^{-1/2}$.  Equivalently, each long graph edge of continuum
length $\alpha$ carries the heat kernel
$p_\alpha(x)=(2\pi\varsigma^2\alpha)^{-1/2}
\exp[-x^2/(2\varsigma^2\alpha)]$, where $\varsigma^2$ is the variance
of the rescaled longitudinal step of the band walk.

The first contribution that is sensitive to the detailed shape of the
localized edge is a genus-one one-point diagram.  In the Liu--Zou
one-point convention it consists of a marked tail attached to a
subdivided theta graph.  Removing the marked tail leaves a four-edge
core with continuum lengths $\alpha_1,\ldots,\alpha_4$.  The heat-kernel
semigroup property reduces its longitudinal integral to
\be
I_{\rm sub}(\bm\alpha)
=
\frac{1}{2\pi\varsigma^2}
\frac{1}{
\sqrt{\alpha_3\alpha_4+
(\alpha_1+\alpha_2)(\alpha_3+\alpha_4)}} .
\label{eq:SMRBMSubTheta}
\ee
The marked tail fixes the total graph length through the polynomial
degree.  Eliminating it introduces a hyperplane Jacobian which cancels
the corresponding lattice-density factor in the Liu--Zou continuum
formula.  After rescaling the four remaining edge lengths by
$\alpha_e=(\tau/2)u_e$, where $\tau$ is the continuum variable
conjugate to the edge coordinate in the one-point transform, the
explicit $1/n$ normalization of the polynomial trace removes one power
of $\tau$.  The contribution $F_\Gamma(\tau)$ of this genus-one graph
therefore takes the form
\be
F_\Gamma(\tau)
=
\frac{\tau^2}{16\pi\varsigma^2}\,J,
\qquad
J=
\frac23\int_0^1
(1-z)\arcsin\sqrt{\frac{z}{4-3z}}\,\dd z
=
\frac{\pi}{27}.
\label{eq:SMRBMFirstDiagram}
\ee
Thus
\be
F_\Gamma(\tau)
=
\frac{\tau^2}{432\varsigma^2}.
\label{eq:SMRBMFirstDiagramCoeff}
\ee
The integral $J=\pi/27$ is the central continuous graph integral in
this calculation.  Its reduction from the subdivided-theta graph,
together with the marked-tail normalization, is derived explicitly in
Sec.~\ref{subsec:SMRBMdiagramDetails}.  For Sodin's sharp-cutoff band
profile the rescaled longitudinal step is uniform on $[-1,1]$, hence
$\varsigma^2=1/3$, and the first regular coefficient of the one-point
transform is $1/144$.

We now relate this graph coefficient to Sodin's one-edge cumulative
intensity $\sigma_2(\xi)$.  Let $\tau$ denote the transform variable
appearing in the edge limit of the Chebyshev polynomial,
Eq.~(B.69) of Ref.~\cite{LiuZou2024}.  Since the matrix trace receives
equal contributions from the upper and lower spectral edges, the
unitary one-point transform is
\be
\phi_1(\tau)
=
2\int_{-\infty}^{\infty}
\frac{\sin(\tau\sqrt{\xi})}{\tau\sqrt{\xi}}\,
\dd\sigma_2(\xi),
\label{eq:SMRBMTransform}
\ee
with analytic continuation of the kernel for $\xi<0$.  The prefactor
$2$ is important: $\sigma_2$ refers to a single spectral edge, whereas
the trace contains both.

Subtract the leading semiclassical edge law by defining
$\xi_+=\max(\xi,0)$ and
\be
\delta\sigma_2(\xi)
=
\sigma_2(\xi)
-\frac{2}{3\pi}\xi_+^{3/2},
\qquad
b_{\rm RBM}
:=
\frac23
\int_{-\infty}^{\infty}\delta\sigma_2(\xi)\,\dd\xi .
\label{eq:SMRBMbRBMDef}
\ee
Here $b_{\rm RBM}$ is the RBM analogue of the regularized moment
$b_\Phi$ defined in Eq.~\eqref{eq:SMRBMbPhiDef}.  Expanding the sinc
kernel for small $\tau$ and integrating by parts gives, for the
regularized transform,
\be
\phi_1^{\rm reg}(\tau)
=
\frac{\tau^2}{3}
\int_{-\infty}^{\infty}\delta\sigma_2(\xi)\,\dd\xi
+o(\tau^2)
=
\frac{b_{\rm RBM}}{2}\tau^2+o(\tau^2).
\label{eq:SMRBMTransformMoment}
\ee
Comparison with
Eq.~\eqref{eq:SMRBMFirstDiagramCoeff} at $\varsigma^2=1/3$ therefore
gives
\be
b_{\rm RBM}=\frac1{72}.
\label{eq:SMRBMbRBM}
\ee
This is the RBM shape coefficient used below to determine the metric
factors.  It is obtained by an explicit evaluation within the
Liu--Zou diagrammatic framework, rather than being a coefficient quoted
from Ref.~\cite{LiuZou2024}.  The full derivation, including the theta
and subdivided-theta graphs, the marked-tail Jacobian cancellation and
the evaluation of $J$, is given in
Sec.~\ref{subsec:SMRBMdiagramDetails}.

\subsection{Metric matching and an overdetermining coefficient test}
\label{subsec:SMRBMMatching}

We can now formulate the one-point universality conjecture precisely.
Allowing independent horizontal and vertical metric factors, we write
\be
\Phi(c\xi)=\mathcal A\,\sigma_2(\xi),
\qquad
c>0,
\qquad
\mathcal A>0 .
\label{eq:SMRBMConjecture}
\ee
Here $c$ converts the RBM edge coordinate $\xi$ to the Riccati scaling
variable $X$, while $\mathcal A$ converts the normalization of the two
cumulative intensities.

The leading allowed-side laws,
$\Phi(X)\sim A_+X^{3/2}$ with
$A_+=4/(3\pi\sqrt3)$ and
$\sigma_2(\xi)\sim2\xi^{3/2}/(3\pi)$, give
\be
\mathcal A
=
\frac{2}{\sqrt3}\,c^{3/2}.
\label{eq:SMRBMLeadingMetric}
\ee
This relation alone cannot determine both metric factors.  The second
condition is supplied by the regularized moments.  Because the leading
backgrounds match under Eq.~\eqref{eq:SMRBMLeadingMetric}, subtracting
them from Eq.~\eqref{eq:SMRBMConjecture} and changing variables
$X=c\xi$ gives
\be
b_\Phi
=
\mathcal A c\,b_{\rm RBM}.
\label{eq:SMRBMMomentMetric}
\ee
Using $b_\Phi=1/288$ and $b_{\rm RBM}=1/72$ therefore yields
\be
c
=
\left(\frac{3}{64}\right)^{1/5},
\qquad
\mathcal A
=
48^{-1/5},
\qquad
\mathcal A c=\frac14 .
\label{eq:SMRBMMetric}
\ee
Numerically,
$c=0.5422358856\ldots$ and
$\mathcal A=0.4610539557\ldots$.
No RBM spectral data are fitted in obtaining these two factors.

Once $c$ and $\mathcal A$ are fixed, further terms in the allowed-side
expansion become genuine tests rather than additional normalization
conditions.  The random-phase ribbon expansion gives
\be
\sigma_2(\xi)
=
\frac{2}{3\pi}\xi^{3/2}
+
C_\sigma
+
a_{\rm RBM}\xi^{-7/2}
+
O(\xi^{-17/2}),
\qquad
a_{\rm RBM}
=
-\frac{7}{2304\pi},
\qquad
\xi\to+\infty .
\label{eq:SMRBMFirstTailCoeff}
\ee
Here $C_\sigma$ is the constant term in the allowed-side expansion.
The coefficient $a_{\rm RBM}$ comes from the first nonanalytic,
genus-two contribution to the one-point transform.  Its detailed
ribbon-graph evaluation is given in
Sec.~\ref{subsec:SMRBMdiagramDetails}; importantly, it is not used in
determining $c$ or $\mathcal A$.

Writing Eq.~\eqref{eq:SMRBMConjecture} equivalently as
$\Phi(X)=\mathcal A\,\sigma_2(X/c)$, the coefficient of $X^{-7/2}$
predicted from the RBM side is
\be
a_\Phi
=
\mathcal A\,a_{\rm RBM}\,c^{7/2}
=
-\frac{7}{24576\pi\sqrt3},
\label{eq:SMRBMCoeffMatch}
\ee
in exact agreement with the independently obtained Painlev\'e-I
coefficient in Eq.~\eqref{eq:SMRBMPhiAllowed}.  This is therefore the
first analytic overdetermining test of
Eq.~\eqref{eq:SMRBMConjecture}, rather than part of the metric
matching.

\subsection{Independent checks, predictions, and status of the conjecture}
\label{subsec:SMRBMStatus}

The normalization entering Eq.~\eqref{eq:SMRBMbRBM} has two checks
independent of the Riccati comparison.  First, the same ribbon-graph
conventions in the supercritical regime reproduce the GUE/Airy
one-point transform: the unitary tadpole vanishes and the first
loopless contribution gives the known coefficient $1/384$.
Second, direct finite-bandwidth measurements of the Liu--Zou one-point
transform for sharp-cutoff complex-Hermitian band matrices with
$W=16$--$48$ give a small-$\tau$ coefficient $0.0065(8)$, consistent
with $1/144=0.006944\ldots$.

A separate numerical test was performed directly on cyclic
complex-Hermitian band matrices with
$(M,W)=(1024,16),(2048,24),(2048,32)$.  Using the metric factors in
Eq.~\eqref{eq:SMRBMMetric}, the three data sets collapse onto the same
crossover shape.  Over approximately $0.8\lesssim\xi\lesssim2$ the raw
intensity lies $2$--$5\%$ below the limiting prediction.  The same
finite-$W$ deficit is seen independently in the fitted allowed-side
coefficient.  Writing the finite-$W$ cumulative intensity in this
region as $\sigma_{2,W}(\xi)\simeq\alpha_W\xi^{3/2}$, one finds
\be
\frac{\alpha_W}{2/(3\pi)}
=
0.904,\quad0.926,\quad0.963
\qquad
(W=16,24,32).
\label{eq:SMRBMFiniteWDeficit}
\ee
Removing this independently measured normalization deficit brings the
crossover curves to the predicted shape at the few-percent level.  We
regard this as supporting numerical evidence, not as a proof of
Eq.~\eqref{eq:SMRBMConjecture}.

The conjecture also makes two sharp predictions not used in fixing the
metric.  First, since the global calculation of
Sec.~\ref{sec:phi-numerics} gives $C_\Phi=0$ and
$C_\Phi=\mathcal A C_\sigma$ under
Eq.~\eqref{eq:SMRBMConjecture}, it predicts
\be
C_\sigma=0 .
\label{eq:SMRBMConstantPrediction}
\ee
Second, the forbidden-side Painlev\'e action
\eqref{eq:SMRBMPhiForbidden} fixes the sharp RBM tail rate:
\be
-\log\sigma_2(\xi)
\sim
C_{\rm RBM}|\xi|^{5/4},
\qquad
C_{\rm RBM}
=
A_-c^{5/4}
=
\frac{8\sqrt3}{5}
=
2.771281292\ldots ,
\qquad
\xi\to-\infty .
\label{eq:SMRBMForbiddenPrediction}
\ee
The vertical metric $\mathcal A$ contributes only an additive $O(1)$
term to $-\log\sigma_2$ and therefore does not affect this rate.
Sodin's Theorem~1.2 establishes a stretched-exponential upper bound
on the $|\xi|^{5/4}$ scale, but not the coefficient in
Eq.~\eqref{eq:SMRBMForbiddenPrediction}.

The logical status of the comparison is therefore precise.  On the
Riccati side, $\Phi$, its Painlev\'e-I asymptotics, the regularized
moment $b_\Phi=1/288$, and its global normalization are fixed by the
preceding sections.  On the RBM side, Sodin proves the localized-edge
$4/5$ scaling and a forbidden-side upper bound on the $5/4$ scale,
while the Liu--Zou theory supplies the trivalent diagram expansion from
which $b_{\rm RBM}=1/72$ is evaluated.  The leading $3/2$ law and these
two regularized moments fix the metric factors without fitting.  With
those factors fixed, the exact agreement of the independent
$\xi^{-7/2}$ coefficient and the finite-bandwidth crossover data
provide additional support.  What remains conjectural is the one-point
identity \eqref{eq:SMRBMConjecture} itself; equality of the full RBM
and Riccati point processes is a stronger statement and is not claimed.

\subsection{Details of the random-band diagram calculations}
\label{subsec:SMRBMdiagramDetails}

For completeness, we give here the diagram calculations entering
Eqs.~\eqref{eq:SMRBMbRBM} and \eqref{eq:SMRBMFirstTailCoeff}.  Besides
providing the numerical coefficients, this makes explicit the
normalization conventions of the Liu--Zou expansion and, in
particular, the factor of two associated with the two spectral edges.

\paragraph{From averaged polynomial traces to ribbon graphs.---}
As described in Sec.~\ref{subsec:SMRBMDiagrams}, Liu and Zou probe the
edge through disorder averages of normalized traces of high-degree
Chebyshev-type polynomials, which in the complex Gaussian case reduce
to $\aver{n^{-1}\Tr U_n(H/2)}$.  Expanding a matrix polynomial expresses
its trace as a sum over closed band walks.  Since the entries of $H$
are centered and independent, disorder averaging eliminates most walks;
in the complex-Hermitian class the surviving terms pair each matrix
element with its complex conjugate.  Long chains of degree-two vertices
may then be contracted to single edges of integer lengths $w_e$,
leaving a finite orientable ribbon graph $\Gamma$.

The resulting construction may be summarized schematically as
\begin{center}
\begin{tikzpicture}[
  node distance=0.42cm,
  box/.style={
    draw,
    rounded corners,
    align=center,
    inner xsep=4pt,
    inner ysep=3pt,
    font=\small
  },
  >=stealth
]
\node[box] (poly) {averaged\\polynomial trace};
\node[box, right=of poly] (walk) {long band\\walks};
\node[box, right=of walk] (ribbon) {ribbon\\graphs};
\node[box, right=of ribbon] (diff) {diffusive\\graph integrals};
\node[box, right=of diff] (edge) {edge\\statistics};
\draw[->] (poly) -- (walk);
\draw[->] (walk) -- (ribbon);
\draw[->] (ribbon) -- (diff);
\draw[->] (diff) -- (edge);
\end{tikzpicture}
\end{center}
The cyclic ordering inherited from the trace supplies the ribbon
structure.  If $V$, $E$ and $F$ denote respectively the vertex, edge
and face sets of the orientable ribbon graph, and $g$ its genus, the
diagrams are organized by Euler's relation
$|V|-|E|+|F|=2-2g$.  After contraction of the long degree-two chains,
the generic critical vertices are trivalent, which is the origin of
the $\phi^3$ terminology in Ref.~\cite{LiuZou2024}.

\paragraph{Diffusive graph weights and the theta graph.---}
The variance profile $S_{xy}=\aver{|H_{xy}|^2}$ is the transition
kernel of the underlying band walk.  A long graph edge therefore
becomes, after rescaling, a one-dimensional heat kernel
\be
p_\alpha(x)
=
\frac{1}{\sqrt{2\pi\varsigma^2\alpha}}
\exp\!\left[-\frac{x^2}{2\varsigma^2\alpha}\right],
\label{eq:SMRBMdetailHeatKernel}
\ee
where $\alpha$ is the continuum edge length and $\varsigma^2$ is the
variance of the rescaled longitudinal step.  For a graph $\Gamma$ the
dependence on its edge lengths is encoded by the first Symanzik
polynomial
\be
\mathcal U_\Gamma(\bm\alpha)
=
\sum_{T\ {\rm spanning\ tree}}
\prod_{e\notin T}\alpha_e ,
\label{eq:SMRBMdetailSymanzik}
\ee
which is Eq.~(3.5) of Ref.~\cite{LiuZou2024}.  In one spatial
dimension the Gaussian integrations over internal vertices produce
$\mathcal U_\Gamma^{-1/2}$.

It is useful to see this explicitly for the theta graph:

\begin{center}
\begin{tikzpicture}[scale=0.92,every node/.style={font=\small}]
  \coordinate (A) at (0,0);
  \coordinate (B) at (4,0);
  \draw[thick] (A) -- node[below] {$\alpha_1$} (B);
  \draw[thick] (A) to[bend left=40]
       node[above] {$\alpha_2$} (B);
  \draw[thick] (A) to[bend right=40]
       node[below] {$\alpha_3$} (B);
  \fill (A) circle (2pt);
  \fill (B) circle (2pt);
  \node[left] at (A) {$A$};
  \node[right] at (B) {$B$};
\end{tikzpicture}
\end{center}

Fixing the translational zero mode by placing $A$ at the origin and
denoting by $x$ the coordinate of $B$ gives
\bea
I_\theta
&=&
\int_{\mathbb R}\dd x\,
p_{\alpha_1}(x)p_{\alpha_2}(x)p_{\alpha_3}(x)
\nonumber\\
&=&
\frac{1}{2\pi\varsigma^2}
\frac{1}{
\sqrt{\alpha_1\alpha_2+\alpha_1\alpha_3+\alpha_2\alpha_3}},
\label{eq:SMRBMdetailTheta}
\eea
in agreement with
$\mathcal U_\theta=\alpha_1\alpha_2+\alpha_1\alpha_3+
\alpha_2\alpha_3$.

The graph needed for the first shape-sensitive coefficient is obtained
by subdividing one theta branch:

\begin{center}
\begin{tikzpicture}[scale=0.92,every node/.style={font=\small}]
  \coordinate (A) at (0,0);
  \coordinate (C) at (2,0);
  \coordinate (B) at (4,0);
  \draw[thick] (A) -- node[below] {$\alpha_1$} (C);
  \draw[thick] (C) -- node[below] {$\alpha_2$} (B);
  \draw[thick] (A) to[bend left=40]
       node[above] {$\alpha_3$} (B);
  \draw[thick] (A) to[bend right=40]
       node[below] {$\alpha_4$} (B);
  \fill (A) circle (2pt);
  \fill (B) circle (2pt);
  \fill (C) circle (2pt);
  \node[left] at (A) {$A$};
  \node[right] at (B) {$B$};
  \node[above] at (C) {$C$};
\end{tikzpicture}
\end{center}

If $y$ is the coordinate of $C$, the heat-kernel semigroup identity
$\int\dd y\,p_{\alpha_1}(y)p_{\alpha_2}(x-y)
=p_{\alpha_1+\alpha_2}(x)$ immediately gives
\be
I_{\rm sub}
=
\frac{1}{2\pi\varsigma^2}
\frac{1}{
\sqrt{\alpha_3\alpha_4+
(\alpha_1+\alpha_2)(\alpha_3+\alpha_4)}} ,
\label{eq:SMRBMdetailSubTheta}
\ee
so that
$\mathcal U_{\rm sub}
=\alpha_3\alpha_4+
(\alpha_1+\alpha_2)(\alpha_3+\alpha_4)$.

\paragraph{Marked tail and the coefficient $b_{\rm RBM}$.---}
In the Liu--Zou one-point convention the subdivided-theta core carries
one additional marked tail:

\begin{figure}[h]
\centering
\begin{tikzpicture}[scale=0.88,every node/.style={font=\small}]
  \coordinate (A) at (0,0);
  \coordinate (C) at (2,0);
  \coordinate (B) at (4,0);
  \coordinate (L) at (2,1.65);
  \draw[thick] (A) -- node[below] {$w_1$} (C);
  \draw[thick] (C) -- node[below] {$w_2$} (B);
  \draw[thick] (A) to[bend left=42]
       node[above] {$w_3$} (B);
  \draw[thick] (A) to[bend right=42]
       node[below] {$w_4$} (B);
  \draw[thick] (C) -- node[right] {$w_0$} (L);
  \fill (A) circle (2pt);
  \fill (B) circle (2pt);
  \fill (C) circle (2pt);
  \fill (L) circle (2pt);
  \node[left] at (A) {$A$};
  \node[right] at (B) {$B$};
  \node[above left] at (C) {$C$};
  \node[above] at (L) {\scriptsize marked leaf};
\end{tikzpicture}
\caption{First shape-sensitive one-point ribbon diagram in the
Liu--Zou convention.  Removing the marked tail $w_0$ leaves the
subdivided-theta core.}
\label{fig:SMRBMMarkedTail}
\end{figure}
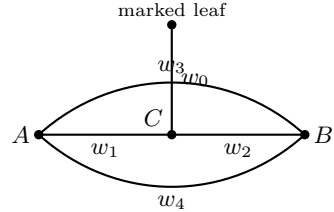

The full graph has four vertices, five edges and one face, hence genus
$g=1$.  For even polynomial degree $n$, the single face traverses
every edge twice, and therefore
\be
2(w_0+w_1+w_2+w_3+w_4)=n,
\qquad
w_0=\frac n2-\sum_{i=1}^4w_i.
\label{eq:SMRBMdetailLengthConstraint}
\ee
Eliminating $w_0$ leaves
$\sum_{i=1}^4w_i\le n/2$.  Let $D\subset\mathbb R^5$ denote the
four-dimensional affine hyperplane defined by the equality in
Eq.~\eqref{eq:SMRBMdetailLengthConstraint}.  Its induced Euclidean
measure is
$\dd\operatorname{Vol}_D=\sqrt5\,\dd w_1\cdots\dd w_4$, whereas
the lattice-density factor associated with $D$ in the Liu--Zou
continuum formula is $C_D=1/\sqrt5$.  Hence
\be
C_D\,\dd\operatorname{Vol}_D
=
\dd w_1\,\dd w_2\,\dd w_3\,\dd w_4 .
\label{eq:SMRBMdetailJacobian}
\ee
Thus the hyperplane factor cancels exactly: the marked tail introduces
no additional graph multiplicity.

Passing to continuum lengths and using
Eq.~\eqref{eq:SMRBMdetailSubTheta}, the core contribution before the
explicit $1/n$ normalization of the trace is
\be
\mathcal I_\Gamma(\tau)
=
\frac{1}{2\pi\varsigma^2}
\int_{\substack{\alpha_e\ge0\\
2\sum_{e=1}^4\alpha_e\le\tau}}
\frac{\dd^4\alpha}{
\sqrt{\alpha_3\alpha_4+
(\alpha_1+\alpha_2)(\alpha_3+\alpha_4)}} .
\label{eq:SMRBMdetailCoreIntegral}
\ee
With $\alpha_e=(\tau/2)u_e$,
\be
\mathcal I_\Gamma(\tau)
=
\frac{\tau^3}{16\pi\varsigma^2}\,J,
\qquad
J=
\int_{\substack{u_i\ge0\\\sum_i u_i\le1}}
\frac{\dd^4u}{
\sqrt{u_3u_4+(u_1+u_2)(u_3+u_4)}} .
\label{eq:SMRBMdetailJdef}
\ee
The factor $1/n$ in the normalized trace removes one power of $\tau$,
so that this graph contributes
$\tau^2J/(16\pi\varsigma^2)$ to the one-point transform.

The remaining integral is the only nontrivial continuous integral in
this calculation.  Set $s=u_1+u_2$, $r=u_3+u_4$ and
$u_3=rv$, $u_4=r(1-v)$.  Integrating successively over the splitting of
$s$, the overall radial variable and $v$ reduces the four-dimensional
simplex integral to
\bea
J
&=&
\frac23\int_0^1
(1-z)\arcsin\sqrt{\frac{z}{4-3z}}\,\dd z
\nonumber\\
&=&
\frac{\pi}{27},
\label{eq:SMRBMdetailJexact}
\eea
where the last equality follows by integration by parts and
$z=\sin^2\theta$.  Consequently the contribution of this graph is
\be
F_\Gamma(\tau)
=
\frac{\tau^2}{432\varsigma^2}
+o(\tau^2).
\label{eq:SMRBMdetailGraphCoeff}
\ee
For Sodin's sharp-cutoff profile the rescaled longitudinal step is
uniform on $[-1,1]$, hence
$\varsigma^2=\frac12\int_{-1}^{1}x^2\dd x=1/3$ and
\be
F_\Gamma(\tau)
=
\frac{\tau^2}{144}+o(\tau^2).
\label{eq:SMRBMdetailGraph144}
\ee
At this order $F_\Gamma$ supplies the $\tau^2$ coefficient of the
regularized one-point transform $\phi_1^{\rm reg}$.

We finally convert this transform coefficient into the one-edge
regularized area.  Because a full matrix trace contains equal
contributions from the upper and lower spectral edges, the transform
is Eq.~\eqref{eq:SMRBMTransform}, with its explicit prefactor $2$.
After subtracting the leading edge background, integration by parts
and
$\sin(\tau\sqrt\xi)/(\tau\sqrt\xi)
=1-\tau^2\xi/6+O(\tau^4\xi^2)$ give
\be
[\tau^2\text{ coefficient of }\phi_1^{\rm reg}]
=
\frac13
\int_{-\infty}^{\infty}
\delta\sigma_2(\xi)\,\dd\xi
=
\frac12\,b_{\rm RBM}.
\label{eq:SMRBMdetailAreaRelation}
\ee
Equation~\eqref{eq:SMRBMdetailGraph144} therefore yields
\be
b_{\rm RBM}=\frac1{72},
\label{eq:SMRBMdetailbExact}
\ee
which is the result used in Eq.~\eqref{eq:SMRBMbRBM}.  Thus
$J=\pi/27$ is the central integral evaluation, whereas
$b_{\rm RBM}=1/72$ is the physical shape coefficient entering the
metric matching.

\paragraph{Two independent normalization checks.---}
Because the metric factors depend directly on
$b_{\rm RBM}$, its normalization can be checked without reference to
the Riccati problem.  First, applying the same ribbon-graph conventions
in the supercritical regime reproduces the known GUE/Airy transform:
the unitary tadpole vanishes and the first loopless class gives
\be
c_1=0,
\qquad
c_2=\frac1{384}.
\label{eq:SMRBMdetailGUEcheck}
\ee
This tests the graph multiplicity, the double traversal of every edge
and the two-edge convention.  Second, direct measurements of the
Liu--Zou transform for sharp-cutoff complex-Hermitian band matrices
with $W=16$--$48$ give
\be
[\tau^2\text{ coefficient of }\phi_1^{\rm reg}]_{\rm num}
=
0.0065(8),
\qquad
\frac1{144}=0.006944\ldots ,
\label{eq:SMRBMdetailNumericalCheck}
\ee
consistent with Eq.~\eqref{eq:SMRBMdetailGraph144}.

\paragraph{The first nonanalytic coefficient.---}
The preceding genus-one calculation supplies the regular coefficient
used to fix the metric.  A logically independent test comes from the
first nonanalytic term of the unitary ribbon expansion.  Write this
term in the regularized one-point transform as
$\phi_1^{\rm reg}(\tau)\supset C_4|\tau|^7$.  In the
random-phase class the orientation constraint eliminates the odd
diagram sectors, and the nonanalytic powers of the one-point transform
are $|\tau|^7,|\tau|^{17},|\tau|^{27},\ldots$, corresponding to the
allowed-side powers
$\xi^{-7/2},\xi^{-17/2},\xi^{-27/2},\ldots$ of the cumulative edge
intensity.

For the first such contribution the reduced one-point diagram has
\be
V=8,\qquad E=11,\qquad F=1,
\qquad
2-2g=V-E+F=-2,
\label{eq:SMRBMdetailGenusTwo}
\ee
and hence $g=2$.  Removing the rooted degree-one vertex and suppressing
the resulting degree-two vertex reduces the problem to rooted cubic
unicellular maps of genus two.  There are $105$ such maps, falling
into five underlying cubic multigraph classes with multiplicities
$9,18,36,36,6$.

For $k\ge1$ let
$\Delta_k=\{(x_1,\ldots,x_{k+1})\in\mathbb R_+^{k+1}:
\sum_{e=1}^{k+1}x_e=1\}$ denote the standard $k$-simplex, equipped
with its induced $k$-dimensional measure.  For a reduced one-point
diagram $\mathscr D$ with eleven edges, the continuum embedding
integral has the form
\be
J_{\mathscr D}
=
\int_{\Delta_{10}}
\frac{\dd^{10}x}{\sqrt{\mathcal U_{\mathscr D}(x)}} .
\label{eq:SMRBMdetailJD}
\ee
The root edge is a bridge and may be integrated explicitly.  After
suppressing the resulting series pair one is left with a six-vertex,
nine-edge cubic multigraph $G$.  Let
$\mathfrak M_2^{(3)}$ denote the set of the $105$ rooted genus-two
cubic unicellular maps, and let $G(\mathcal M)$ be the resulting
six-vertex multigraph associated with a map
$\mathcal M\in\mathfrak M_2^{(3)}$.  Then
\be
I_G
=
\int_{\Delta_8}
\mathcal U_G(\bm\alpha)^{-1/2}\,\dd^8\alpha,
\qquad
\sum_{\mathscr D}J_{\mathscr D}
=
\frac1{72}
\sum_{\mathcal M\in\mathfrak M_2^{(3)}}
I_{G(\mathcal M)} .
\label{eq:SMRBMdetailIG}
\ee

These one-dimensional graph integrals can be evaluated particularly
simply in position space.  Introduce four independent loop momenta
$p_1,\ldots,p_4$; momentum conservation expresses the momentum
$q_e$ carried by each of the nine graph edges as a linear combination
of the $p_\ell$.  The corresponding massive scalar integral is
\be
\mathcal I_G
=
\int\prod_{\ell=1}^{4}\frac{\dd p_\ell}{2\pi}
\prod_{e=1}^{9}\frac1{q_e^2+1}.
\label{eq:SMRBMdetailFeynman}
\ee
Schwinger parametrization gives
$\mathcal I_G=(45/\pi^2)I_G$, while the one-dimensional propagator
$\int\dd p\,e^{ipx}/[2\pi(p^2+1)]=e^{-|x|}/2$ gives, after fixing one
vertex coordinate, say $x_6=0$, to remove the overall translational
zero mode,
\be
I_G
=
\frac{\pi^2}{23040}K_G,
\qquad
K_G
=
\int_{\mathbb R^5}
\exp\!\left[
-\sum_{(uv)\in E(G)}|x_u-x_v|
\right]\dd x_1\cdots\dd x_5 .
\label{eq:SMRBMdetailKG}
\ee
Ordering the six vertex coordinates, including the fixed coordinate
$x_6=0$, removes the absolute values.  If $c_j(\pi)$ is the number of
graph edges crossing the cut between the first $j$ and the last $6-j$
vertices in an ordering $\pi\in S_6$, then
\be
K_G
=
\sum_{\pi\in S_6}
\prod_{j=1}^{5}\frac1{c_j(\pi)} .
\label{eq:SMRBMdetailCuts}
\ee
Thus every graph integral reduces to a finite rational sum.  To make
the five graph classes explicit, choose vertices $1,\ldots,6$ and
represent each class by the following edge multiset; repeated pairs
denote parallel edges:
\be
\begin{aligned}
E_1={}&\{15,16,16,23,24,25,34,34,56\},\\
E_2={}&\{15,16,16,24,25,26,34,34,35\},\\
E_3={}&\{15,16,16,23,24,26,34,35,45\},\\
E_4={}&\{14,15,16,23,25,26,34,36,45\},\\
E_5={}&\{14,15,16,24,25,26,34,35,36\}.
\end{aligned}
\label{eq:SMRBMdetailGraphRepresentatives}
\ee
Here, for example, $16$ denotes an edge joining vertices $1$ and $6$.
For these five classes one obtains
\be
\begin{array}{c|c|c}
\text{class}&\text{multiplicity}&K_G\\
\hline
1&9&128/81\\
2&18&1316/1215\\
3&36&379/405\\
4&36&104/135\\
5&6&298/405
\end{array},
\qquad
\sum_{\mathcal M\in\mathfrak M_2^{(3)}}
K_{G(\mathcal M)}
=
\frac{896}{9}.
\label{eq:SMRBMdetailFiveGraphs}
\ee
Combining the graph multiplicities, the continuum normalization and
Eq.~\eqref{eq:SMRBMdetailKG} gives the coefficient $C_4$ of the
$|\tau|^7$ term,
\be
C_4
=
\frac{9}{1024\pi^2}
\frac1{72}
\frac{\pi^2}{23040}
\frac{896}{9}
=
\frac{7}{13271040}.
\label{eq:SMRBMdetailC4}
\ee

To convert this two-edge transform coefficient to the one-edge
cumulative intensity, suppose
$\delta\sigma_2(\xi)\supset a_{\rm RBM}\xi^{-7/2}$.
The nonanalytic part of Eq.~\eqref{eq:SMRBMTransform} then gives
\be
C_4
=
-\frac{\pi}{5760}\,a_{\rm RBM},
\label{eq:SMRBMdetailC4a}
\ee
where the denominator $5760$ includes the factor $2$ from the two
equivalent spectral edges.  Hence
\be
a_{\rm RBM}
=
-\frac{7}{2304\pi}.
\label{eq:SMRBMdetailaRBM}
\ee
This is the coefficient quoted in
Eq.~\eqref{eq:SMRBMFirstTailCoeff}.  Since the metric factors
$c$ and $\mathcal A$ were already fixed independently by the leading
$3/2$ law and by $b_\Phi/b_{\rm RBM}$, the exact agreement of
Eq.~\eqref{eq:SMRBMdetailaRBM} with the Painlev\'e-I
$X^{-7/2}$ coefficient is a genuine overdetermining test of the
one-point universality conjecture.

\section{Finite-length propagator and the lowest-eigenvalue distribution}
\label{sec:SMsurvival}

Sections~\ref{sec:SMmodel} and~\ref{sec:SMprojective} already identify
the finite-length Dirichlet level count with the total projective
winding count of the Riccati eigenvalues, and describe the cyclic
reinjection at $\lambda=\pm\infty$.  We now use these results to
construct the full finite-length counting propagator.

The central object is a transition density in which every projective
passage is assigned a factor $z$.  It contains three cases of direct
interest.  For $z=1$ it is the ordinary Riccati transition density;
for $z=e^{\ii\chi}$ it generates the finite-length spectral counting
statistics; and for $z=0$ it retains only trajectories with no
projective passage.  In the Dirichlet problem the last event is
precisely the event that the lowest eigenvalue lies above $E$.
At $\beta=2$, bulk fermionization remains valid for all $z$: the
counting field changes only the projective boundary matching.  The
special value $z=0$ removes reinjection altogether and turns the
problem into $N$ noninteracting fermions in the self-adjoint quartic
Hamiltonian appearing below.

\subsection{Counting-weighted propagator}
\label{subsec:SMcountingPropagator}

We first construct the finite-length object from which both the
spectral counting statistics and the lowest-eigenvalue distribution
follow.  Fix $E$, let
$\bm\mu=(\mu_1,\ldots,\mu_N)$ be an interior point of the ordered
chamber $\mathcal C_N$, and let $K_x(E)$ denote the total number of
projective passages of the Riccati eigenvalues during $(0,x]$, counted
with multiplicity.  The relation between these passages and the
Dirichlet spectral count was established by matrix oscillation in
Sec.~\ref{sec:SMmodel}.

For a complex counting parameter $z$, define the weighted transition
density by
\be
P_x^{[z]}(\bm\lambda\mid\bm\mu)\,\dd^N\lambda
=
\mathbb E_{\bm\mu}\!\left[
z^{K_x(E)}
\mathbf 1_{\{\bm\lambda(x)\in\dd^N\lambda\}}
\right].
\label{eq:SMweightedTransitionDefinition}
\ee
Here $\mathbb E_{\bm\mu}$ denotes expectation for the Riccati process
started from $\bm\lambda(0)=\bm\mu$.  Thus $z=1$ gives the ordinary
transition density, $z=e^{\ii\chi}$ assigns the phase
$e^{\ii\chi}$ to every projective passage, while $z=0$ retains only
trajectories with no passage.  The latter case will give the
lowest-eigenvalue survival probability below.

The bulk dynamics is unchanged by the factor $z^{K_x}$, since $K_x$
changes only when a Riccati eigenvalue crosses the projective point.
We may therefore use the same exact $\beta=2$ fermionization as in
Sec.~\ref{sec:SMfermionization}.  With
\be
V_N(\bm\lambda)
=
\prod_{i<j}(\lambda_j-\lambda_i),
\qquad
\mathcal C_N
=
\{\lambda_1<\cdots<\lambda_N\},
\label{eq:SMfiniteVandermonde}
\ee
the transformation $P=V_N\Psi$ reduces the many-body bulk evolution
to $N$ copies of the one-body operator
\be
\ell_N
=
T\partial_\lambda^2
+(E+\lambda^2)\partial_\lambda
+(N+1)\lambda ,
\label{eq:SMfiniteOneBodyOperator}
\ee
which is the operator already obtained in
Eq.~\eqref{eq:SMfermionization}.

Let $G_x^{[z]}(\lambda,\mu)$ denote its forward kernel with the
counting-weighted projective boundary condition specified below:
\be
\partial_xG_x^{[z]}(\lambda,\mu)
=
\ell_N^{(\lambda)}G_x^{[z]}(\lambda,\mu),
\qquad
G_0^{[z]}(\lambda,\mu)
=
\delta(\lambda-\mu).
\label{eq:SMcountingOneBodyKernel}
\ee
Then the weighted $N$-body propagator is
\be
P_x^{[z]}(\bm\lambda\mid\bm\mu)
=
\frac{V_N(\bm\lambda)}{V_N(\bm\mu)}
\det_{1\leq i,j\leq N}
\left[
G_x^{[z]}(\lambda_i,\mu_j)
\right].
\label{eq:SMweightedKM}
\ee
The determinant solves the fermionized bulk equation, vanishes when
two Riccati eigenvalues collide, and together with the Vandermonde
ratio reproduces the delta-function initial condition in the ordered
chamber.  What remains is to specify how the counting factor $z$ is
implemented at the projective boundary.

\paragraph{Counting weight at the projective boundary.---}
The global geometry needed here has already been established in
Sec.~\ref{sec:SMprojective}.  Recall only that
$\lambda=\cot\theta$, with $\theta=0$ and $\theta=\pi$ representing
the same point of $\mathbb{RP}^1$, and that a passage through infinity
is followed by the cyclic relabelling
Eq.~\eqref{eq:SMcyclicReinjection}.  The corresponding compactified
one-body orbitals carry the fermionic factor
\be
\sigma_N=(-1)^{N-1}
\label{eq:SMfiniteSigma}
\ee
under one circuit of the projective circle, as shown in
Eq.~\eqref{eq:SMfermionicTwist}.

To state the same condition for the propagator, write
$\lambda=\cot\theta$ and $\mu=\cot\varphi$, and define its compactified
form by
\be
\widehat G_x^{[z]}(\theta,\varphi)
=
\sin^{-(N+1)}\theta\,
G_x^{[z]}(\cot\theta,\cot\varphi)\,
\sin^{N-1}\varphi .
\label{eq:SMcompactCountingKernel}
\ee
For fixed initial point $\varphi$, assigning an additional factor $z$
to a forward passage through the projective point changes the
untwisted matching condition to
\be
\widehat G_x^{[z]}(0,\varphi)
=
\sigma_N z\,
\widehat G_x^{[z]}(\pi,\varphi),
\qquad
\sigma_N=(-1)^{N-1}.
\label{eq:SMforwardtwist}
\ee
No modification of the differential operator
Eq.~\eqref{eq:SMcountingOneBodyKernel} is involved: the counting field
enters entirely through this global boundary condition.

It remains to check that Eq.~\eqref{eq:SMforwardtwist} indeed assigns
the factor $z$ to the physical many-body trajectory.  Under one
projective passage, restoring the ordered Riccati labels cyclically
permutes the rows of the Slater determinant in
Eq.~\eqref{eq:SMweightedKM}.  This produces a second factor
$(-1)^{N-1}=\sigma_N$.  The fermionic sign in
Eq.~\eqref{eq:SMforwardtwist} therefore cancels the permutation sign,
while the circular Vandermonde is continuous under the same cyclic
gluing, as shown in Eq.~\eqref{eq:SMcircularGluing}.  The complete
many-body propagator consequently acquires exactly
\be
\sigma_N^2z=z
\ee
at each passage.  After $K_x(E)$ passages its weight is therefore
$z^{K_x(E)}$, which proves the probabilistic interpretation
Eq.~\eqref{eq:SMweightedTransitionDefinition} of
Eq.~\eqref{eq:SMweightedKM}.

The three boundary realizations needed below are consequently
transparent.  At $z=1$ one recovers the physical reinjecting Riccati
process of Sec.~\ref{sec:SMprojective}.  At
$z=e^{\ii\chi}$ one obtains the counting-field kernel used in the
Main Text,
\be
G_x^\chi(\lambda,\mu)
:=
G_x^{[e^{\ii\chi}]}(\lambda,\mu).
\label{eq:SMGchiDefinition}
\ee
Finally, at $z=0$ there is no incoming component after a projective
passage: the process is killed at its first passage through infinity.
This zero-count realization is the starting point of
Sec.~\ref{subsec:SMzeroCount} below.

\subsection{Dirichlet entrance and the finite-length level count}

For the spectral problem the fundamental matrix obeys
$Y(0)=0$ and $Y'(0)=I_N$, so that
$Z(x)=Y'(x)Y^{-1}(x)\simeq I_N/x$ as $x\downarrow0$.
Thus all $N$ Riccati eigenvalues enter from $+\infty$ together.
Since the determinantal formula
Eq.~\eqref{eq:SMweightedKM} is written for distinct points in the
ordered chamber, this degenerate entrance must be understood as a
confluent limit.

More precisely, for fixed $M$ take
\be
\mu_i=M+\varepsilon\xi_i,
\qquad
\xi_1<\cdots<\xi_N,
\label{eq:SMDirichletConfluent}
\ee
so that $\bm\mu\in\mathcal C_N$ for every $\varepsilon>0$, and first
let $\varepsilon\to0^+$.  This produces the confluent limit
$\mu_1,\ldots,\mu_N\to M$ while preserving the ordering throughout
the regularization.  The resulting initial condition is
$Z(0)=MI_N$, equivalently $Y(0)=I_N$, $Y'(0)=MI_N$.
After the harmless rescaling $Y\mapsto M^{-1}Y$, the subsequent limit
$M\to+\infty$ gives the Dirichlet normalization
$Y(0)=0$, $Y'(0)=I_N$.  We denote the combined limiting procedure by
$\lim_{\bm\mu\to+\infty}^{\rm D}$.

With the convention that the initial point itself is not counted,
matrix oscillation, Sec.~\ref{sec:SMmodel}, gives
\be
K_L(E)=\mathcal N_L(E),
\label{eq:SMfiniteCountEquality}
\ee
where $\mathcal N_L(E)$ is the number of Dirichlet eigenvalues of
$\mathcal H_L$ below $E$.  Integrating
Eq.~\eqref{eq:SMweightedKM} over the final Riccati configuration
therefore gives
\bea
\mathcal F_N(z;E,L)
&:=&
\mathbb E\!\left[z^{\mathcal N_L(E)}\right]
\nonumber\\
&=&
\lim_{\bm\mu\to+\infty}^{\rm D}
\int_{\mathcal C_N}
\frac{V_N(\bm\lambda)}{V_N(\bm\mu)}
\det_{i,j}\!
\left[
G_L^{[z]}(\lambda_i,\mu_j)
\right]
\,\dd^N\lambda .
\label{eq:SMfiniteLengthGenerating}
\eea
For $z=e^{\ii\chi}$ this is Eq.~\eqref{eq:FCS} of the Main Text.
Equivalently,
\be
\mathcal F_N(z;E,L)
=
\sum_{m\geq0}
z^m
\Pr\{\mathcal N_L(E)=m\}.
\label{eq:SMcountseries}
\ee
Thus the same one-body kernel generates the complete finite-length
Dirichlet level-counting distribution.

If the Dirichlet eigenvalues are ordered as
$\mathcal E_0(L)<\mathcal E_1(L)<\cdots$, then
\be
\Pr\{\mathcal E_k(L)\leq E\}
=
\Pr\{\mathcal N_L(E)\geq k+1\}.
\label{eq:SMfiniteEigenvalueCount}
\ee
In particular the lowest-eigenvalue distribution is obtained from the
zero-count sector.

\subsection{Zero count and the killed quartic problem}
\label{subsec:SMzeroCount}

Setting $z=0$ in Eq.~\eqref{eq:SMcountseries} selects trajectories with
no projective passage:
\be
S_N(E,L)
:=
\mathcal F_N(0;E,L)
=
\Pr\{\mathcal N_L(E)=0\}
=
\Pr\{\mathcal E_0(L)>E\}.
\label{eq:SMgroundevent}
\ee
Thus the survival probability of the Riccati process before its first
projective passage is exactly the probability that the finite
Schr\"odinger operator has no eigenvalue below $E$.

For $z=0$ reinjection is suppressed.  In the physical Riccati
dynamics a particle that reaches $\lambda=-\infty$ normally reappears
at $\lambda=+\infty$; the factor $z$ multiplies precisely this
reinjected contribution.  Setting $z=0$ therefore removes it:
probability reaching $\lambda=-\infty$ is lost, and the trajectory is
killed at its first passage through infinity.  The two ends of the
Riccati line are no longer identified.  Thus the killed problem has an
exit boundary at $\lambda=-\infty$ and the usual regular entrance
behaviour at $\lambda=+\infty$.

The relation to an ordinary Hermitian problem becomes transparent
after the standard gauge transformation.  With
$U_E(\lambda)=\lambda^3/3+E\lambda$,
\be
e^{U_E/(2T)}\ell_Ne^{-U_E/(2T)}
=
-H_N,
\qquad
H_N
=
-T\partial_\lambda^2
+\frac{(\lambda^2+E)^2}{4T}
-N\lambda .
\label{eq:SMkilledgauge}
\ee
This identity concerns the bulk differential expression; the boundary
realization still has to be specified.  For the reinjecting Riccati
process the values at $\lambda=-\infty$ and $\lambda=+\infty$ are tied
by the projective matching, so the transformed problem is not the
usual self-adjoint realization of $H_N$.

For $z=0$ this matching is absent.  The absorbing condition at
$\lambda=-\infty$, together with regularity at the entrance
$\lambda=+\infty$, selects under the gauge transformation the
decaying solution of the Schr\"odinger equation at each end.
Equivalently, the transformed eigenfunctions belong to
$L^2(\mathbb R)$.  Since
\be
\frac{(\lambda^2+E)^2}{4T}-N\lambda
\sim
\frac{\lambda^4}{4T},
\qquad |\lambda|\to\infty ,
\ee
the transformed potential is confining, so $H_N$ has the standard
self-adjoint realization on $L^2(\mathbb R)$ and a discrete real
spectrum,
\be
H_N\phi_n=h_n\phi_n,
\qquad
h_0<h_1<\cdots ,
\qquad
\phi_n(\lambda)\longrightarrow0
\quad(\lambda\to\pm\infty),
\label{eq:SMkilledEigenbasis}
\ee
with the eigenfunctions chosen orthonormal,
$\int_{\mathbb R}\phi_n(\lambda)\phi_m(\lambda)\,\dd\lambda
=\delta_{nm}$.

Transforming the spectral resolution back to the original Riccati
variables gives the killed one-body kernel
\be
G_x^{[0]}(\lambda,\mu)
=
e^{-[U_E(\lambda)-U_E(\mu)]/(2T)}
\sum_{n\geq0}
\phi_n(\lambda)\phi_n(\mu)e^{-xh_n}.
\label{eq:SMkilledkernel}
\ee

The many-body meaning of this spectral resolution is particularly
simple.  The killed $N$-body propagator is a determinant of the
one-body kernels~\eqref{eq:SMkilledkernel}.  Expanding this determinant
by Cauchy--Binet organizes the result into the antisymmetric
eigenstates of the $N$-fermion Hamiltonian
\be
\mathbb H_N
=
\sum_{i=1}^N H_N^{(i)} .
\label{eq:SMmanyBodyKilledHamiltonian}
\ee
Since the fermions are noninteracting, an $N$-particle eigenstate is
obtained by occupying $N$ distinct one-body levels.  We label such an
occupation by
\be
J=\{j_1<\cdots<j_N\},
\qquad
\Phi_J(\bm\lambda)
=
\det_{1\leq i,a\leq N}
[\phi_{j_a}(\lambda_i)] ,
\label{eq:SMkilledSlaters}
\ee
with many-body energy
\be
H_J(E)
=
\sum_{j\in J}h_j(E).
\label{eq:SMmanyBodyKilledEnergy}
\ee
Thus $J$ is simply the set of occupied one-body levels of the Slater
state $\Phi_J$.

Substituting the one-body expansion
\eqref{eq:SMkilledkernel} into the determinantal propagator
\eqref{eq:SMweightedKM} gives a sum over these Slater states.  After
integrating over the final Riccati configuration and taking the
Dirichlet confluent limit at the initial point, the zero-count
probability therefore has the spectral representation
\be
S_N(E,L)
=
\sum_{J:\,|J|=N}
A_J(E)e^{-L H_J(E)} .
\label{eq:SMsurvivalSlater}
\ee
The coefficient $A_J(E)$ is the product of the overlap of
$\Phi_J$ with the final integration measure and its overlap with the
Dirichlet entrance.  Its explicit form is
\be
A_J=B_JC_J^{\rm D},
\label{eq:SMsurvivalAmplitude}
\ee
where, with
\be
m(\bm\lambda)
=
V_N(\bm\lambda)
e^{-\sum_iU_E(\lambda_i)/(2T)},
\label{eq:SMkilledMeasureFactor}
\ee
one has
\be
B_J
=
\int_{\mathcal C_N}
m(\bm\lambda)\Phi_J(\bm\lambda)\,\dd^N\lambda,
\qquad
C_J^{\rm D}
=
\lim_{\bm\mu\to+\infty}^{\rm D}
\frac{\Phi_J(\bm\mu)}{m(\bm\mu)} .
\label{eq:SMboundaryoverlaps}
\ee
These coefficients are boundary overlaps, not equilibrium occupation
probabilities.

The lowest antisymmetric $N$-fermion state is obtained by filling the
$N$ lowest one-body levels,
\be
J_0=\{0,1,\ldots,N-1\}.
\label{eq:SMfilledKilledGroundState}
\ee
Hence at large $L$ it dominates the sum
\eqref{eq:SMsurvivalSlater}, giving
\bea
S_N(E,L)
&=&
A_0(E)e^{-L\kappa_N(E)}
\left[
1+
O\!\left(
e^{-L[h_N(E)-h_{N-1}(E)]}
\right)
\right],
\nonumber\\
\kappa_N(E)
&=&
\sum_{n=0}^{N-1}h_n(E)>0 .
\label{eq:SMsurvivalLargeL}
\eea
where $A_0$ is the overlap corresponding to
$J=\{0,\ldots,N-1\}$.  This is the statement quoted in the Main Text:
the probability of finding no level below $E$ is governed
by the filled ground-state energy of $N$ free fermions in the quartic
Hamiltonian $H_N$.

It is worth emphasizing the difference between this problem and the
stationary one studied in the preceding sections.  The differential
expression is the same quartic one, but the global boundary condition
is different.  The stationary problem has projective reinjection and
a nonzero winding current; the zero-count problem suppresses
reinjection and is therefore an absorbing, or killed, problem.  Its
decay rate $\kappa_N(E)$ is consequently not, in general, equal to the
stationary mean level current $\nu_N(E)$.

\subsection{Collective edge and companion calculation}

Equations~\eqref{eq:SMfiniteLengthGenerating}--\eqref{eq:SMsurvivalLargeL}
are the starting point of the finite-length analysis developed in the
companion work~\cite{FLDGap}.  There the Dirichlet confluent limit and
the boundary overlaps are evaluated explicitly, the finite-$N$
survival probability is reduced to a determinant representation, and
the collective edge limit is taken.

At fixed disorder strength, with
$T=J^2/(2N)$ and
$s_J=(2J^2)^{1/3}$, the critical finite-length variables are
\be
E
=
E^*
+s_J^2N^{-4/5}X,
\qquad
E^*
=
-\frac34s_J^2,
\qquad
\ell
=
s_JN^{-1/5}L .
\label{eq:SMgapscaling}
\ee
In this limit the companion calculation constructs a universal
survival function $\mathcal S(\ell,X)$ for the lowest eigenvalue,
represented there as a Fredholm determinant.  Its long-length form is
\be
\mathcal S(\ell,X)
\sim
\mathcal A(X)e^{-\ell\mathcal K(X)},
\qquad
\ell\to\infty ,
\label{eq:SMcriticalSurvival}
\ee
with the amplitude $\mathcal A(X)$ and the scaled escape rate
$\mathcal K(X)$ determined in Ref.~\cite{FLDGap}.  These quantities
contain information beyond the mean density of states studied in the
present Letter: the probability of observing zero levels is a full
counting observable and is not fixed by the first counting cumulant
alone.

\section{Counting statistics, generalized Lyapunov exponents, and joint tilts}
\label{sec:SMFCS}

The finite-length fermionic formulation of
Sec.~\ref{sec:SMsurvival} naturally extends to two broader fluctuation
problems.  A counting tilt probes the full statistics of the number
$\mathcal N_L(E)$ of Dirichlet levels below $E$, while an amplitude
tilt probes moments of $|\det Y(L)|$ and hence generalized Lyapunov
exponents, in the spirit of Ref.~\cite{FLRT2018}.  In both cases the
bulk $\beta=2$ fermionization survives: the deformation enters either
through the projective boundary matching or through a simple tilted
one-body operator.  We formulate these spectral problems below and
identify the cumulants generated by their dominant eigenvalues.
Except for the quantities already obtained in the Letter, we do not
attempt here to solve the complete tilted boundary-value problems.

\subsection{Spectral counting field}

The finite-length counting generating function was constructed in
Sec.~\ref{sec:SMsurvival}.  To characterize its fluctuations as
$L\to\infty$, it is convenient to use a real counting field $t$,
with $z=e^t$, and define the scaled cumulant-generating function
\be
\Lambda_N^{\rm cnt}(t;E)
:=
\lim_{L\to\infty}\frac1L
\log\mathbb E\!\left[e^{t\mathcal N_L(E)}\right].
\label{eq:SMrealSCGF}
\ee
For large $L$ the tilted propagator is governed by its dominant
spectral branch.  Since the counting field changes only the projective
boundary matching, the bulk fermionization of
Sec.~\ref{subsec:SMcountingPropagator} is unchanged.

Let $\varepsilon_a(t)$, $a=1,\ldots,N$, denote the one-body spectral
branches forming the dominant $N$-fermion state continuously connected
to the physical state at $t=0$.  Then
\be
\Lambda_N^{\rm cnt}(t;E)
=
\sum_{a=1}^N\varepsilon_a(t)
-
\sum_{a=1}^N\varepsilon_a(0).
\label{eq:SMthetaSum}
\ee
The subtraction enforces
$\Lambda_N^{\rm cnt}(0;E)=0$.  The relevant branches are selected by
the counting-weighted projective boundary condition; they are not an
arbitrary choice of $N$ one-body eigenvalues.

Near $t=0$, derivatives of
$\Lambda_N^{\rm cnt}$ generate the cumulants of the level count per
unit length.  In particular,
\be
\left.
\partial_t\Lambda_N^{\rm cnt}(t;E)
\right|_{t=0}
=
\nu_N(E),
\qquad
\left.
\partial_t^2\Lambda_N^{\rm cnt}(t;E)
\right|_{t=0}
=
\lim_{L\to\infty}
\frac{\operatorname{Var}\mathcal N_L(E)}{L},
\label{eq:SMcountcumulants}
\ee
provided the dominant branch is differentiable to the indicated
order.  The present Letter evaluates the first derivative through the
stationary winding current; higher counting cumulants are not
calculated here.

For the characteristic function one sets locally
$t=\ii\chi$.  Thus the counting-field exponent
$\Theta_N(\chi;E)$ is the analytic continuation of the same dominant
branch near $\chi=0$.  Away from the origin, however, changes of the
dominant spectral branch may occur, so a global continuation in
$\chi$ should not be assumed without further analysis.

The zero-count problem of Sec.~\ref{subsec:SMzeroCount} is another
boundary value of the same finite-length generating function.  At
fixed $L$,
\be
\lim_{t\to-\infty}
\mathbb E\!\left[e^{t\mathcal N_L(E)}\right]
=
S_N(E,L)
=
\Pr\{\mathcal N_L(E)=0\}.
\label{eq:SMzeroCountLimit}
\ee
This identity does not require, and should not be confused with, an
interchange of the limits $L\to\infty$ and $t\to-\infty$.  Solving
the killed problem therefore does not by itself determine the full
counting SCGF.  For $N>1$, after one Riccati eigenvalue passes through
infinity the remaining eigenvalues retain their configuration, so
successive passages do not in general form an independent renewal
process.

\subsection{Determinant moments and generalized Lyapunov exponents}

The second deformation concerns the growth of the Cauchy solution
$Y(x)$.  It is the multichannel analogue of the generalized Lyapunov
exponent studied for the scalar white-noise Schr\"odinger problem in
Ref.~\cite{FLRT2018}.  Between zeros of $\det Y$ one has
\be
\partial_x\log|\det Y(x)|
=
\Tr Z(x).
\label{eq:SMlogdetTiltIdentity}
\ee
This suggests defining the determinant generalized Lyapunov exponent
\be
\Lambda_N^{\rm amp}(q;E)
:=
\lim_{L\to\infty}\frac1L
\log
\mathbb E\!\left[|\det Y(L)|^q\right],
\label{eq:SMamplitudeGLE}
\ee
whenever the moment and the limit exist.

Formally, the corresponding Feynman--Kac deformation adds
$q\Tr Z$ to the backward many-body generator.  At $\beta=2$ the bulk
fermionization again survives, and the tilted one-body operator is
\be
\ell_{N,q}^{\dagger}
=
T\partial_\lambda^2
-(E+\lambda^2)\partial_\lambda
+(N-1+q)\lambda .
\label{eq:SMamplitudetilt}
\ee
For $N=1$ this reduces, up to our normalization conventions, to the
tilted Riccati operator entering the scalar generalized-Lyapunov
problem of Ref.~\cite{FLRT2018}.

For an integer $q=m\geq0$,
$\ell_{N,m}^{\dagger}$ preserves the polynomial space
$\mathcal P_{N+m-1}$.  Thus signed integer determinant moments possess
a finite-dimensional algebraic sector, generalizing the corresponding
integer-moment closure of the scalar problem.  This algebraic sector
does not, however, by itself determine the physical absolute-moment
problem for general $q$.

The difficulty is global rather than local.  The identity
\eqref{eq:SMlogdetTiltIdentity} holds between zeros of $\det Y$, where
one or more Riccati eigenvalues pass through infinity.  Continuation
through such a passage must be inherited from the regular matrix
solution $Y$, equivalently from the appropriate projective boundary
matching.  This issue is already essential in the scalar generalized
Lyapunov problem~\cite{FLRT2018}.

For integer $m$ the relation between signed and absolute moments makes
the coupling to the spectral count explicit:
\be
(\det Y(L))^m
=
|\det Y(L)|^m
(-1)^{m\mathcal N_L(E)},
\label{eq:SMsignedAbsoluteMoment}
\ee
with the Dirichlet normalization fixing the initial sign.  Hence for
even $m$ the signed and absolute moments coincide, whereas for odd
$m$ the signed moment contains the additional spectral-parity factor
$(-1)^{\mathcal N_L(E)}$.  For general real $q$, the admissible
boundary realization and the integrability of the absolute moment
must be specified separately.  The amplitude and counting tilts are
therefore naturally coupled at the projective boundary.

\subsection{Joint spectral and amplitude generator}

The counting tilt $t$ and the generalized-Lyapunov tilt $q$ probe,
respectively, spectral winding and the growth of the Cauchy solution.
Their natural joint scaled generating function is
\be
\Lambda_N(q,t;E)
:=
\lim_{L\to\infty}\frac1L
\log\mathbb E\!\left[
|\det Y(L)|^q
e^{t\mathcal N_L(E)}
\right].
\label{eq:SMjoint}
\ee
Where this limit is differentiable at the origin, its first
derivatives recover the two observables calculated in the stationary
problem,
\be
\left.
\partial_q\Lambda_N(q,t;E)
\right|_{q=t=0}
=
\Gamma_N(E),
\qquad
\left.
\partial_t\Lambda_N(q,t;E)
\right|_{q=t=0}
=
\nu_N(E),
\label{eq:SMjointFirstDerivatives}
\ee
where $\Gamma_N(E)$ is the total Lyapunov exponent and $\nu_N(E)$ the
IDOS per unit length.  The second derivatives would give the
asymptotic variance rates of
$\log|\det Y(L)|$ and $\mathcal N_L(E)$, together with their mixed
covariance,
\bea
\left.
\partial_q^2\Lambda_N
\right|_{0}
&=&
\lim_{L\to\infty}
\frac{
\operatorname{Var}\log|\det Y(L)|
}{L},
\nonumber\\
\left.
\partial_t^2\Lambda_N
\right|_{0}
&=&
\lim_{L\to\infty}
\frac{
\operatorname{Var}\mathcal N_L(E)
}{L},
\nonumber\\
\left.
\partial_q\partial_t\Lambda_N
\right|_{0}
&=&
\lim_{L\to\infty}
\frac{
\operatorname{Cov}
\!\left(
\log|\det Y(L)|,\mathcal N_L(E)
\right)
}{L},
\label{eq:SMjointSecondDerivatives}
\eea
whenever these limits exist.

The counting-only boundary problem has been specified in
Sec.~\ref{subsec:SMcountingPropagator}, and the zero-count sector
$q=0$, $t\to-\infty$ is the killed problem analyzed in
Sec.~\ref{subsec:SMzeroCount}.  The general nonzero-$q$ projective
matching, and hence the evaluation of the complete joint generating
function~\eqref{eq:SMjoint}, are not pursued here.  The companion
finite-length calculation~\cite{FLDGap} addresses the zero-count
sector at $q=0$ rather than these broader joint fluctuations.


\begin{thebibliography}{99}
\bibitem{footnote1} With this convention, positive $\mathcal J_i$ is oriented toward decreasing
$\lambda_i$, namely in the physical winding direction
$+\infty\to-\infty$.  The conventional probability current oriented toward
increasing $\lambda_i$ is $-\mathcal J_i$.
\bibitem{footnoteCurrent}
In the ordered chamber every projective winding crosses the single boundary
face $\lambda_1=-\infty$; after reinsertion at $+\infty$ the ordered labels
are cyclically permuted. Hence this boundary flux is already the total
winding rate, with no additional factor of $N$.
\bibitem{footnoteZ}
The normalization on the ordered chamber is
$\cZ_N=\prod_{a=1}^N g_a$, where
$g_a:=\int_{\mathbb R}p_a(\lambda)v_a(\lambda)\,\dd\lambda$.
For the normalized symmetric extension to $\mathbb R^N$ one instead has
$\cZ_N^{\rm sym}=N!\cZ_N$.
\bibitem{footnote2} The Hadamard finite part FP
 is only a compact way of writing a regular function: apparent poles at the zeros of $p_a$ cancel between the primitive and the polynomial prefactor. 
 A no-pole first-order construction, which is also the preferred numerical representation, is given in~\cite{SM}.
\end{thebibliography}
\end{document}